\documentclass{aa}  

\usepackage{twoopt}
\usepackage[breaklinks=true]{hyperref} %% to avoid \citeads line fills
\bibpunct{(}{)}{;}{a}{}{,} %% natbib format for A&A and ApJ
\makeatletter
\newcommandtwoopt{\citeads}[3][][]{\href{http://adsabs.harvard.edu/abs/#3}%
{\def\hyper@linkstart##1##2{}%
\let\hyper@linkend\@empty\citealp[#1][#2]{#3}}}
\newcommandtwoopt{\citepads}[3][][]{\href{http://adsabs.harvard.edu/abs/#3}%
{\def\hyper@linkstart##1##2{}%
\let\hyper@linkend\@empty\citep[#1][#2]{#3}}}
\newcommandtwoopt{\citetads}[3][][]{\href{http://adsabs.harvard.edu/abs/#3}%
{\def\hyper@linkstart##1##2{}%
\let\hyper@linkend\@empty\citet[#1][#2]{#3}}}
\newcommandtwoopt{\citeyearads}[3][][]%
{\href{http://adsabs.harvard.edu/abs/#3}
{\def\hyper@linkstart##1##2{}%
\let\hyper@linkend\@empty\citeyear[#1][#2]{#3}}}
\makeatother   
\usepackage{graphicx}
\usepackage{siunitx}
\usepackage{subcaption}
\usepackage{adjustbox}
\usepackage{makecell}
\usepackage[usenames,dvipsnames]{xcolor}
\usepackage[normalem]{ulem}
\usepackage{multirow}
\usepackage{amsmath}
\usepackage{float}
\usepackage{longtable}
\usepackage{placeins} % FN: missing package for \FloatBarrier
\usepackage{soul}
\usepackage{txfonts}
\graphicspath{{figs/}}
\newcommand{\ttcotwoone}{\mbox{$^{13}$CO\,(2--1)}}
\newcommand{\ttcosixfive}{\mbox{$^{13}$CO\,(6--5)}}
\newcommand{\cosixfive}{\mbox{$^{12}$CO\,(6--5)}}

\newcommand{\ttcoeightseven}{\mbox{$^{13}$CO\,(8--7)}} 
\newcommand{\cetosixfive}{\mbox{C$^{18}$O\,(6--5)}}

\newcommand{\mumeter}{$\mu$m}

\begin{document} 

   \title{ATLASGAL-selected high-mass clumps in the inner Galaxy}
   \subtitle{XI. Morphology and kinematics of warm inner envelopes}

   \author{Thanh Dat Hoang\thanks{Member of the International Max Planck Research School (IMPRS) for Astronomy and Astrophysics at the Universities of Bonn and Cologne.}\inst{1}
          \and
          Min-Young Lee \inst{2,3}
          \and 
          Friedrich Wyrowski \inst{1} 
          \and 
          Agata Karska \inst{1,4,5}
          \and 
          Felipe Navarete \inst{6}
          \and
          Karl M. Menten \inst{1} 
          }

   \institute{Max-Planck-Institut f\"ur Radioastronomie, 
            Auf dem H\"ugel 69, 53121 Bonn, Germany\\
              \email{tdhoang@mpifr-bonn.mpg.de}
         \and
             Korea Astronomy and Space Science Institute, 776 Daedeok-daero, Yuseong-gu, Daejeon 34055, Republic of Korea
        \and 
             Department of Astronomy and Space Science, University of Science and Technology, 217 Gajeong-ro, Daejeon 34113, Republic of Korea
        \and 
            Argelander-Institut für Astronomie, Universität Bonn, Auf dem Hügel 71, 53121 Bonn, Germany
        \and 
            Institute of Astronomy, Faculty of Physics,
    Astronomy and Informatics, Nicolaus Copernicus University, ul. Grudziądzka 5, 87-100 Toruń, Poland
        \and
            SOAR Telescope/NSF's NOIRLab, Avda Juan Cisternas 1500, 1700000, La Serena, Chile
             }

 \date{Received XX; accepted XX}
\titlerunning{Mid-$J$ CO emission in high-mass clumps}
\authorrunning{T.~D.~Hoang et al.} 

% \abstract{}{}{}{}{} 
% 5 {} token are mandatory
 
  \abstract
  % context heading (optional)
   {High-mass stellar embryos are embedded in warm envelopes that provide mass reservoirs for the accretion process onto final stars. Feedback from star formation activities in return impacts on the properties of the envelopes, offering us a unique opportunity to investigate star formation processes.
   }
  % aims heading (mandatory)
   {Our goals are to characterise the properties of warm envelopes of proto- or young stellar objects in different evolutionary stages based on the morphology and kinematics of submillimetre emission from the \ttcosixfive{} line and to examine their relations with star formation processes.
   }
  % methods heading (mandatory)
    {Using the Atacama Pathfinder EXperiment (APEX) telescope, we obtained maps of \ttcosixfive{} emission with an angular size of 80$\arcsec$ $\times$ 80$\arcsec$ (ranging from 0.3\,pc $\times$ 0.3\,pc to 4.9\,pc $\times$ 4.9\,pc in physical size) of 99 massive clumps from the ATLASGAL survey of submillimetre dust continuum emission. Our maps are classified based on morphological complexities, and the radial structure of \ttcosixfive{} emission is characterised for simple single-core sources. In addition, the velocity centroids of \ttcosixfive{} emission are compared to small- and large-scale gas kinematics (traced by \cosixfive{} and \ttcotwoone{} emission, respectively), aiming to shed light on the origin of envelope kinematics.
    }
  % results heading (mandatory)
   {\ttcosixfive{} emission is detected towards sources in all stages of high-mass star formation, with a detection rate of 83\% for the whole sample. The detection rate, line width, and peak brightness temperature increase with evolutionary stage, and the line luminosity is strongly correlated with the bolometric luminosity and the clump mass. These results indicate that the excitation of \ttcosixfive{} emission is closely related to star formation processes. In addition, the radial distributions of \ttcosixfive{} emission for single-core sources can be well fitted by power-law functions, suggesting a relatively simple envelope structure for a majority of our sources (52 out of 99). The slopes of the radial distributions are systematically steeper for the most evolved group of sources (that host H\textsc{ii} regions), which likely results from enhancements in density and/or temperature at the central parts of the warm envelopes. As for the \ttcosixfive{} kinematics, linear velocity gradients are common among the single-core sources (44 out of 52), and the measured mean velocity gradients (MVGs) are on average 3\,km\,s$^{-1}$\,pc$^{-1}$. Our comparison of the \ttcosixfive{}, \cosixfive{}, and \ttcotwoone{} kinematics suggests that the origin of the linear velocity gradients in the warm envelopes is complex and unclear for many sources.
   }
   % conclusions heading (optional), leave it empty if necessary 
   {\ttcosixfive{} emission is ubiquitous in a wide variety of massive clumps, ranging from young sources where protostars have not yet been formed to evolved sources with fully developed H\textsc{ii} regions. The excitation of \ttcosixfive{} emission in warm envelopes is likely impacted by different processes at different epochs of high-mass star formation, while the origin of the \ttcosixfive{} kinematics remains elusive and needs further investigation.
   }


   \maketitle
%
%________________________________________________________________

\section{Introduction}
High-mass stars ($M$ $>$ 8 $M_{\odot}$) play an essential role in the evolution of galaxies \citep[e.g.,][]{kennicutt05}. They provide a substantial amount of radiative and mechanical feedback to the surrounding interstellar medium (ISM) through ultraviolet (UV) radiation fields, stellar outflows/winds, and supernova explosions, regulating star formation in galaxies \citep{hopkins14}. In addition, they produce and release heavy elements throughout their lifetime, driving the chemical enrichment of galaxies \citep{matteucci21}.

Despite their importance for the evolution of galaxies, how high-mass stars form and interact with their surroundings remains poorly understood \citep[e.g.,][]{ zinnecker07,motte18}. Current competing theoretical concepts of high-mass star formation include collapsing turbulent cores \citep{mckee03}, competitive accretion in stellar clusters \citep{bonnell01,bonnell06}, global hierarchical collapse \citep{vazquez19}, and converging inertial flows \citep{padoan20}. Evaluating these different theories rigorously requires a statistically significant sample of high-mass star-forming regions, covering a wide range of evolutionary stages. Such a sample has been provided by the APEX Telescope Large Area Survey of the Galaxy (ATLASGAL; \citealt{schuller2009atlasgal}). ATLASGAL is a survey of 870\,$\mu$m emission in the inner Galaxy (covering $|l| < 60^{\circ}$ at $|b| < 1.5^{\circ}$ and $280^{\circ} < l < 300^{\circ}$ at $-2^{\circ} < b < 1^{\circ}$) conducted with the Atacama Pathfinder EXperiment \citep[APEX\footnote{This publication is based on data acquired with the Atacama Pathfinder Experiment (APEX). APEX is a collaboration between the Max-Planck-Institut fur Radioastronomie, the European Southern Observatory, and the Onsala Space Observatory.} ;][]{apex} 12 m submillimetre telescope in Chile. With an excellent sensitivity of 50--70 mJy beam$^{-1}$ and a large coverage of $\sim$420 deg$^{2}$, ATLASGAL provides a comprehensive census of high-mass star-forming regions in the inner Galaxy.

Within high-mass star-forming regions, dense envelopes are of particular interest as they provide mass reservoirs that feed the mass build up of newly formed protostellar objects at their centres through accretion \citep{shu1987star,wyrowski2016infall,beltran2016accretion,pillai23}. Once young stellar objects (YSOs) have formed, stellar feedback significantly changes the properties of the surrounding envelopes, so much so that they can be entirely dispersed at the end of the evolution, stopping the accretion process \citep[e.g.,][for low-mass star-forming regions]{arce2006evolution}. Among other tracers, mid-$J$ ($5 < J < 10$) rotational transitions of isotopologues of carbon monoxide (CO) are especially suitable to probe the warm ($T \gtrsim 50$\,K) inner parts of the envelopes. Previous studies have suggested that the warm envelopes surrounding low- and high-mass YSOs can be heated by various physical processes that are related to star formation activities. For example, passive heating and UV radiation from outflow cavity walls have been proposed to explain \ttcosixfive{} emission in low-mass star-forming regions \citep{vankempen09I,yildiz12}. Similarly, UV photons from photodissociation regions (PDRs) have been shown to play an important role in the excitation of \ttcosixfive{} emission in several high-mass star-forming clumps with strong H\textsc{ii} regions \citep{graf93,koe94,leurini2013distribution}.

While there have been certainly a number of dedicated studies of the warm envelopes, mid-$J$ transitions of CO isotopologues still remain difficult to access observationally due to the required dry atmospheric conditions, and there is particularly a lack of studies across a wide range of high-mass star-forming regions (including those in the early stages before the development of H\textsc{ii} regions). In this paper, we examine the properties of \ttcosixfive{}-traced warm envelopes for a sample of 99 ATLASGAL-selected high-mass star-forming regions based on APEX observations with high spatial and spectral resolutions. In particular, we focus on probing the morphology and kinematics of \ttcosixfive{} emission and how they change along evolutionary stages, with an aim of providing insight into how high-mass stars form and their impact on their environments.

This paper is organised as follows: In Sect.\,\ref{s:obs}, we describe our sample, the APEX \ttcosixfive{} observations, and the derivation of moment maps. In Sect.\,\ref{s:results}, we present \ttcosixfive{} line properties and our classification of the sources based on the integrated intensity maps. Subsequently, we investigate the morphology and kinematics of the warm envelopes traced by \ttcosixfive{} emission (Sect.\,\ref{s:analyses}) and discuss our results in the context of previous studies (Sect.\,\ref{s:discussion}). Finally, we finish this paper by presenting a summary of our main results (Sect.\,\ref{s:conclusions}).

%__________________________________________________________________
\section{Observations and data reduction} \label{s:obs}
\subsection{Top100 sample}
The ATLASGAL survey has identified a collection of $\sim$10$^{4}$ candidates for high-mass star-forming regions in the inner part of the Milky Way \citep{schuller2009atlasgal,urq22}. Among these sources, a sample of the $\sim$110 brightest objects (Top100 sample hereafter) was selected by \citet{giannetti2014atlasgal} for follow-up studies using various dust and gas tracers, namely dust continuum emission: \citealt{konig2017atlasgal}; shocked gas traced by SiO \citealt{cgeri16}; outflows traced by mid-$J$ CO lines \citealt{navarete2019atlasgal}; neutral carbon ([C~\textsc{i}])-traced gas \citealt{MY22}. The 99 sources selected for this study are part of the Top100 sample and can be categorised into four evolutionary groups based on their infrared (IR) and radio continuum emission \citep{konig2017atlasgal}:

\begin{itemize} 
\item 14 in the starless/pre-stellar (far-IR 70\,\mumeter{} weak group, 70w): these comprise sources in the earliest phase, in which collapse might have already started, but no protostellar object has yet been formed, according to the lack of compact 70\,\mumeter{} far-IR emission detected in the Hi-GAL survey \citep{molinari2010hi}.

\item 30 in the protostellar (near/mid IR weak group, IRw): these objects are identified by compact 70\,\mumeter{} emission. However, they are not yet associated with a strong and compact mid-IR emission source, implying that they are dominated by cool gas. 

\item 35 in the high-mass protostellar (mid IR bright group, IRb): the gas around protostellar objects becomes hotter as the sources appear luminous at 8\,\mumeter{} and 24\,\mumeter{}. 

\item 20 in the compact H\textsc{ii} region group (HII): these can be considered to be in the most evolved phase, in which stellar objects start to disperse their natal envelopes. Intense stellar radiation ionises hydrogen atoms and forms ultra- and, later, compact H\textsc{ii} regions which are associated with compact radio continuum emission.
\end{itemize}

The physical properties of our 99 objects were determined by fitting dust spectral energy distributions from mid-IR to sub-mm wavelengths \citep{konig2017atlasgal}. The results show that the dust temperature ($T_{\mathrm{dust}}$) varies from 11 to 41\,K from the coldest to the warmest objects. In addition, the range of bolometric luminosity, $L_{\mathrm{bol}}$, spans five orders of magnitude from 57 $L_{\odot}$ to $3.7 \times 10^6$ $L_{\odot}$, and the clump mass, $M_{\textrm{clump}}$, ranges from 17 to \mbox{$4.3 \times 10^4$ $M_{\odot}$}. Most of our sources can form at least one massive star \citep{konig2017atlasgal} according to the mass-size limit proposed by \citet{kauffmann2010mass}, indicating that they constitute a statistically significant sample of massive star-forming regions. Finally, we note that the distance distribution ($\sim$1--13 kpc) is comparable amongst the different evolutionary groups \citep{konig2017atlasgal}, implying no distance-related bias in our comparison of the groups (Sects. \ref{s:results} and \ref{s:analyses}).

\subsection{CHAMP$^+$ observations and data reduction}
\begin{figure}
    \centering
    \resizebox{0.7\hsize}{!}{\includegraphics{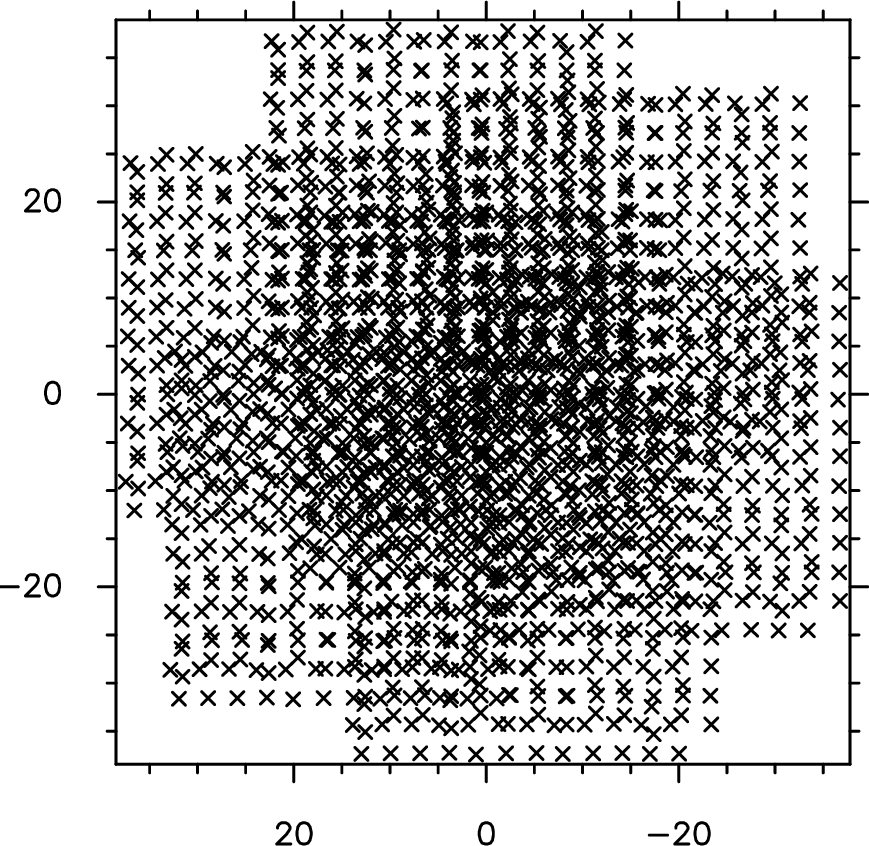}}
    \caption{Example of pointing positions for a single source in our OTF observations. The image is centred on the source and is in the units of arcsecond.}
    \label{fig:dump-pos}
\end{figure}

\begin{table*}
\begin{center} 
\caption{Parameters of the CHAMP$^{+}$ observations} 
\label{t:observation}     
\renewcommand{\footnoterule}{}  
\begin{tabular}{c c c c c c c c c c}     
\hline\hline       
Molecule & Transition\tablefootmark{a} & Frequency\tablefootmark{a} & $E_\mathrm{u}$/$k_\mathrm{B}$\tablefootmark{a} & $A_\mathrm{u}$\tablefootmark{a} & $g_\mathrm{u}$\tablefootmark{a}  & Beam\tablefootmark{b}   & $\Delta V$\tablefootmark{b}\\
~ & $J_\mathrm{u}-J_\mathrm{l}$ & (GHz) & (K) & (10$^{-5}$ s$^{-1}$) & ~ & ('') & (km s$^{-1}$)\\
\hline          
$^{13}$CO & $6-5$ & 661.067277  & 111.05 & 1.9 & 13 & 9 & 0.33 \\
\hline
\end{tabular}
\begin{flushleft}
\tablefoot{
\tablefoottext{a}{Molecular data adopted from the Cologne Database for Molecular Spectroscopy\footnote{https://cdms.astro.uni-koeln.de} \citep[CDMS,][]{muller2001-cdms,muller2005}. The frequency has been determined by \citep{Klapper2000}. $E_\mathrm{u}$ is the energy of the upper level, $k_\mathrm{B}$ is the Boltzmann constant, $A_\mathrm{u}$ is the Einstein A coefficient, and $g_\mathrm{u}$ is the statistical weight of the upper level.}
\tablefoottext{b}{Original angular resolution (full width at half maximum (FWHM) beam width) and spectral resolution.}
} 
\end{flushleft}
\end{center} 
\end{table*}

The observations of \ttcosixfive{} emission towards our sample were carried out using the Carbon Heterodyne Array of the MPIfR \citep[CHAMP$^+$;][]{kasemann2006champ, gusten2008submillimeter} on the APEX telescope in 2013 and 2014. The data were obtained under the project ``Probing the warm and dense envelopes through the evolution of massive protostars'' (project IDs: M0035-91 from June to August 2013, M0010-92 in October 2013, and M0027-93 in May 2014).

CHAMP$^+$ is a $2\times7$ element dual-colour heterodyne receiver array that can simultaneously observe tunings in the frequency ranges of 620--720 GHz in the low (LFA) and 780--950 GHz in the high frequency array (HFA). The Array Fast Fourier Transform Spectrometer (A-FFTS) \citep{gusten2008submillimeter} serves as the backend for CHAMP$^+$, and the CHAMP$^+$/A-FFTS IF system is set up to process 2.8\,GHz of instantaneous bandwidth over 16.4\,k channels that are 212\,kHz each \citep{gusten2008submillimeter}. The LFA was tuned to cover both the \ttcosixfive{} and \cetosixfive{} transitions, while the HFA was tuned to cover the \ttcoeightseven{} line. The excellent weather with low precipitable water vapour content (0.24--1.07 mm) allowed us to observe these high frequency lines. The sources were scanned in the on-the-fly (OTF) mode, resulting in data cubes with spatial dimensions of $80\arcsec$ $\times$ $80\arcsec$ with an original angular resolution of 9$\arcsec$ at 661 GHz. The 7-pixel hexagonal array enables fast mapping and improves the signal-to-noise ratio towards the map centre where the pixels' coverages overlap (Fig.\,\ref{fig:dump-pos}). The parameters of our observations are presented in Table \ref{t:observation}.

For the present study, we focused on the \ttcosixfive{} observations and analysed the \ttcosixfive{} data using the CLASS package of the GILDAS\footnote{https://www.iram.fr/IRAMFR/GILDAS/} software. Specifically, the obtained spectra were converted into the main-beam temperature, $T_{\text{MB}}$, scale via \mbox{$T_{\text{MB}}$ = ($F_{\mathrm{eff}}/B_{\mathrm{eff}}$)/$T_{\mathrm{A}}^{\ast}$,} where $F_{\mathrm{eff}}$ is the forward efficiency of 0.95, $B_{\mathrm{eff}}$ is the main-beam efficiency of 0.43\footnote{https://www.mpifr-bonn.mpg.de/4480868/efficiencies} obtained from a Mars observation, and $T_{\mathrm{A}}^{\ast}$ is the corrected antenna temperature. The spectra were smoothed to a velocity resolution of 0.35 km s$^{-1}$. Data cubes were then created by the CLASS gridding routine \texttt{XY\_Map} with a pixel size of 4.8\arcsec, where a Gaussian profile of one-third of the beam size is convolved with the gridded data to yield a final angular resolution of 10\arcsec.

The \ttcosixfive{} spectra of five sources exhibit narrow negative features in velocity channels close to the peaks at sources' velocities, which indicates the presence of emission in the reference positions. To correct for this contamination, we estimated average spectra within 20\arcsec of outer parts of the maps where only those negative features were observed and added their reversed profiles to the original spectra. This approach of utilising the average spectrum, as opposed to a single spectrum, minimises the noise one would add to the data during the correction process. The five sources affected by the contaminated reference positions are as follows: G305.21$+$0.2, G320.88$-$0.4, G337.17$-$0.0, G338.78$+$0.4, and G351.13$+$0.7.

\subsection{Moment maps} \label{s:mmtmaps}
We produced zeroth and first moment maps from the \ttcosixfive{} data cubes, which show the distributions of the integrated intensity and velocity centroid on the sky. These maps are valuable tools for examining the morphology and kinematics of warm gas in the inner envelopes. For each pixel, zeroth and first moment values were calculated as follows: 
\begin{align}
M_0 &= \int T_i \, dv,\\
M_1 &= \frac{\int T_i v_i \, dv}{\int T_i \, dv},
\end{align}
where $T_i$ and $v_i$ are the main-beam brightness temperature and velocity at each channel. To determine the velocity range for the integration of $M_0$ and $M_1$, an average spectrum was estimated from the whole data cube for each source using CLASS (individual spectra were weighted by observing times), and the range of velocity channels over which emission is clearly detected above 3$\sigma$ ($\sigma$ is the spectral noise) was determined based on visual inspection. During the integration for $M_0$ and $M_1$, velocity channels below 3$\sigma$ were not included. Similarly, pixels whose peak-to-noise ratios are lower than three were considered as non-detections and were masked in our final maps.

%__________________________________________________________________
\section{Results} \label{s:results}
\subsection{Detection statistics} \label{s:detection_rate}
In this section, we analyse the detection of \ttcosixfive{} emission in our sources. For our analysis, we derived an average spectrum over the central region with a size of 20\arcsec{} (close to one ATLASGAL beam and noise level is improved (Fig.\,\ref{fig:dump-pos})) for each source and considered a line to be detected if the peak main-beam brightness temperature was higher than 3$\sigma$.

\begin{figure}
    \centering
     \resizebox{0.85\hsize}{!}{\includegraphics{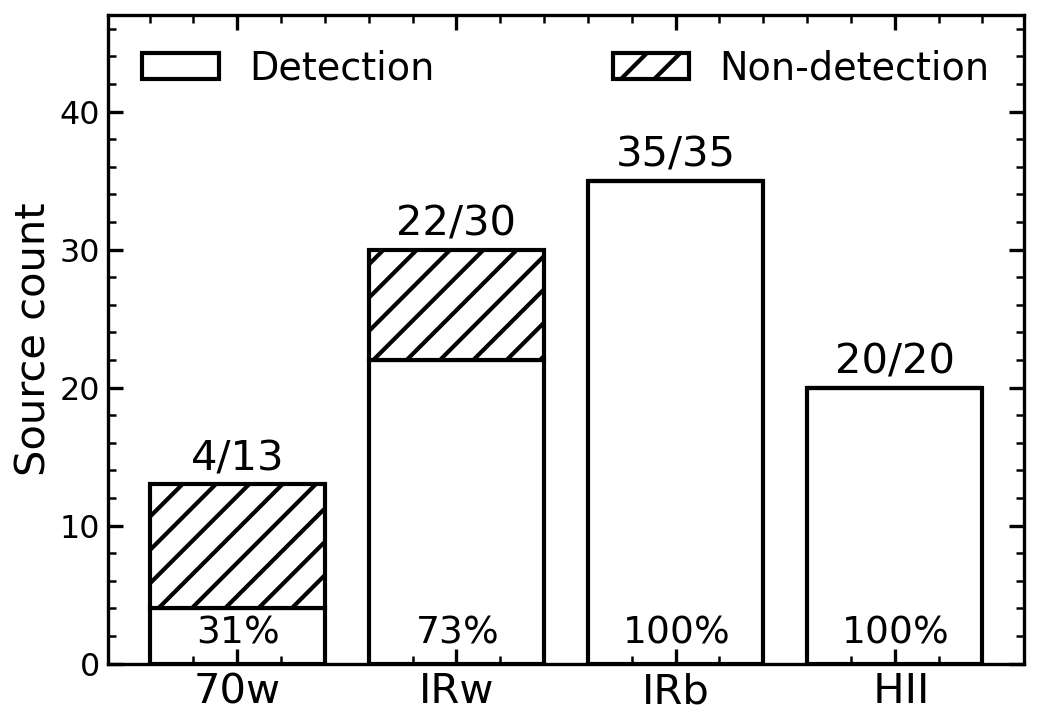}}
    \caption{Detection statistics of \ttcosixfive{} emission for the evolutionary groups. The exact number of the detected sources for the observed clumps is presented at the top of each bar, and the corresponding detection rates are shown in percentage terms.}
    \label{fig:detect13co65}
\end{figure}

\begin{figure}
    \centering
     \resizebox{0.85\hsize}{!}{\includegraphics{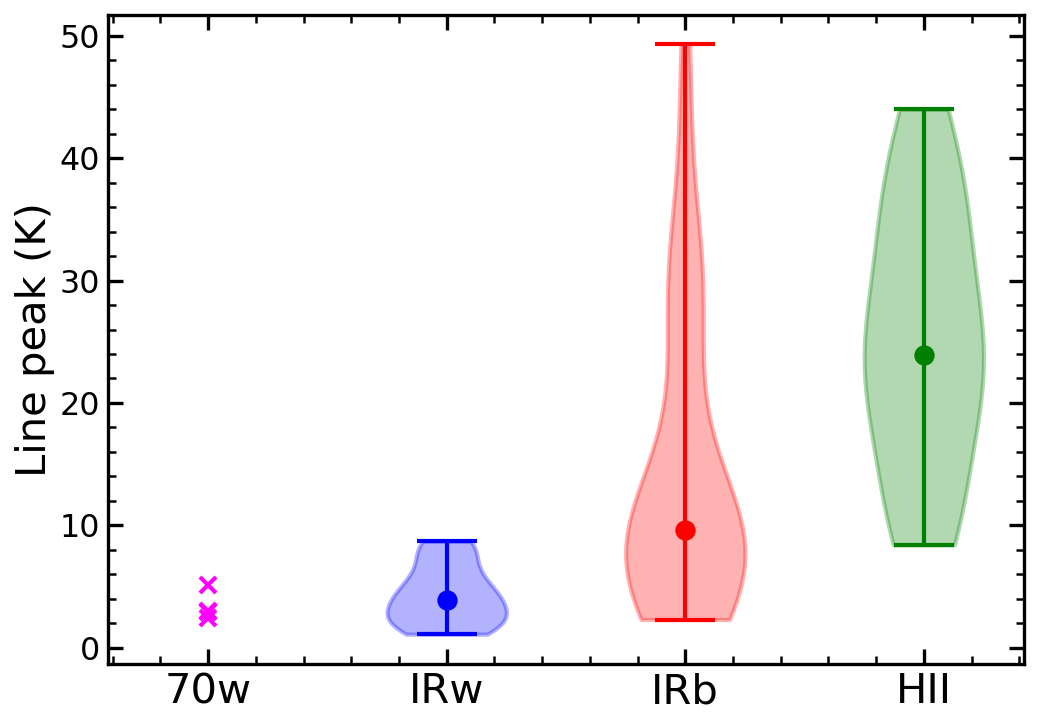}}
    \caption{Distributions of the peak brightness temperatures. These temperatures were obtained from the average spectra within the 20\arcsec{}-size central regions. For each evolutionary group, the minimum and maximum are shown as bars, while the median values are indicated as filled circles. For the 70w group that has only a few entries, the values for the individual sources are presented as crosses.}
    \label{fig:linepeaks}
\end{figure}

We found that \ttcosixfive{} emission is detected towards 81 clumps with varying detection rates among the evolutionary groups (Fig. \ref{fig:detect13co65}). Specifically, the detection rate increases from 31\%\footnote{G353.42$-$0.0 (70w) shows \ttcosixfive{} emission at $-$17.2 km\,s$^{-1}$, which is different from the source velocity of $-$54.4 km\,s$^{-1}$, and was excluded from our analyses.} for the youngest group (70w) to 70--100\% for more evolved groups (IRw, IRb and HII) and is 83\% for the whole sample. Similarly, the peak main-beam brightness temperature tends to increase towards the evolved source groups (Fig. \ref{fig:linepeaks}). This increase in the detection rate and peak main-beam brightness temperature could indicate increasing density and/or temperature along the evolutionary sequence of high-mass star formation, considering that the enhancement in density and/or temperature is conducive to more easily exciting $^{13}$CO molecules to mid-$J$ rotational levels. At the same time, the warmer environment could allow more frozen $^{13}$CO on dust grains to sublimate, enhancing the abundance of $^{13}$CO molecules in the gas and the chance to detect mid-$J$ $^{13}$CO emission. Indeed, \citet{giannetti2017atlasgal} noted a difference between the IRw and IRb groups, finding CO to be  mainly locked on dust grains in the former stage, while in the latter it exists mostly in the gas phase.

\subsection{Line profiles} \label{s:line_profiles}
In this section, we examine the general shape of \ttcosixfive{} line profiles based on the representative spectra of the different evolutionary groups. To produce these representative spectra, we shifted the average spectra of the central 20\arcsec{} regions to 0\,km\,s$^{-1}$ according to C$^{17}$O\,(3$-$2)-based source velocities \citep{giannetti2014atlasgal} and calculated the average of all sources in each evolutionary group. Four clumps without C$^{17}$O\,(3$-$2) data (two IRw, one IRb, and one HII) and one 70w source whose \cosixfive{} spectra heavily affected by a contaminated reference position were excluded from our calculation. For comparison, we also produced representative \cosixfive{} spectra in the same manner by using APEX \cosixfive{} observations from \citet{navarete2019atlasgal}.

\begin{figure*}
  \centering
   \begin{subfigure}{0.25\textwidth}
    \centering
    \includegraphics[width=\textwidth]{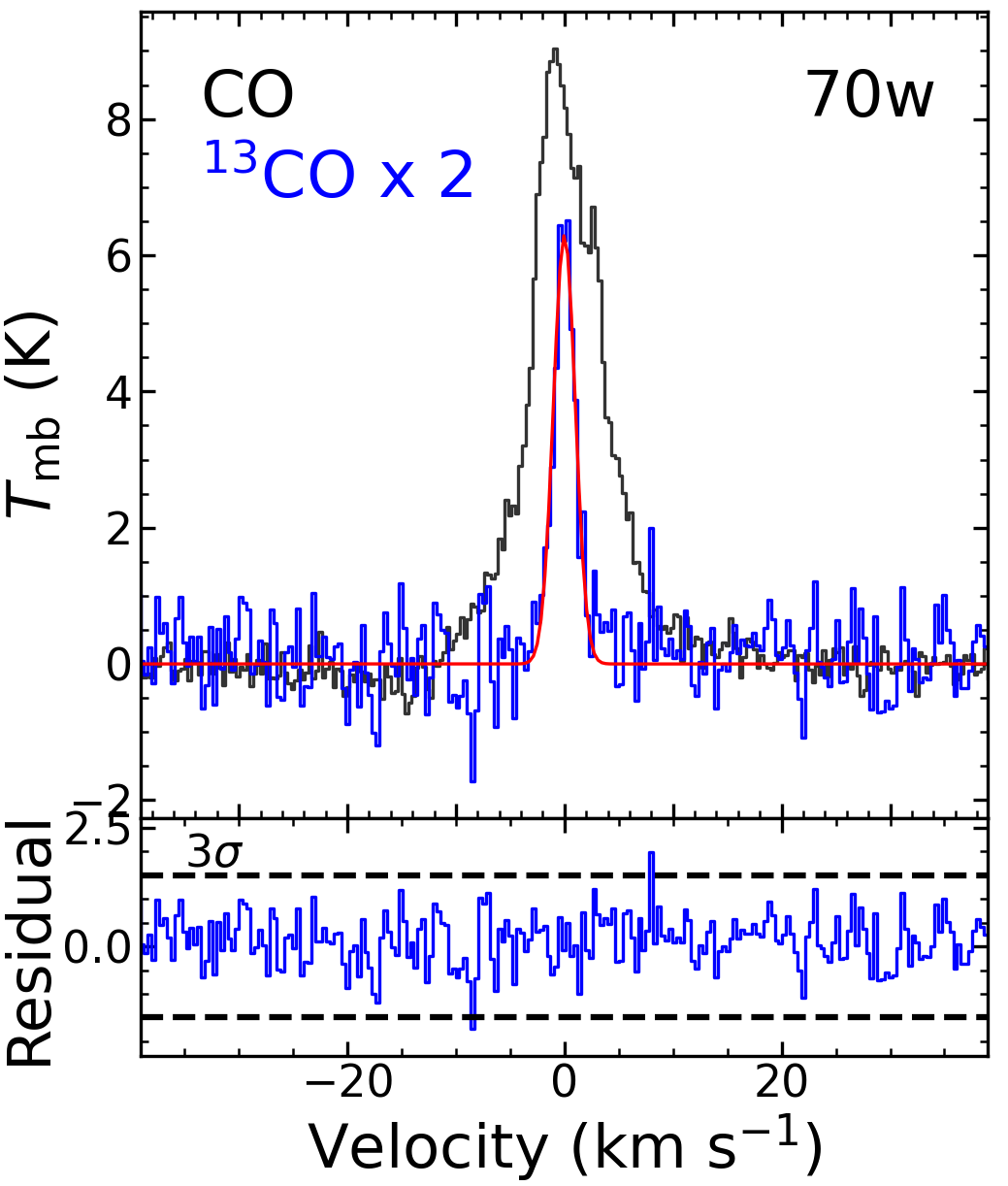}
  \end{subfigure}%
  \begin{subfigure}{0.25\textwidth}
    \centering
    \includegraphics[width=\textwidth]{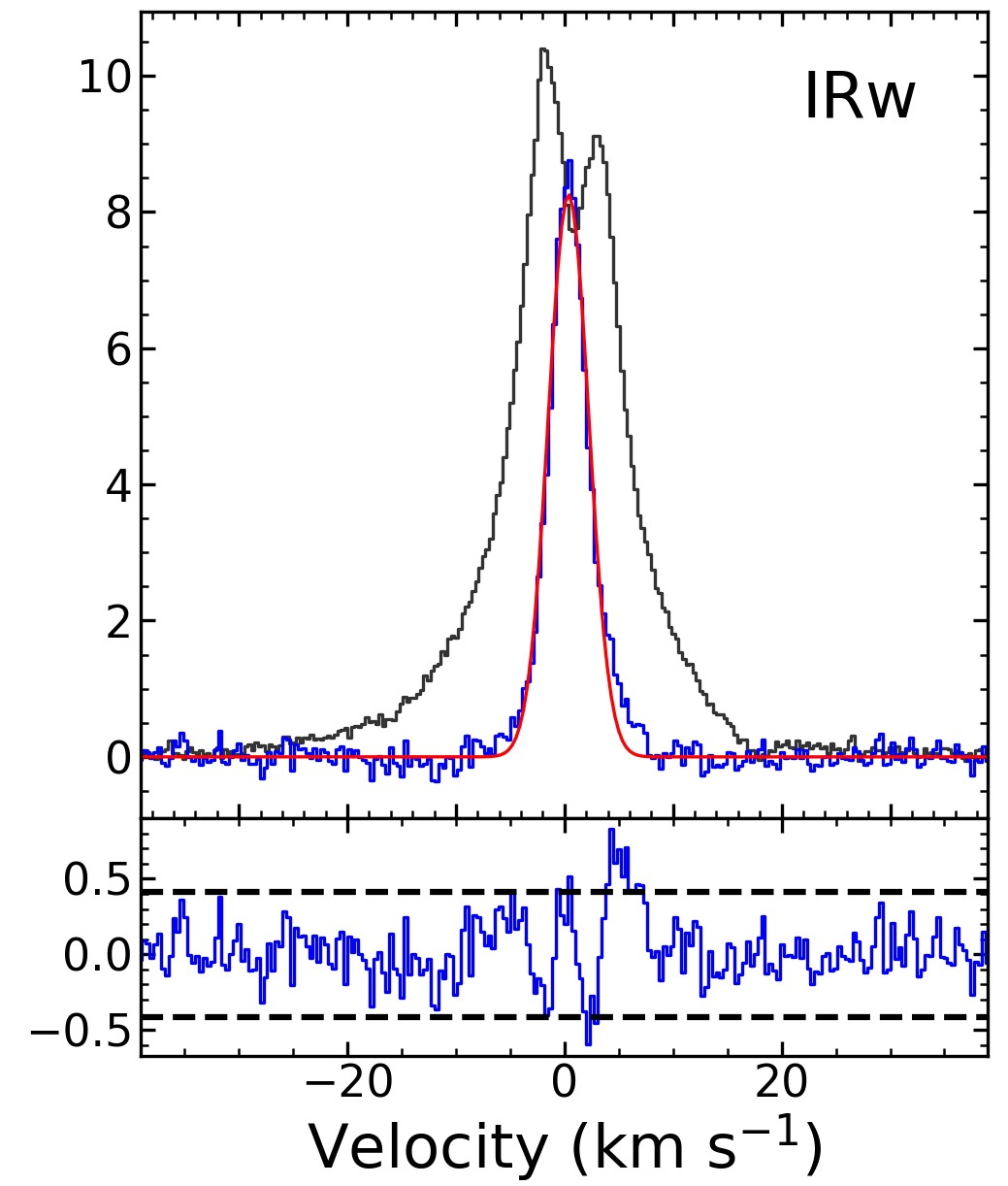}
  \end{subfigure}%
  \begin{subfigure}{0.25\textwidth}
    \centering
    \includegraphics[width=\textwidth]{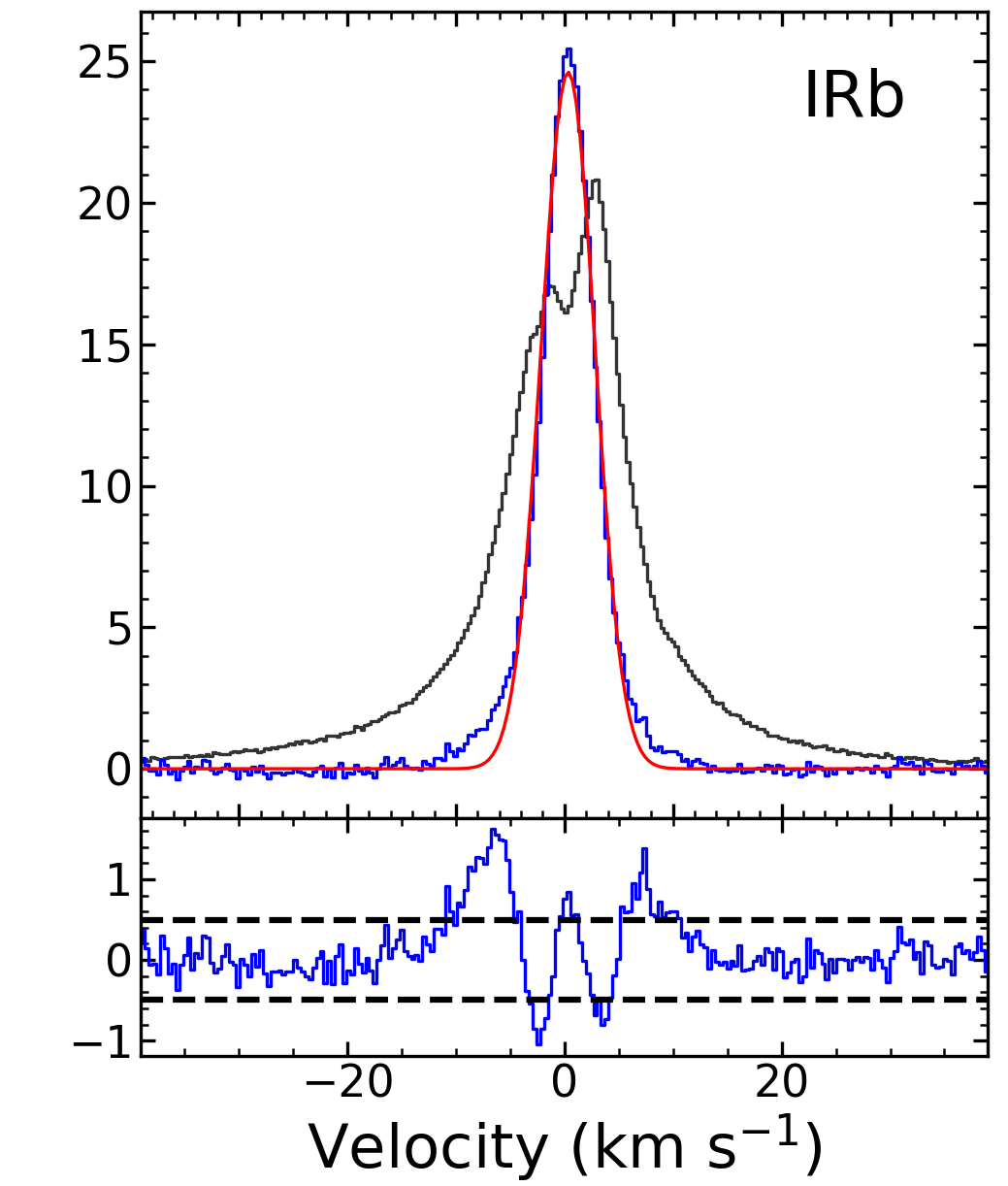}
  \end{subfigure}%
  \begin{subfigure}{0.25\textwidth}
    \centering
    \includegraphics[width=\textwidth]{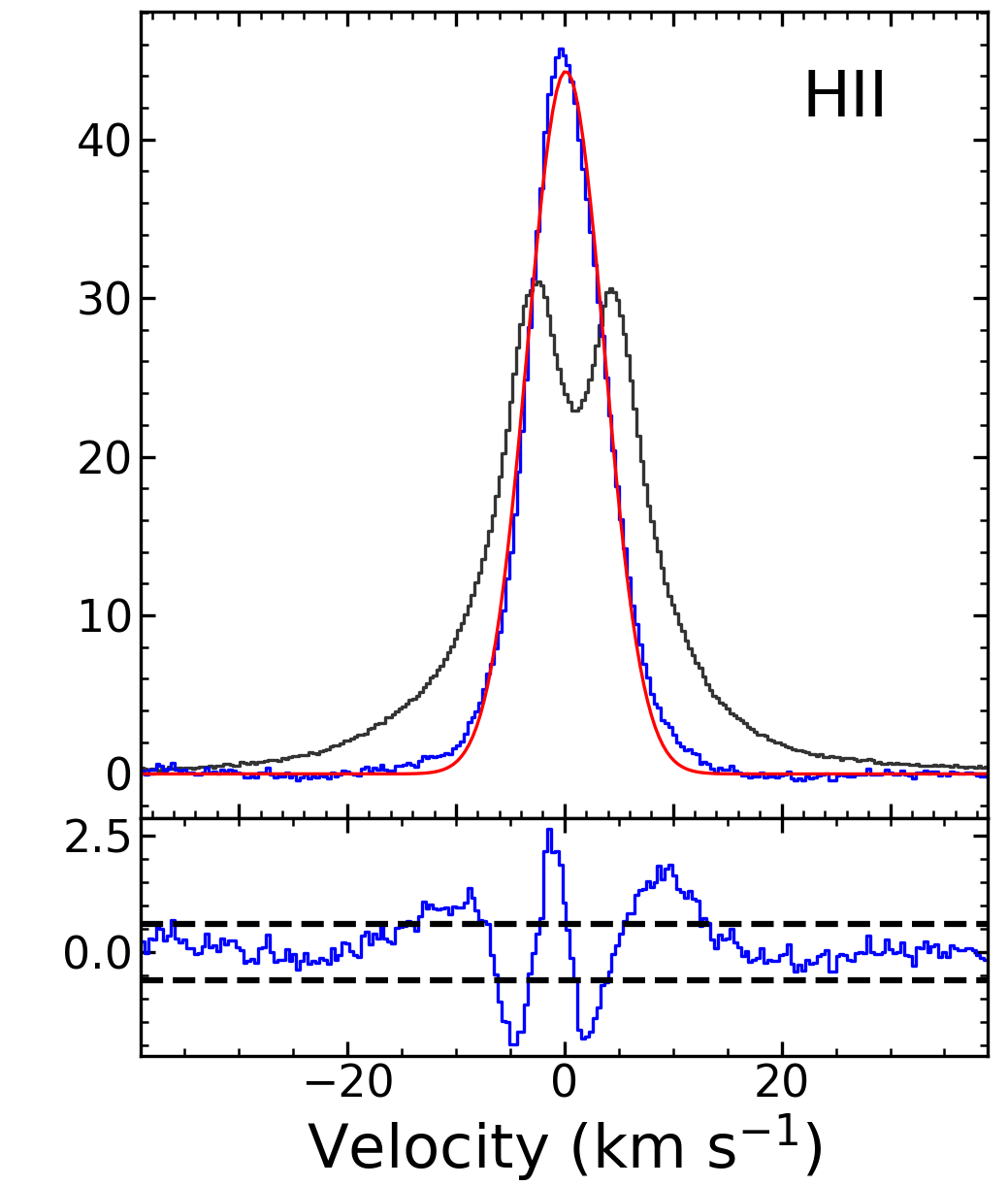}
  \end{subfigure}
  \caption{Representative \cosixfive{} (black) and \ttcosixfive{} (blue) spectra of the four evolutionary groups. The \ttcosixfive{} spectra are scaled up by a factor of two for an easier comparison, and the fitted Gaussian functions are overlaid as red curves. In the bottom panels, the residuals from Gaussian fitting are shown with $\pm$3$\sigma$ noise levels (dashed lines).}
  \label{fig:avgspectra}
\end{figure*}

Figure\,\ref{fig:avgspectra} shows that the \ttcosixfive{} and \cosixfive{} spectra have systematically different shapes. In general, the \ttcosixfive{} spectra are much narrower and are not severely affected by self-absorption, indicating that \ttcosixfive{} and \cosixfive{} emission likely traces different regions of (or different conditions within) the envelopes. For 81 sources with \ttcosixfive{} detection in our sample, \citet{navarete2019atlasgal} indeed estimated that only 15 sources are not affected by self-absorption in their \cosixfive{} spectra.
We fitted a single Gaussian function to each of the \ttcosixfive{} representative spectra (Fig.\,\ref{fig:avgspectra}) to quantify the line width and found FWHM values of 2.4, 4.4, 5.8, and 8.3\,km\,s$^{-1}$ for the 70w, IRw, IRb, and HII spectra, respectively. The increase in the FWHM is less likely affected by the opacity broadening effect as \ttcosixfive{} emission is mostly optically thin (Appendix\,\ref{a:tau13co}). Instead, it is closely linked to the increase in the source size (Sect.\,\ref{s:size_apr}; Larson's size--line width relation) and suggests that non-thermal motions become stronger in evolved sources.

While a single Gaussian represents the overall shape of the \ttcosixfive{} spectra reasonably well, residuals are visible at velocities of $\sim$5--10$\,\rm km\,s^{-1}$ for the IRw, IRb, and HII groups, hinting at the existence of a secondary component. In the case of low-mass YSOs, \ttcosixfive{} emission is considered to trace quiescent envelopes with an additional contribution from outflow cavity walls that are heated by UV photons, and/or from bow shocks or accretion disks \citep[e.g.,][]{spaans95,vankempen09II,vankempen09I}. Our finding is consistent with the low-mass YSO results in that it indicates a complex origin of \ttcosixfive{} emission, and we will further discuss possible heating mechanisms for \ttcosixfive{} emission in Sect.\,\ref{s:warm_gas_tracer}.

\subsection{Line luminosities} \label{s:line_luminosity}
\begin{figure}[h!]
    \centering
    \resizebox{0.85\hsize}{!}{\includegraphics{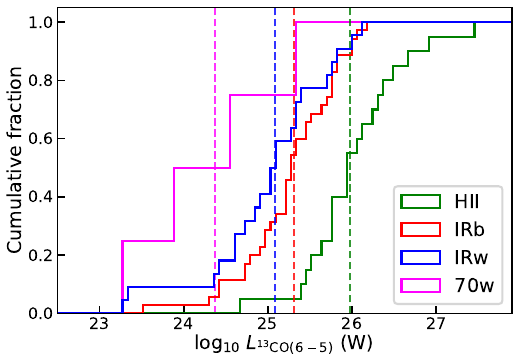}}
   \caption{Cumulative distributions of the \ttcosixfive{} line luminosities. These luminosities were derived from the spectra averaged over the central $20^{\prime\prime}$ regions. The median values of the four evolutionary groups are shown as dashed lines in different colours.}
   \label{fig:CDF_13CO_lum}
\end{figure}

\begin{figure*}
   \begin{subfigure}{0.5\hsize}
      \centering
      \resizebox{0.9\hsize}{!}{\includegraphics{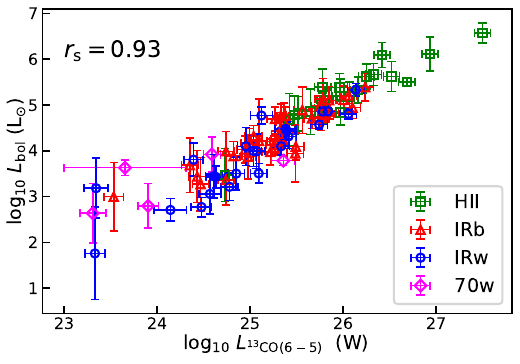}}
   \end{subfigure}%
   \begin{subfigure}{0.5\hsize}
      \centering
      \resizebox{0.9\hsize}{!}{\includegraphics{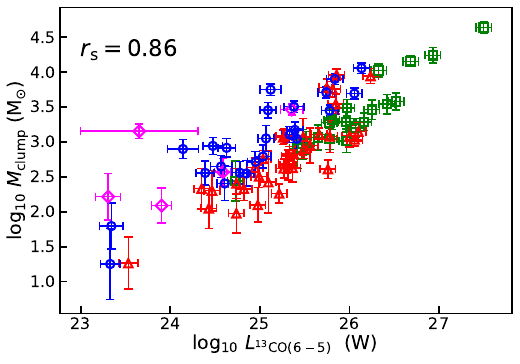}}
   \end{subfigure}
   \caption{\ttcosixfive{} line luminosity as a function of bolometric luminosity (left) and clump mass (right). The sources in the HII, IRb, IRw, and 70w groups are shown as green squares, red triangles, blue circles, and magenta diamonds, respectively. The Spearman's rank correlation coefficient ($r_{\mathrm{s}}$) between the two quantities is presented on each plot.}
    \label{fig:Lco_vs_clump_prop}
\end{figure*}

\begin{figure}[h!]
    \centering
    \resizebox{0.9\hsize}{!}{\includegraphics{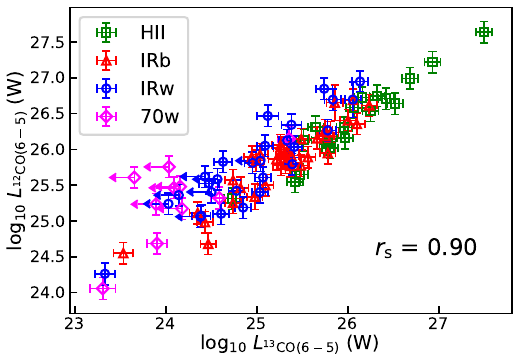}}
   \caption{Relation between the \ttcosixfive{} and \cosixfive{} line luminosities. The sources without \ttcosixfive{} detections are indicated as leftward arrows whose locations represent 3$\sigma$ values. The colours and markers are the same as in Fig.\,\ref{fig:Lco_vs_clump_prop}.}
   \label{fig:12CO_vs_13CO_lum}
\end{figure}

To investigate further how the properties of \ttcosixfive{} emission change with the evolution of high-mass star formation, we computed the line luminosity $L_{\mathrm{CO}}$ and examined correlations between $L_{\mathrm{CO}}$ and some of the key clump characteristics such as $L_{\mathrm{bol}}$ and $M_{\mathrm{clump}}$. The line luminosity was calculated from the average spectra of the central $20^{\prime\prime}$ regions based on the following equation: 
\begin{equation}
L_{\mathrm{CO}} = 4 \pi D^2 F_{\lambda},
\end{equation}
where $D$ is the distance to each source and $F_{\lambda}$ is the integrated intensity in W\,m$^{-2}$. The conversion from K km s$^{-1}$ to W m$^{-2}$ for $F_{\lambda}$ was done following Eq.\,(2) in \citet{indriolo2017co}: 
\begin{equation}
\frac{F_{\lambda}}{(\mathrm{W}\,\mathrm{m}^{-2})} = 
1.0248 \times 10^{-18} 
\left(\frac{\Omega}{\mathrm{sr}} \right) \left(\frac{\nu}{\mathrm{GHz}} \right)^3 
\left( \frac{M_{\rm 0}}{\mathrm{K}\,\mathrm{km}\, \mathrm{s}^{-1}} \right),
\end{equation}
where $\nu$ is the line frequency, $\Omega = \pi \theta^2/4\mathrm{ln}(2)$ is the solid angle occupied by a Gaussian beam with a size $\theta$ = $20^{\prime\prime}$. For the sources with detections, $M_{0}$ values were calculated over the same velocity ranges used in Sect. \ref{s:mmtmaps}. On the other hand, for the sources without detections, upper limits on $L_{\mathrm{CO}}$ were estimated based on the 3$\sigma$ values and the median velocity range for the sources with detections (27 km\,s$^{-1}$).

Figure \ref{fig:CDF_13CO_lum} presents the cumulative distributions of $L_{\mathrm{CO}}$ for the detected sources. From these distributions, we found that the median $L_{\mathrm{CO}}$ increases from the youngest to the most evolved group: $2.3\times10^{24}$\,W, $1.2\times10^{25}$\,W, $2.1\times10^{25}$\,W, and $9.3\times10^{25}$\,W for the 70w, IRw, IRb, and HII group, respectively. When \mbox{$k$-sample} Anderson-Darling tests were performed on each pair of the groups with sufficient numbers of sources (i.e., IRw, IRb, and HII) with a significance level of 0.05, all combinations except for the IRw-IRb were found to be drawn from different populations. These results suggest that the \ttcosixfive{} luminosities are closely related to the evolutionary state  of a source. As such, $L_{\mathrm{CO}}$ is also correlated with two of the other evolutionary sequence indicators, $L_{\mathrm{bol}}$ and $M_{\mathrm{clump}}$ \citep{konig2017atlasgal} (Fig. \ref{fig:Lco_vs_clump_prop}) with Spearman's rank correlation coefficients of 0.93 and 0.86, respectively.

In addition, we found that our \ttcosixfive{} luminosities are strongly correlated with the \cosixfive{} luminosities from \citet{navarete2019atlasgal} (Fig. \ref{fig:12CO_vs_13CO_lum}; Spearman's rank correlation coefficient of 0.90). The correlation spans several orders of magnitude and remain continuous out to the high luminosity regime, which is surprising as one may expect it to flatten once the \cosixfive{} emission becomes saturated due to its high opacity\footnote{The \cosixfive{} profiles often suffer from self-absorption (Fig.\,\ref{fig:avgspectra}).} while the \ttcosixfive{} emission is still optically thin (Appendix\,\ref{a:tau13co}). One plausible explanation for the continuous increase of the \cosixfive{} luminosity could be a non-trivial contribution of optically thin high-velocity emission arising from outflows to the total emission. If this were the case, the strong correlation between the \ttcosixfive{} and \cosixfive{} luminosities would suggest that similar mechanisms would partly power the emission in both transitions (Sect.~\ref{s:warm_gas_tracer}).

In summary, we found that \ttcosixfive{} emission is ubiquitous in massive clumps, including in the youngest sources that do not yet harbour protostars. The increases in the detection rate, peak brightness temperature, and line width determined for evolved sources indicate systematic changes in physical conditions, such as density, temperature, CO column density, and non-thermal motions,  along the sequence of high-mass star formation. Furthermore, the tight correlation of the \ttcosixfive{} luminosity with the \cosixfive{} luminosity, bolometric luminosity, and clump mass implies that the origin of \ttcosixfive{} emission is likely linked to star formation processes.

\subsection{Classification of our sources} \label{s:M0_classification}
\begin{figure}[h!]
    \centering
     \resizebox{0.9\hsize}{!}{\includegraphics{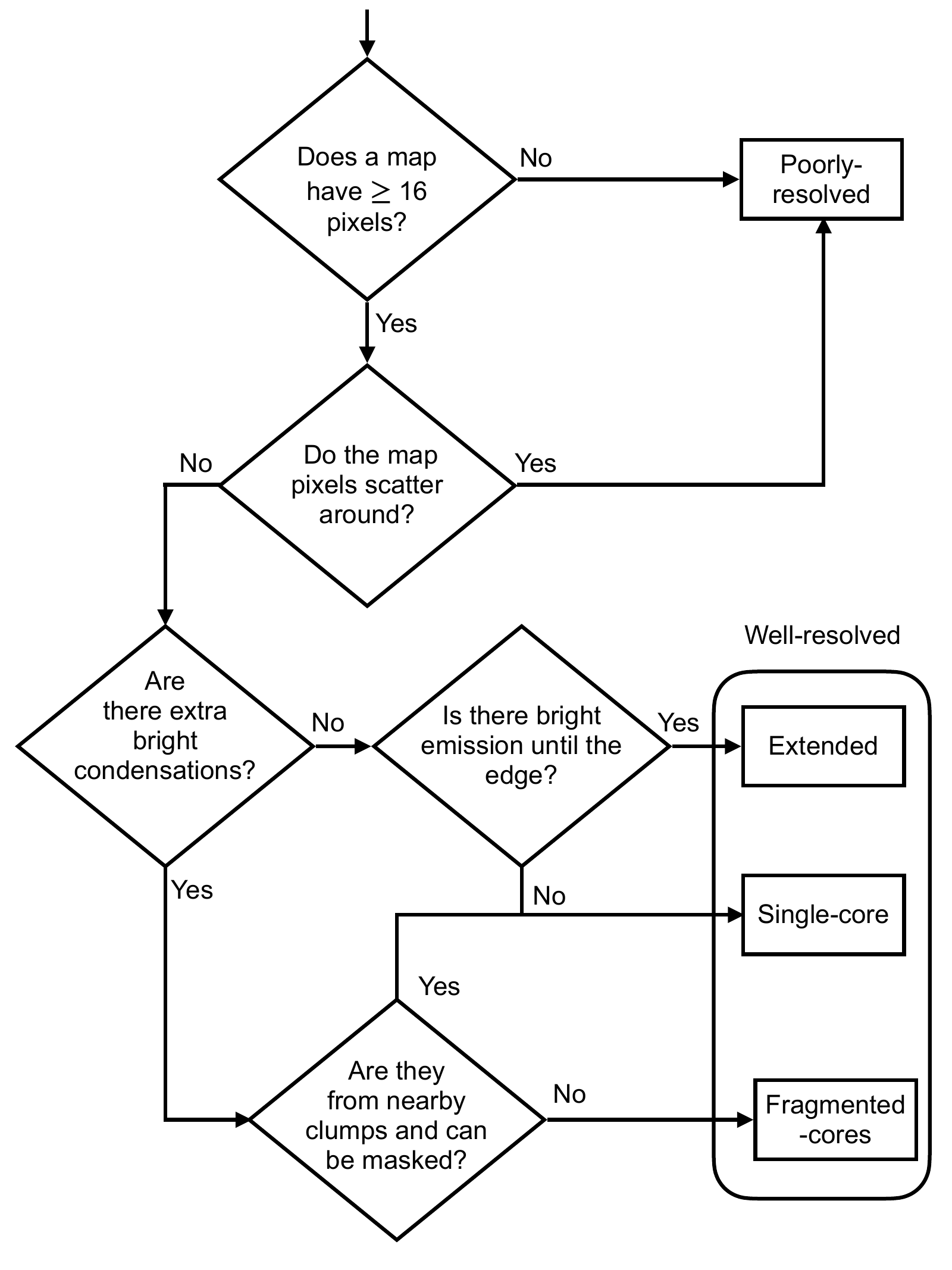}}
    \caption{Block diagram for our classification of sources based on our \ttcosixfive{} integrated intensity maps.}
    \label{fig:morpho_classification}
\end{figure}

\begin{figure*}[h!]
  \centering
  \begin{subfigure}{0.5\textwidth}
    \centering
    \includegraphics[width=\textwidth]{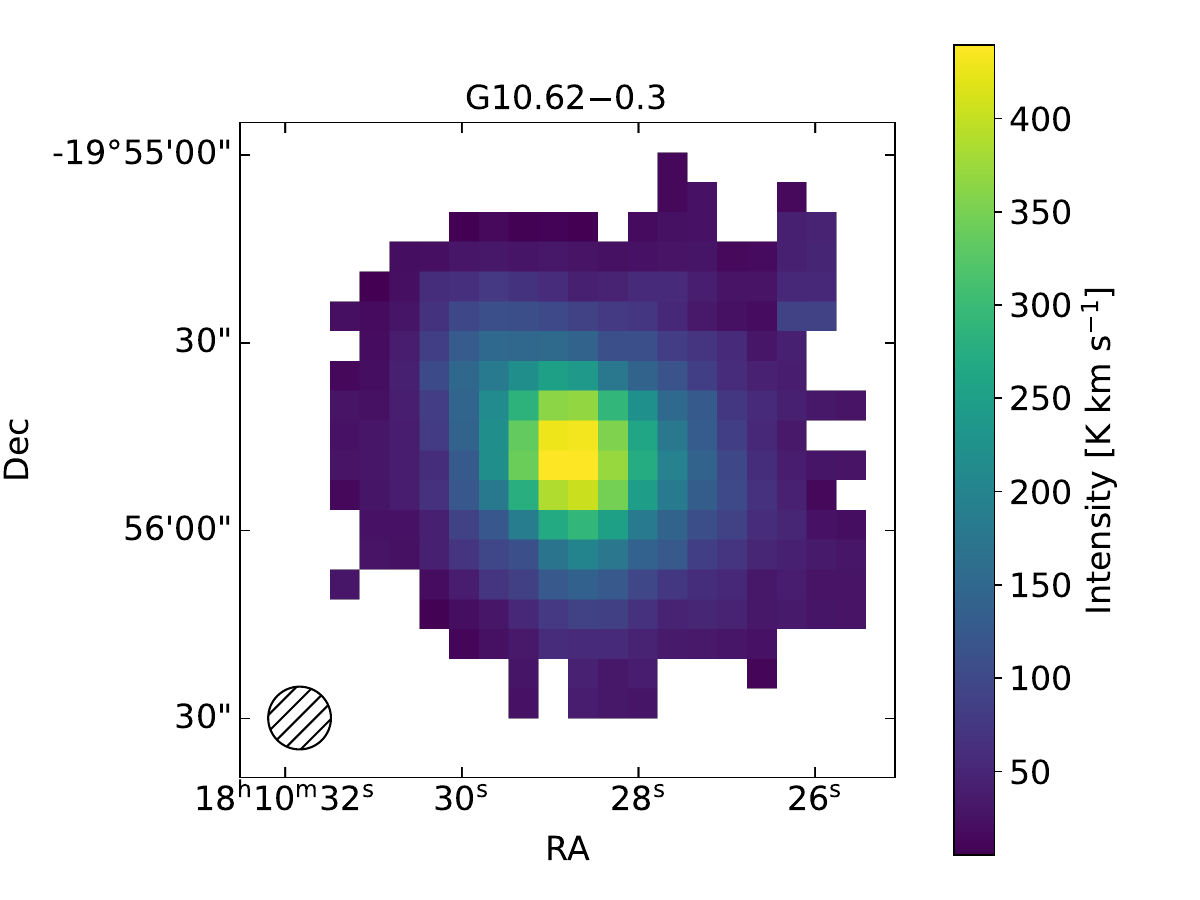}
  \end{subfigure}%
  \begin{subfigure}{0.5\textwidth}
    \centering
    \includegraphics[width=\textwidth]{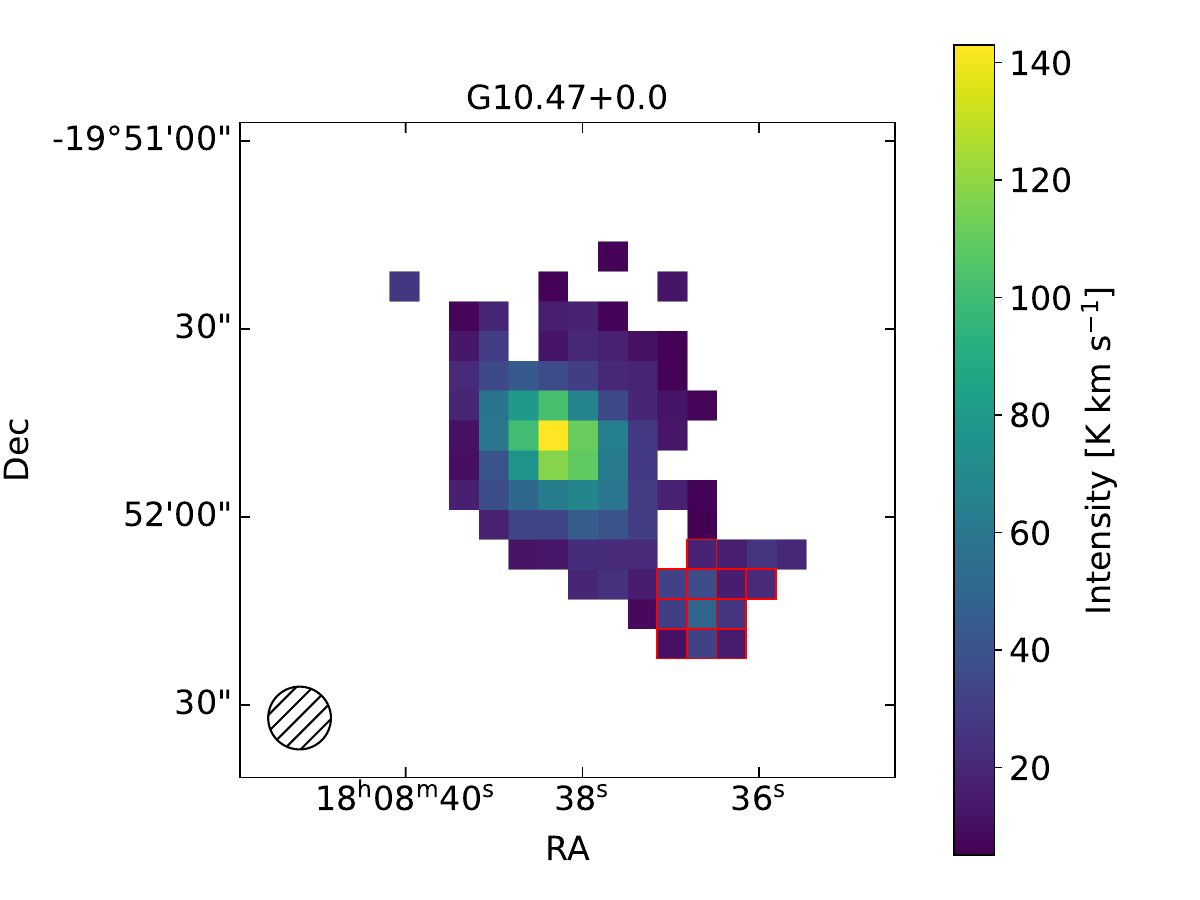}
  \end{subfigure}
  \begin{subfigure}{0.5\textwidth}
    \centering
    \includegraphics[width=\textwidth]{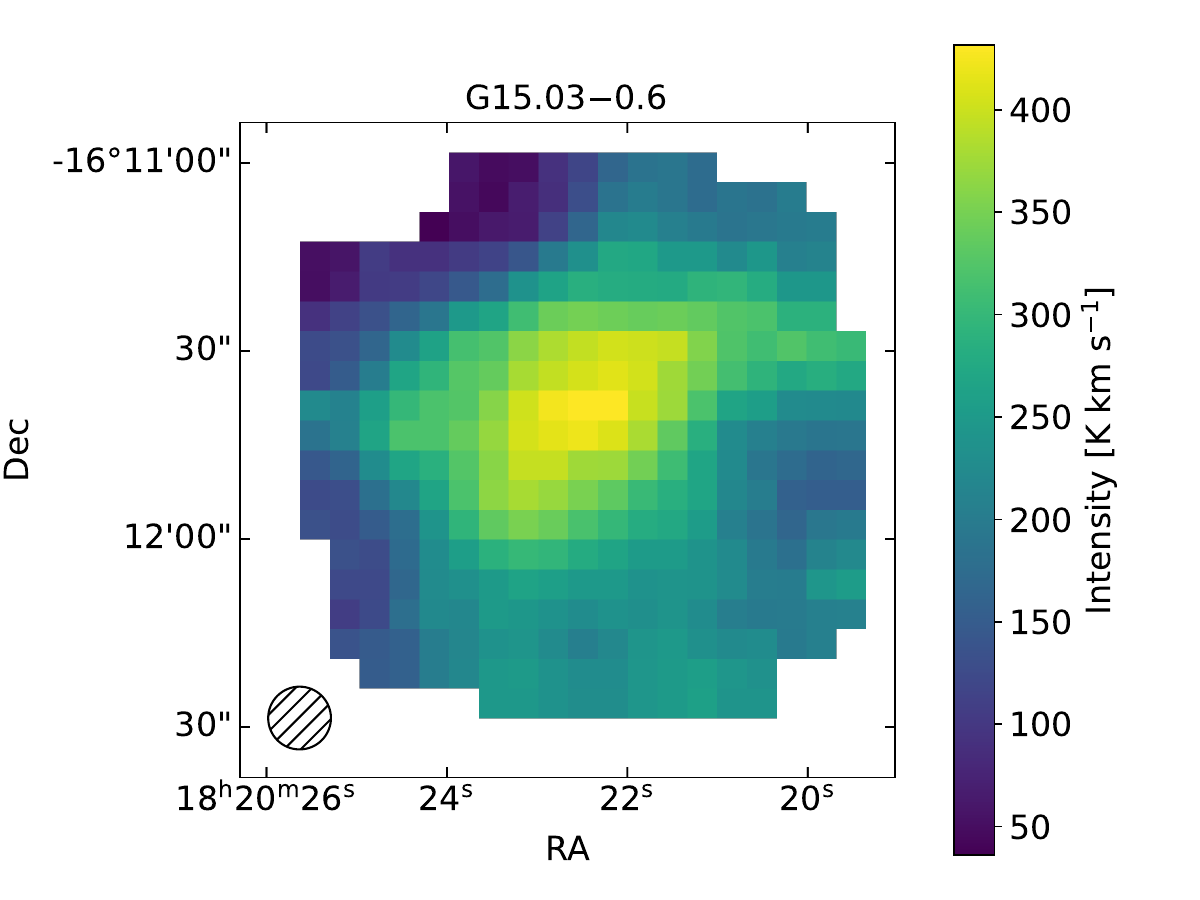}
  \end{subfigure}%
  \begin{subfigure}{0.5\textwidth}
    \centering
    \includegraphics[width=\textwidth]{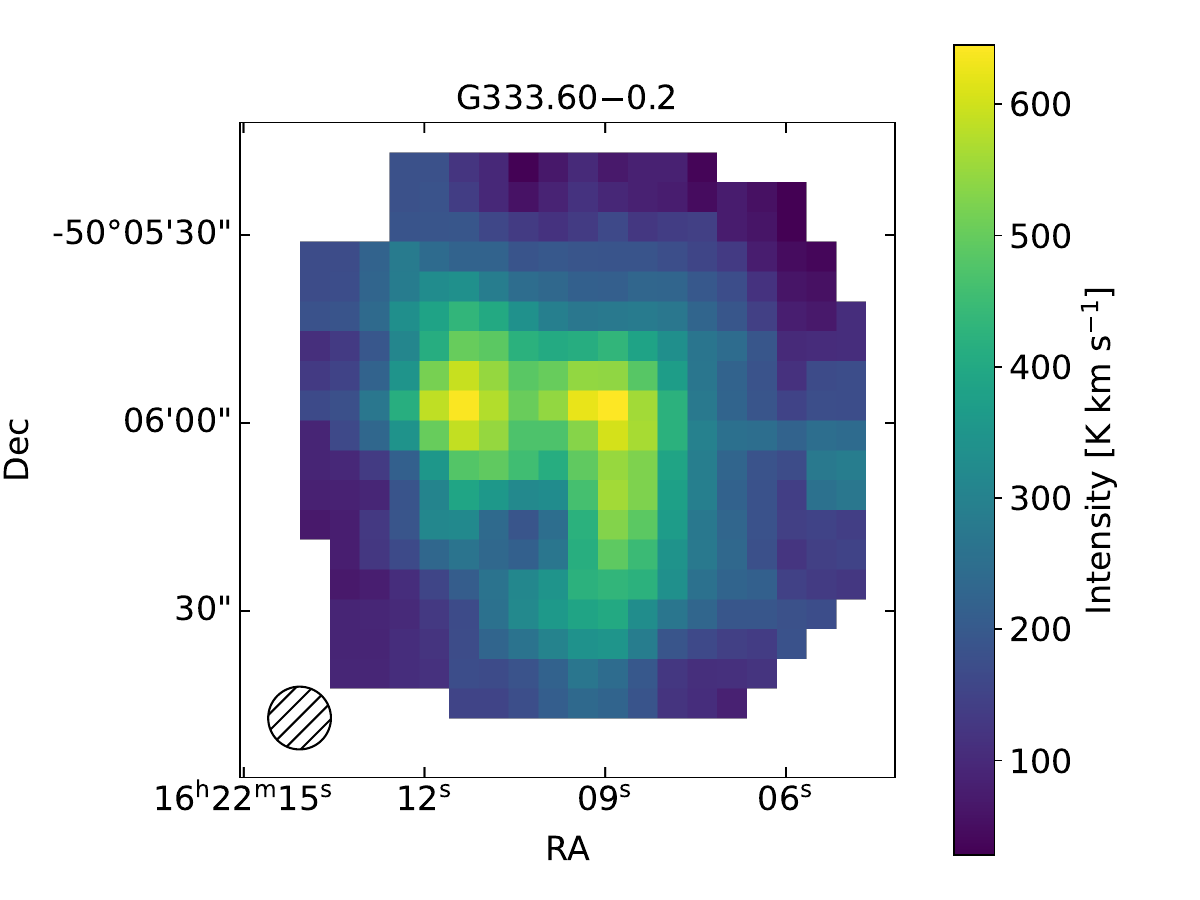}
  \end{subfigure}
  \begin{subfigure}{0.5\textwidth}
    \centering
    \includegraphics[width=\textwidth]{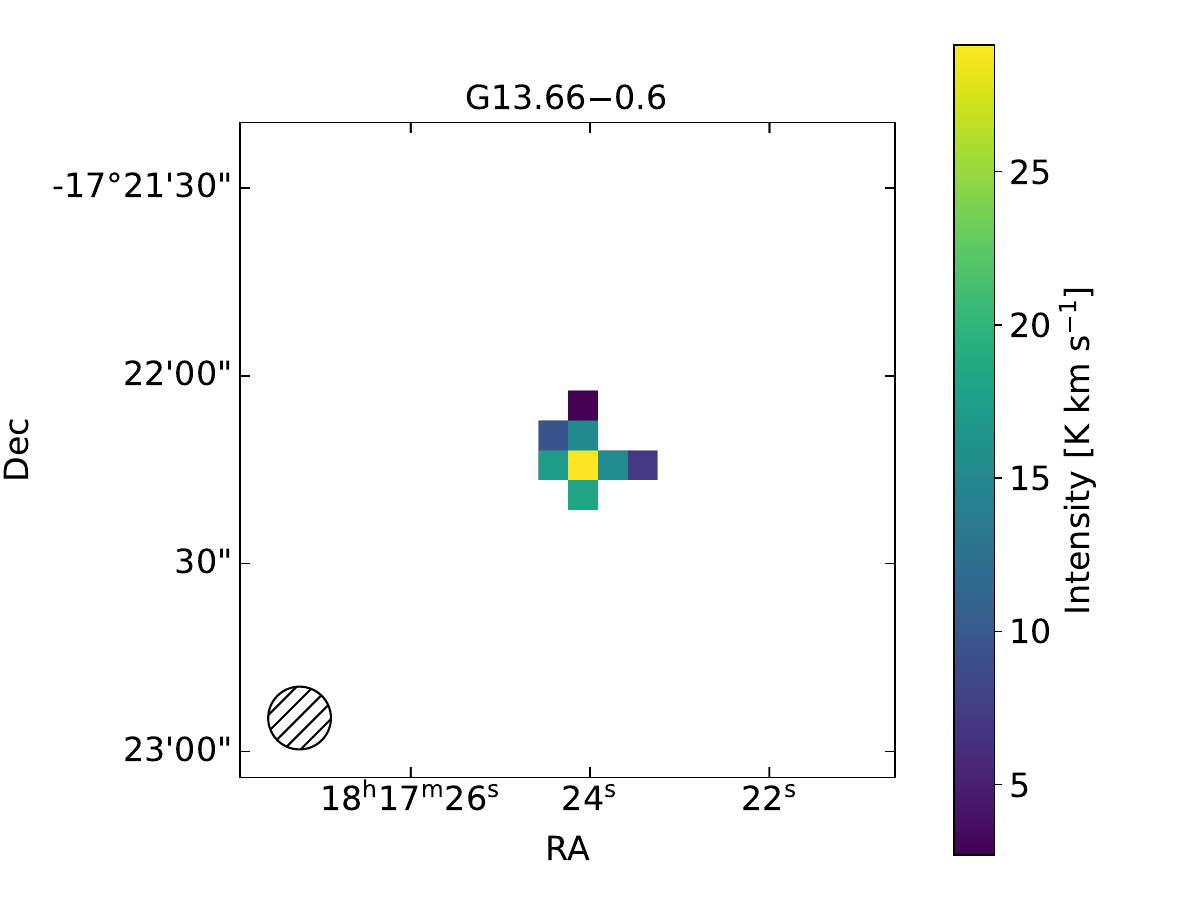}
  \end{subfigure}%
  \begin{subfigure}{0.5\textwidth}
    \centering
    \includegraphics[width=\textwidth]{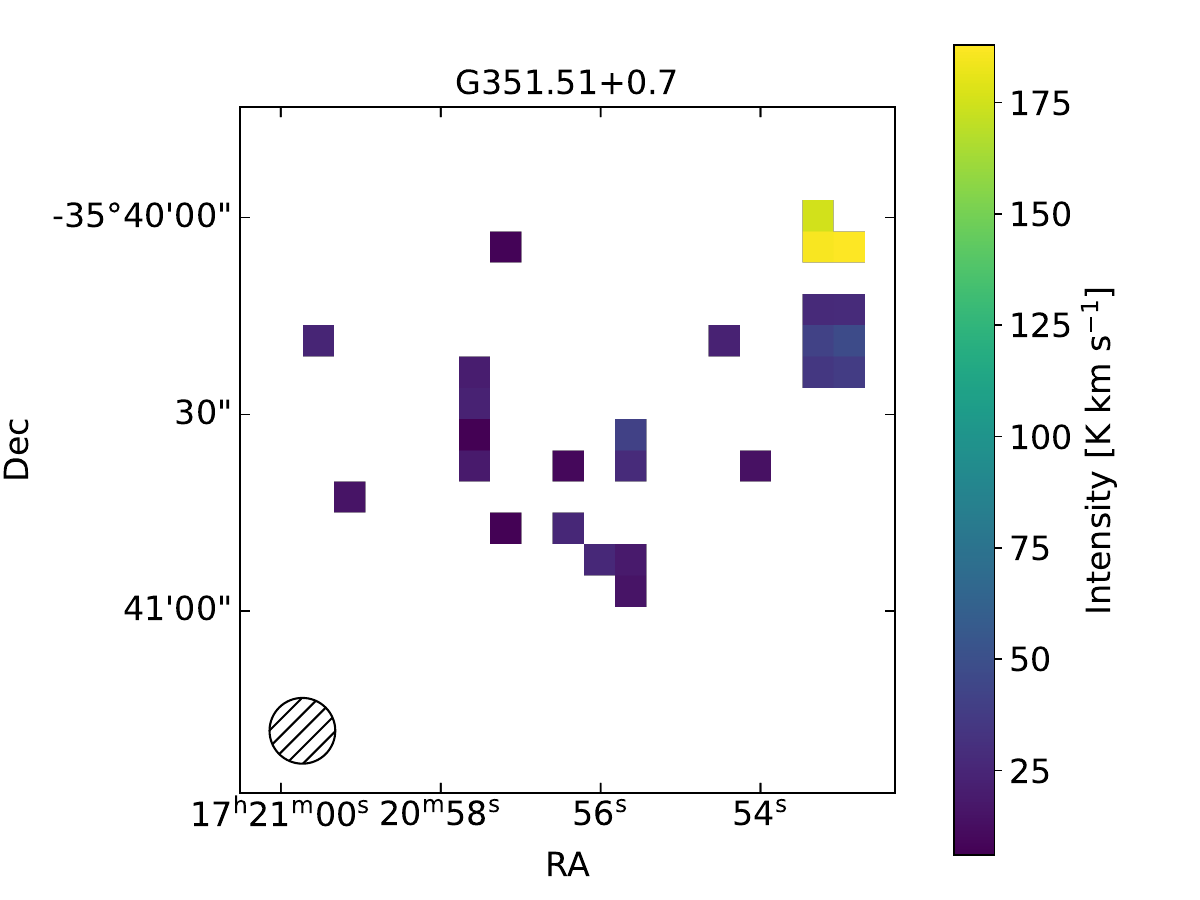}
  \end{subfigure}
  \caption{Examples of the different morphological types in our sample. The top row panels shows two single-core sources: the left one is an isolated source, while the right one has a neighbouring clump whose pixels outlined in red were masked to exclude them from further analyses. The middle panels presents an extended source on the left and fragmented cores on the right. Finally, the bottom panel illustrates two poorly-resolved clumps characterised by compact emission in neighbouring pixels (left) and distributed pixels (right). The hatched circle at the bottom left corner of each figure indicates the FWHM of telescope beam.}
  \label{fig:exampleM0maps}
\end{figure*}

Before probing the distribution and kinematics of warm molecular gas in detail (Sect. \ref{s:analyses}), we classified our sources based on their morphologies in the \ttcosixfive{} integrated intensity maps. The block diagram for our classification is shown in Fig. \ref{fig:morpho_classification}, and examples of the such classified sources are presented in Fig. \ref{fig:exampleM0maps}.

In essence, our classification is based on the extent and morphology of \ttcosixfive{} emission. First, we considered sources with fewer than 16 detected pixels (corresponding to four telescope beams) as ``poorly-resolved'' and the rest as ``well-resolved''. An exception is G351.51$+$0.7, which is categorised as a poorly-resolved clump in spite of its significant number of detected pixels as the pixels are scattered across the map. Among the well-resolved sources, we classified those with a single bright core in the map centre as ``single-core'' or ``extended''. Single-core clumps have bright emission well confined in the map centre, while extended clumps show emission extending from the centre to the edge of the maps, implying that our maps capture only a part of larger structures. ATLASGAL 870\,\mumeter{} maps of these regions indeed reveal large complexes of dust condensations, whose structures are generally consistent with the observed \ttcosixfive{} distributions (Appendix \ref{a:extM0}). In addition to the simple single-core case described above, we also considered sources that have central cores along with additional peaks at the map outskirts as ``single-core'', as the peripheral spots arise from neighbouring objects and can be masked out (Appendix \ref{a:mask}). Finally, we classified sources that have multiple central cores without extended emission as ``fragmented-cores''.

Table \ref{t:classifications} presents the number of sources in each category and evolutionary stage. In total, 59 sources are considered as well-resolved. The majority of them (52) are single-core sources, while the rest (7) show extended emission or signs of fragmentation. Interestingly, 50 out of the 59 well-resolved clumps (85\%) have evolved to IRb and HII phases, while 17 out of the 22 poorly-resolved clumps (77\%) are in younger stages (70w and IRw). This result suggests that the four evolutionary groups are distinctive in terms of the spatial distribution of their \ttcosixfive{} emission, as well as their peak brightness temperature.

\begin{table*}
\centering
\caption{Number of sources in each category determined by the \ttcosixfive{} integrated intensity maps.}
\label{t:classifications}
\begin{tabular}{ c|c | c c c c }
    \hline
    \hline 
     \multicolumn{2}{c|}{Category} & 70w & IRw & IRb & HII \\
    \hline
     \multicolumn{2}{c|}{Poorly-resolved} & 3 & 14 & 5  & 0  \\
    \hline 
    \multirow{3}{*}{Well-resolved} & Single-core     & 1 & 8  & 26 & 17 \\ \cline{2-2}
    & Fragmented-cores  & 0 & 0 & 2  & 2  \\  \cline{2-2}
    & Extended       & 0 & 0 & 2 & 1 \\  \cline{2-2}
    \hline
\end{tabular}
\end{table*}

%____________________________________________
\section{Analyses} \label{s:analyses}
In this section, we investigate the distribution and kinematics of \ttcosixfive{} emission in the envelopes of massive clumps by considering 52 single-core sources (Sect. \ref{s:mmtmaps}). While the simplest morphology of these sources makes our analyses straightforward, we note that we are biased in favour of evolved envelopes as the majority of the single-core sources are in the IRb or HII group. As such, we do not put too much emphasis on the information on evolutionary trends derived here.

\subsection{Size and aspect ratio} \label{s:size_apr}
\begin{figure}[h!]
    \centering
    \begin{subfigure}{0.5\textwidth}
      \centering
      \resizebox{0.9\hsize}{!}{\includegraphics{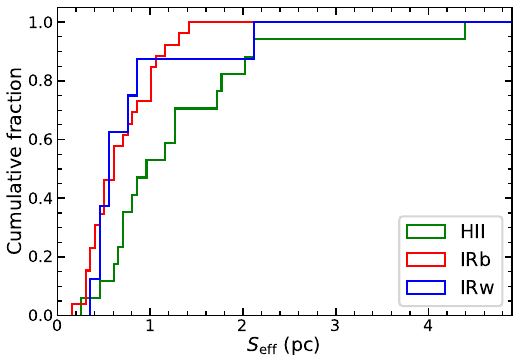}}
   \end{subfigure}
   \caption{Cumulative distribution functions of the effective size. The 70w group is not presented here, since it comprises only a single source.}
   \label{fig:size-lum}
\end{figure}

We started our analyses by measuring the size of each source based on the total number of detected pixels. Specifically, we calculated the effective size, $S_{\mathrm{eff}}$, as the square root of the total emitting area and considered it as a representative size. The resultant cumulative distribution functions (Fig. \ref{fig:size-lum}) suggest that the $^{13}$CO emission regions associated with HII sources are generally larger than those found for IRw and IRb sources. For example, the median sizes of the three groups (IRw, IRb, and HII) are 0.59, 0.63, and 0.99\,pc, respectively, while $S_{\mathrm{eff}}$ of the only 70w source is 0.37\,pc. When a $k$-sample Anderson-Darling test was performed on the IRb and HII groups for which we have sufficient numbers of sources, the sizes of the HII envelopes were indeed found to be systematically larger at a significance level of 0.05. Interestingly, the largest source, with $S_{\mathrm{eff}}$ of 2.13\,pc, is an IRw source, G342.48$+$0.1.

In addition, we estimated the aspect ratio (APR) of each source by fitting a two-dimensional Gaussian function to the detected pixels in the integrated intensity image and found a range of 1.0--2.6 with a median of 1.3 and a standard deviation of 0.4. Considering that APRs vary with inclination angles and the members of our statistically significant sample of single-core sources likely have a wide range of inclination angles, the measured small range of the APR implies that the envelopes are less likely filamentary structures whose minimum APR is defined as three \citep{mattern2018sedigism}.

\subsection{Radial intensity profiles} \label{s:13co_radprofile}
\begin{figure*}[h!]
  \centering
  \begin{subfigure}{0.45\textwidth}
    \centering
    \includegraphics[width=\textwidth]{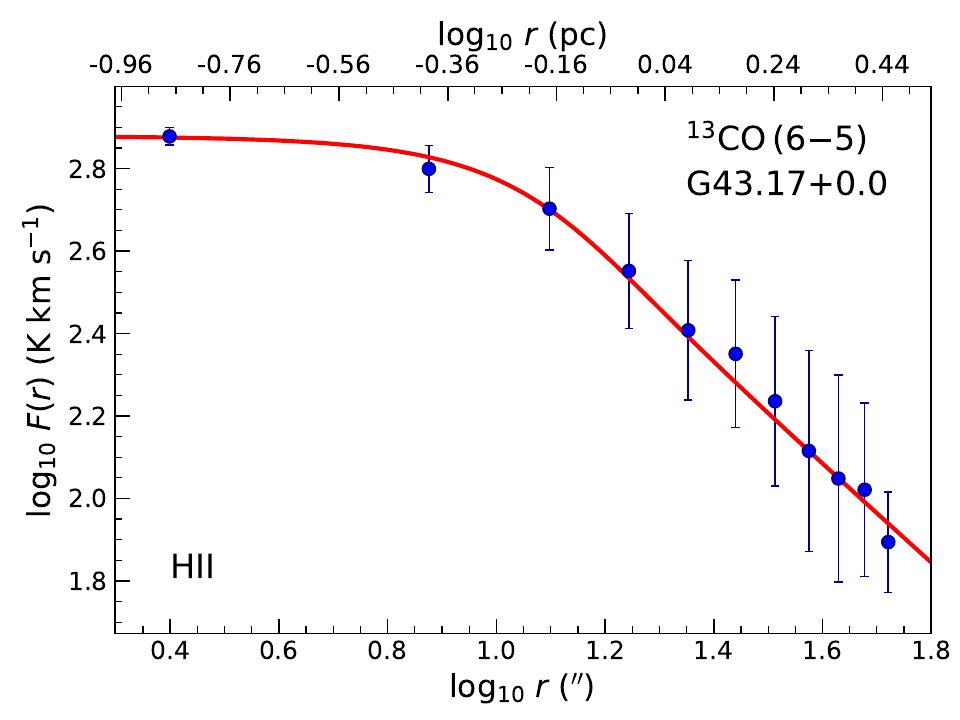}
  \end{subfigure}%
  \begin{subfigure}{0.45\textwidth}
    \centering
    \includegraphics[width=\textwidth]{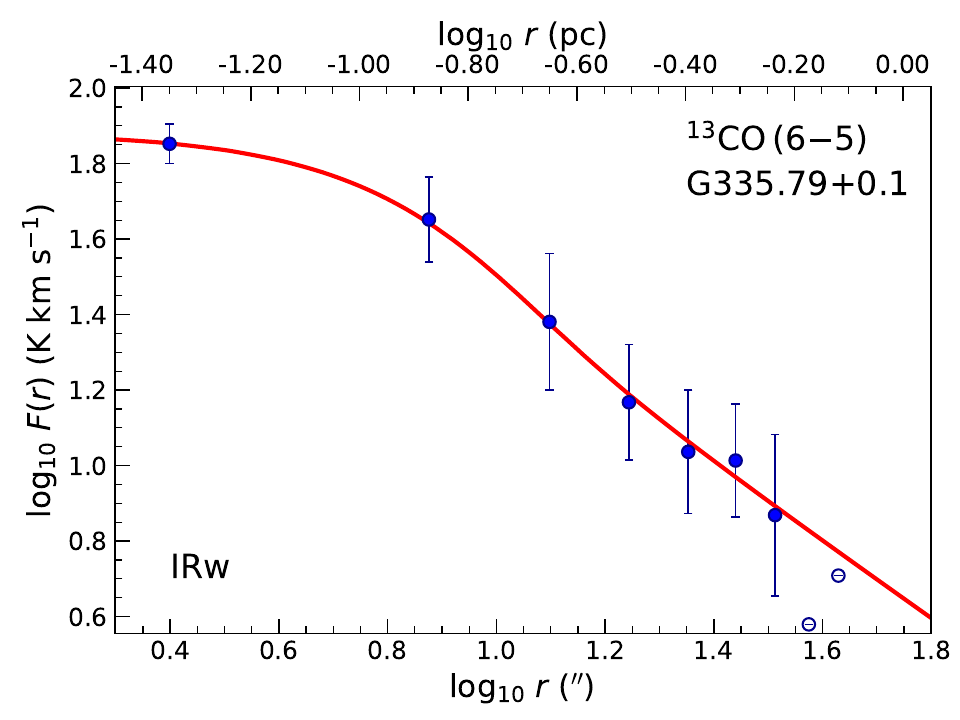}
  \end{subfigure}%
  \caption{Example \ttcosixfive{} radial profiles for an HII source (G43.17$+$0.0) (left) and an IRw source (G335.79$+$0.1) (right). At 5$\arcsec$-wide individual annuli, the median and standard deviation values are presented as filled blue circles and associated ``error'' bars. Annuli with an insufficient number of detected pixels are indicated as open blue circles and were excluded from the determination of best-fit curves (red).}
  \label{fig:M0-radprof}
\end{figure*}

Next, we examined the spatial distribution of \ttcosixfive{} emission by constructing the radial intensity profiles of our single-core sources. To do so, we divided each $M_0$ map into 5$\arcsec$-wide individual annuli centred on the peak location of \ttcosixfive{} emission and calculated median and standard deviation values. Example radial profiles are presented in Fig. \ref{fig:M0-radprof}.

The constructed radial profiles were analysed with power-law functions following  \citet{beuther2002high}: 
\begin{equation}
\label{eq:CO_radial}
    F(r) \propto 
    \begin{cases}
        r^{-m} \quad \text{if} \quad r > r_{\mathrm{b}} \\
        r_{\mathrm{b}}^{-m} \quad \text{if} \quad r \leq r_{\mathrm{b}}, 
    \end{cases}
\end{equation}
where $F(r)$ is the integrated intensity at a given radius $r$, $m$ is the power-law index, and $r_{\rm b}$ is the break radius within which the integrated intensity becomes constant (introduced to prevent $F(r)$ from going to infinity as $r$ approaches zero). This power-law function was devised based on the theoretical expectations for the structures of collapsing cores \citep[e.g.,][]{shu1987star,mclaughlin96,motte01} and was convolved with a 10$\arcsec$-size Gaussian beam (comparable to our \ttcosixfive{} angular resolution) to be fitted to our radial profiles.

To constrain the power-law index and break radius, we performed non-linear least squares fitting using the Python function \texttt{scipy.optimize.curve\_fit} on the model and observed radial profiles. We excluded data points at large radii for which the number of detected pixels is less than 5\% of the total pixels (e.g., blue open circles in Fig. \ref{fig:M0-radprof}) from fitting, as they are less likely statistically representative. The best-fit curves were found to reproduce the observed radial profiles reasonably well within 1$\sigma$ uncertainties (hence obviating the need for a second power-law function), and their parameters are presented in Appendix\,\ref{a:table}.

\begin{figure}
   \begin{subfigure}{\hsize}
      \centering
      \resizebox{0.9\hsize}{!}{\includegraphics{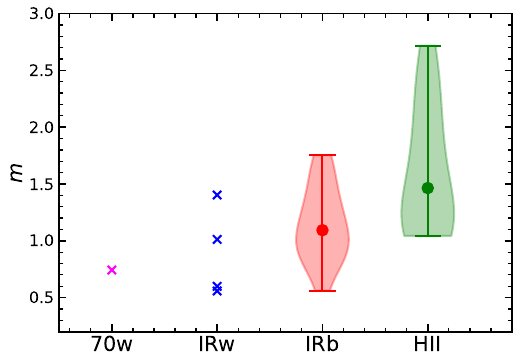}}
   \end{subfigure}
   \caption{Distributions of the power-law index, $m$, for the different evolutionary groups: 70w, IRw, IRb, and HII in magenta, blue, red, and green, respectively. The colours and symbol schemes are the same as in Fig.\,\ref{fig:linepeaks}.}
   \label{fig:M0-fitparas}
\end{figure}

We examined the constrained power-law indices and break radii in detail by focusing on 36 sources (one 70w, four IRw, 17 IRb, and 14 HII sources) whose the best-fit parameters are statistically significant (higher than their 1$\sigma$ uncertainties). The distributions of the power-law index for these sources are presented in Fig.\,\ref{fig:M0-fitparas}. We find that the power-law index ranges from 0.5 to 2.7 across the entire sample and varies less than a factor of three within each group, indicating relatively comparable slopes between the sources in the same evolutionary stage. Interestingly, there is a slight tendency for the power-law index to increase towards more evolved sources: %For example,
median $m$ values are 0.8, 1.1, and 1.5 for the IRw, IRb, and HII group, respectively. A $k$-sample Anderson-Darling test on the IRb and HII groups indeed suggests that the HII sources have distinctly steeper slopes than the IRb sources. These steeper slopes could result from a significant increase of the brightness of \ttcosixfive{} emission in the central parts of the most evolved sources (e.g., Fig.~\ref{fig:linepeaks}), which is in turn most likely due to the strong enhancement in density and/or temperature.

\begin{figure}
    \begin{subfigure}{\hsize}
      \centering
      \resizebox{0.9\hsize}{!}{\includegraphics{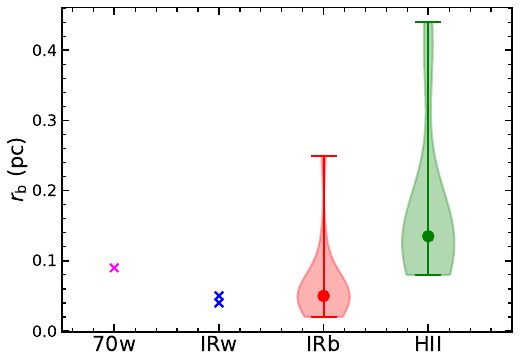}}
   \end{subfigure}
   \begin{subfigure}{\hsize}
      \centering
      \resizebox{0.9\hsize}{!}{\includegraphics{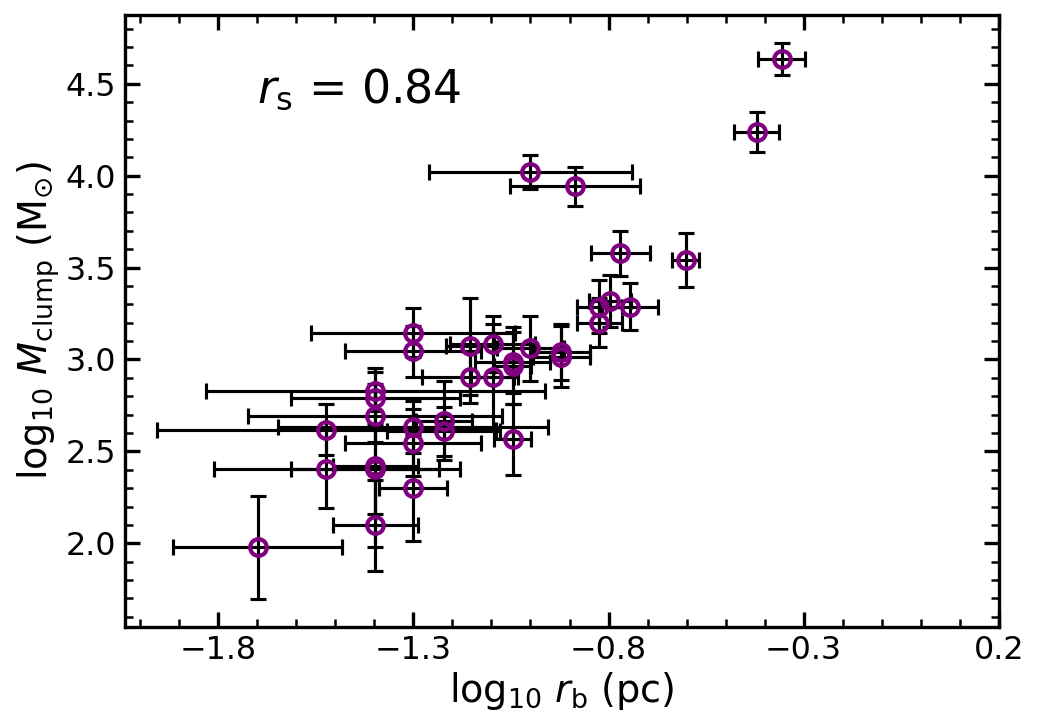}}
   \end{subfigure}
   \caption{(Top) Distributions of the break radius,  $r_{\mathrm{b}}$, for the different evolutionary groups. The colours and symbol schemes are the same as in Fig.\,\ref{fig:linepeaks}. (Bottom) Scatter plot between the clump mass and the break radius. The Spearman's rank correlation coefficient is presented on the top left corner of the plot.}
   \label{fig:breakradius}
\end{figure}

In addition, we probed the constrained break radii and found that $r_{\mathrm{b}}$ changes from 0.02\,pc to 0.44\,pc across the entire sample (Fig.\,\ref{fig:breakradius}). The variations in $r_{\mathrm{b}}$ (e.g., a factor of 22 for the whole sample and a factor of 13 for the IRb sources) are much more significant than those in $m$, which partly reflects the sources' wide range of distances. As in the case of $m$, $r_{\mathrm{b}}$ tends to increase towards the most evolved sources (e.g., median $r_{\mathrm{b}}$ of 0.05, 0.05, and 0.14\,pc for the IRw, IRb, and HII group, respectively), and a $k$-sample Anderson-Darling test on the IRb and HII groups confirms this tendency.

While the break radius was introduced for a mathematical reason, it could manifest underlying physical conditions in the central parts of our sources, such as fragmentation on small scales and high opacity of the \ttcosixfive{} emission. As for the fragmentation scenario, interferometric observations of massive star-forming clumps, including several of our sources, have indeed identified fragmented cores with sizes smaller than or comparable to our spatial resolution ($\lesssim$ 0.1\,pc) \citep{cesaroni17, beltran2021fragmentation}. The convolution of these fragmented cores with our single-dish beam would smooth out small-scale structures and result in the uniform distribution of \ttcosixfive{} emission in the source centres. Interestingly, \citet{beuther2002high} argued that more massive clumps tend to produce larger fragmented cores or clusters, which implies larger $r_{\mathrm{b}}$ for more massive clumps. We found indeed such a positive correlation between $r_{\mathrm{b}}$ and $M_{\mathrm{clump}}$ (Fig.\,\ref{fig:breakradius}, bottom), supporting the fragmentation scenario for the break radius. However, we also note that $M_{\mathrm{clump}}$ is related to $D$ as $M_{\mathrm{clump}} \propto D^{2}$ \citep{konig2017atlasgal}, which could partially contribute to the correlation between $r_{\mathrm{b}}$ and $M_{\mathrm{clump}}$.

Another possible explanation for the break radius is high opacity of \ttcosixfive{} emission. When the \ttcosixfive{} emission is optically thick, its brightness temperature approaches the gas kinetic temperature, resulting in the saturation of the integrated intensity. Our preliminary analysis based on local thermodynamic equilibrium (LTE) suggests that most of our sources likely have optically thin \ttcosixfive{} emission (Appendix.\,\ref{a:tau13co}). Besides, only a weak correlation exists between $\tau_{^{13}\mathrm{CO}}$ and $r_{\mathrm{b}}$ (Spearman's rank correlation coefficient of 0.32). These results imply that the break radius is less likely to be determined by the effect of opacity.

In conclusion, we found that the radial intensity profiles of single-core clumps can be reasonably well described as power-law functions, indicating relatively simple structures of the envelopes of massive star-forming regions. The observed radial profiles retain comparable slopes within the same evolutionary group, while the most evolved sources tend to have steeper slopes likely related to star formation processes (higher densities and/or temperatures). In addition, fragmentation could begin in the early stage of high-mass star formation (70w) and be responsible for the uniform intensity distribution in the central parts of our sources.

\subsection{Comparison between dust and warm molecular gas distributions} \label{s:comparison_radprofile}
In this section, we examine how the distribution of warm molecular gas in high-mass star-forming regions compares to that of the colder medium by comparing \ttcosixfive{} and 160\,\mumeter{} radial intensity profiles. For our examination, we employed Hi-GAL 160\,\mumeter{} maps as their angular resolution (11\arcsec $\times$ 13\arcsec) is comparable to that of the \ttcosixfive{} data \citep{molinari2010hi}.

\begin{figure*}
  \centering
  \begin{subfigure}{0.5\textwidth}
    \centering
    \includegraphics[width=\textwidth]{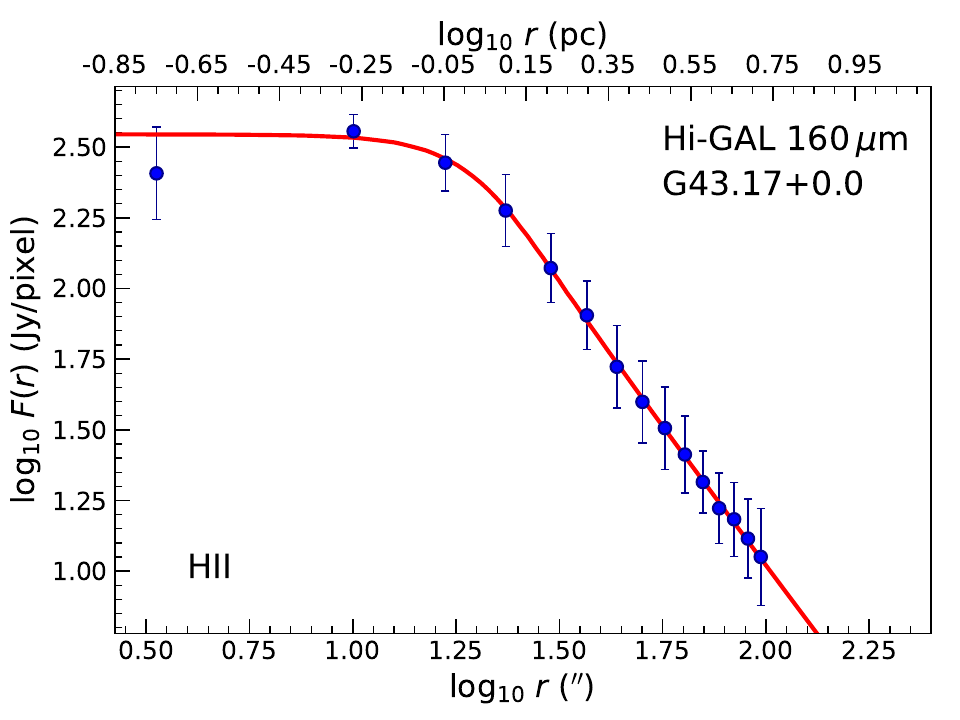}
  \end{subfigure}%
  \begin{subfigure}{0.5\textwidth}
    \centering
    \includegraphics[width=\textwidth]{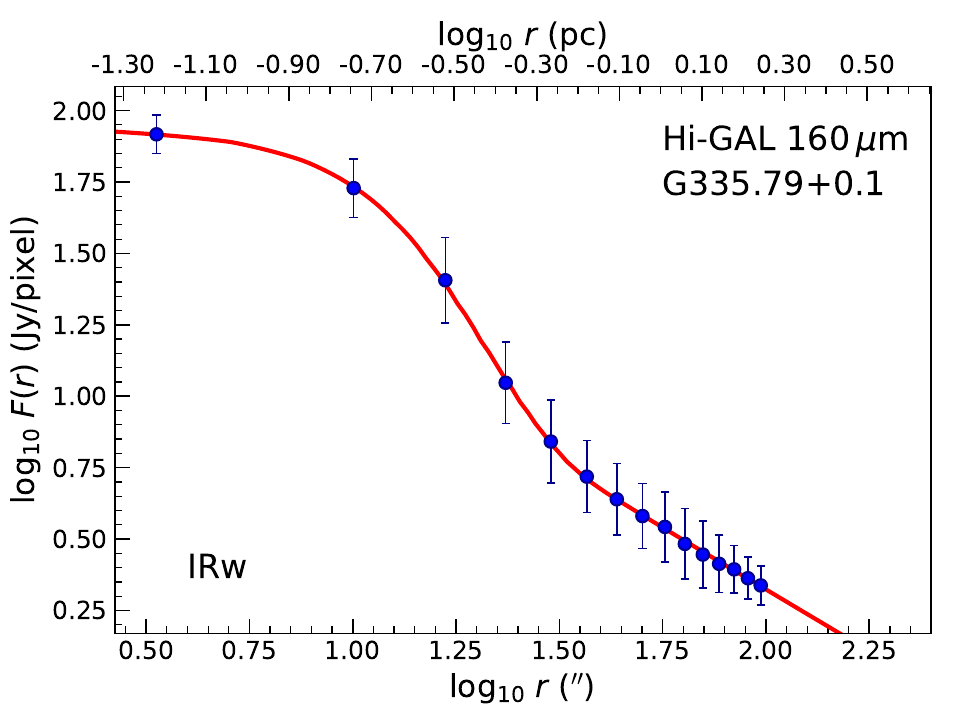}
  \end{subfigure}%
  \caption{Example 160\,\mumeter{} radial profiles of G43.17$+$0.0 (left) and G335.79$+$0.1 (right). As in Fig. \ref{fig:M0-radprof}, the median and standard deviation values at 5\arcsec-wide individual annuli are shown as blue circles and associated ``error'' bars. Best-fit curves (singular and two power-law model for G43.17$+$0.0 and G335.79$+$0.1, respectively) are overlaid in red.}
  \label{fig:160micron-radprof}
\end{figure*}

To derive radial profiles, Hi-GAL 160\,\mumeter{} maps with a size of $5^{\prime} \times 5^{\prime}$ were first converted to the equatorial coordinate system to match the \ttcosixfive{} data. Once the pixels with emission from nearby objects were masked (Appendix \ref{a:mask}), 160\,\mumeter{} radial profiles were constructed in the same manner as the \ttcosixfive{} radial profiles (Sect.\,\ref{s:13co_radprofile}) and were cut off at 100\arcsec\ to meet the angular extent of the \ttcosixfive{} radial profiles. In total, we produced the 160\,\mumeter{} radial profiles for a sample of 36 sources for which we found reliable fitting parameters for the \ttcosixfive{} radial profiles (Sect.\,\ref{s:13co_radprofile}).

Visual inspections of the derived radial profiles indicated the need for more than one power-law for some sources (e.g., Fig.\,\ref{fig:160micron-radprof}, right), prompting us to fit both single (Eqn.\,\ref{eq:CO_radial}) and two power-law functions to the radial profiles. The two power-law model can be written as follows: 
\begin{equation}
    F(r) \propto 
    \begin{cases}
        r_{\mathrm{b}_1}^{-m_{\mathrm{i}}} \quad \text{if} \quad r \leq r_{\mathrm{b}_1} \\
        r^{-m_{\mathrm{i}}} \quad \text{if} \quad r_{\mathrm{b}_1} < r \leq r_{\mathrm{b}_2} \\
        r^{-m_{\mathrm{o}}} \quad \text{if} \quad r > r_{\mathrm{b}_2},
    \end{cases}
\end{equation}
where $m_{\mathrm{i}}$ and $m_{\mathrm{o}}$ are the indices for the inner and outer power-law functions, $r_{\mathrm{b}_1}$ is the break radius within which the central intensity distribution becomes constant, and $r_{\mathrm{b}_2}$ is the transition radius from the inner to the outer power-law function. Both the single and two power-law models were convolved with a 12\arcsec-size Gaussian beam (average of the major and minor axes of the 160\,\mumeter{} beam) to be fitted to our radial profiles.

We performed the same non-linear least squares fitting as in Sect.\,\ref{s:13co_radprofile} and found that the two power-law model significantly improves fitting results for 27 sources by lowering the reduced $\chi^2$ by more than a factor of two compared to the single power-law model. The radial profiles of the remaining nine sources were better described with the single power-law model, considering that the reduced $\chi^2$ was improved by less than a factor of two or the width of the inner power-law distribution was smaller than the 160\,\mumeter{} beam; in other words, the inner power-law distribution is not well constrained. The fitting results for all 36 sources were deemed reliable (fitting parameters above their 1$\sigma$ uncertainties) and are summarised in Appendix\,\ref{a:table}.

For 26 out of the 27 sources for which the two power-law model is favoured, we found $m_{\mathrm{i}} > m_{\mathrm{o}}$, which suggests that central compact cores are surrounded by extended diffuse structures. Interestingly, \citet{beuther2002high} found the opposite result (shallower inner gradients) from their analysis of 1.2\,mm dust continuum emission in a number of massive star-forming regions. It is difficult to make a one-to-one comparison with our result, however, since the sources in \citet{beuther2002high} suffer from distance ambiguities and we cannot make sure if the 1.2\,mm and 160\,\mumeter{} profiles probe comparable spatial extents.

\begin{figure}
   \begin{subfigure}{\hsize}
      \centering
      \resizebox{0.9\hsize}{!}{\includegraphics{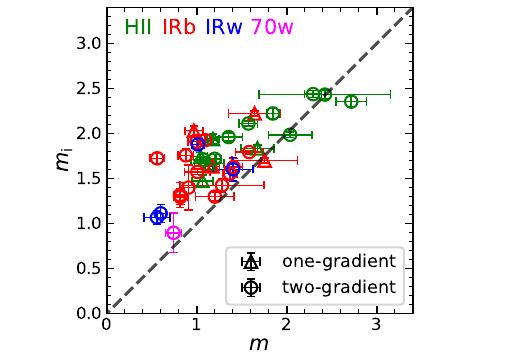}}
   \end{subfigure}
   \begin{subfigure}{\hsize}
      \centering
      \resizebox{0.9\hsize}{!}{\includegraphics{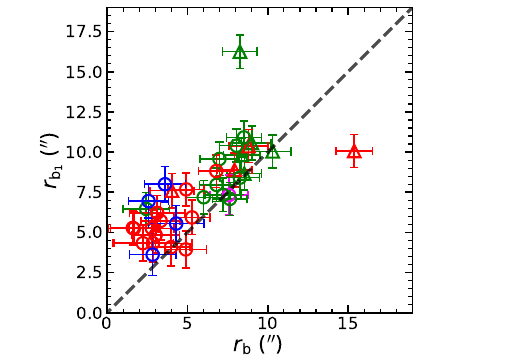}}
   \end{subfigure}
   \caption{Comparison of the 160\,\mumeter{} and \ttcosixfive{} radial profiles in terms of the power-law index (top) and break radius (bottom). The 27 sources whose 160\,\mumeter{} radial profiles were fitted with the two power-law model are shown as circles, and their inner power-law distributions are used for the comparison. The remaining nine sources with the single power-law distributions in 160\,\mumeter{} emission are indicated as triangles. Both circles and triangles are colour coded based on evolutionary stage: 70w, IRw, IRb, and HII in magenta, blue, red, and green, respectively. One-to-one relations are overlaid as dashed lines.} 
   \label{fig:higal160vsmidjco}
\end{figure}

The inner power-law distributions of the 160\,\mumeter{} radial profiles have on average similar extents as the \ttcosixfive{} radial profiles ($\sim$40\arcsec), indicating that they probe comparable physical scales. Considering this, we directly compared the properties of the inner power-law distributions of the 160\,\mumeter{} radial profiles (power-law index and break radius) to those of the \ttcosixfive{} radial profiles. To increase the sample size, we also considered the nine sources whose 160\,\mumeter{} radial profiles were fitted with the single power-law function. Figure\,\ref{fig:higal160vsmidjco} shows a positive correlation between the power-law indices of the 160\,\mumeter{} and \ttcosixfive{} radial profiles. On closer inspection, the 160\,\mumeter{} profiles generally have steeper slopes, implying that 160\,\mumeter{} and \ttcosixfive{} emission could trace different physical conditions or processes (e.g., 160\,\mumeter{} emission likely traces colder structures than \ttcosixfive{} emission). Alternatively, the discrepancy in the power-law indices could result from the different dependencies of the two tracers on the gas temperature and density. Interestingly, the 160\,\mumeter{} radial profiles tend to be steeper towards more evolved sources, which is in line with what we found from the \ttcosixfive{} radial profiles (Sect.\,\ref{s:13co_radprofile}). On the other hand, the break radii of the 160\,\mumeter{} and \ttcosixfive{} radial profiles seem to be consistent in most cases, suggesting that the same mechanism (fragmentation) is likely responsible for the uniform distribution of 160\,\mumeter{} and \ttcosixfive{} emission in the central parts of our sources.

In conclusion, we found that the 160\,\mumeter{} emission from our sources typically shows two distinct structures: compact cores and their surrounding diffuse halos. Compared to 160\,\mumeter{} emission, \ttcosixfive{} emission from warm gas is more compact and probes mostly the core structures with its radial distribution decreasing less steeply. The emission from both tracers shows uniform intensity distributions in the centre of our sources, which likely originates from fragmentation on small scales.

\subsection{Mean Velocity Gradients (MVGs)} \label{s:mvg}
\begin{figure*}[h!]
  \centering
  \begin{subfigure}{0.5\textwidth}
    \centering
    \includegraphics[width=\textwidth]{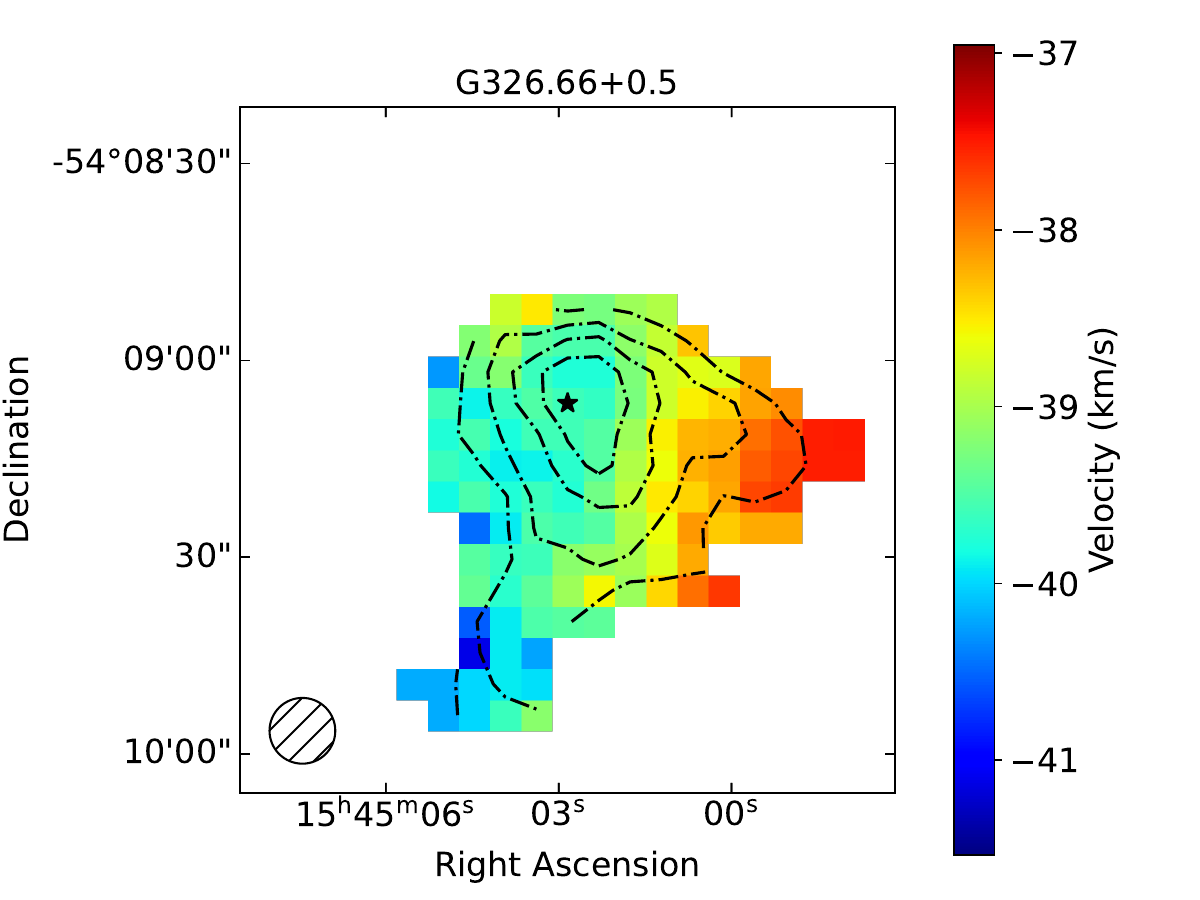}
  \end{subfigure}%
  \begin{subfigure}{0.5\textwidth}
    \centering
    \includegraphics[width=\textwidth]{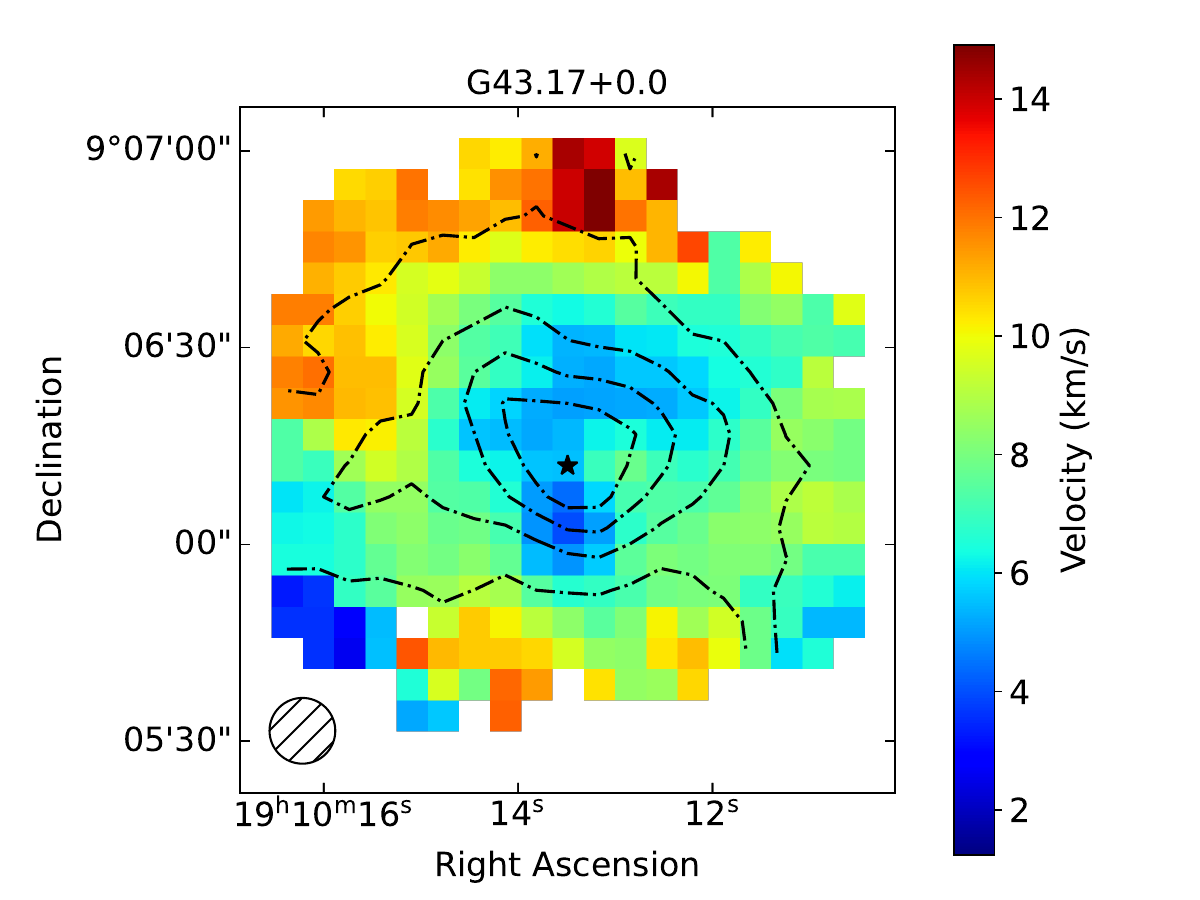}
  \end{subfigure}
  \begin{subfigure}{0.5\textwidth}
    \centering
    \includegraphics[width=\textwidth]{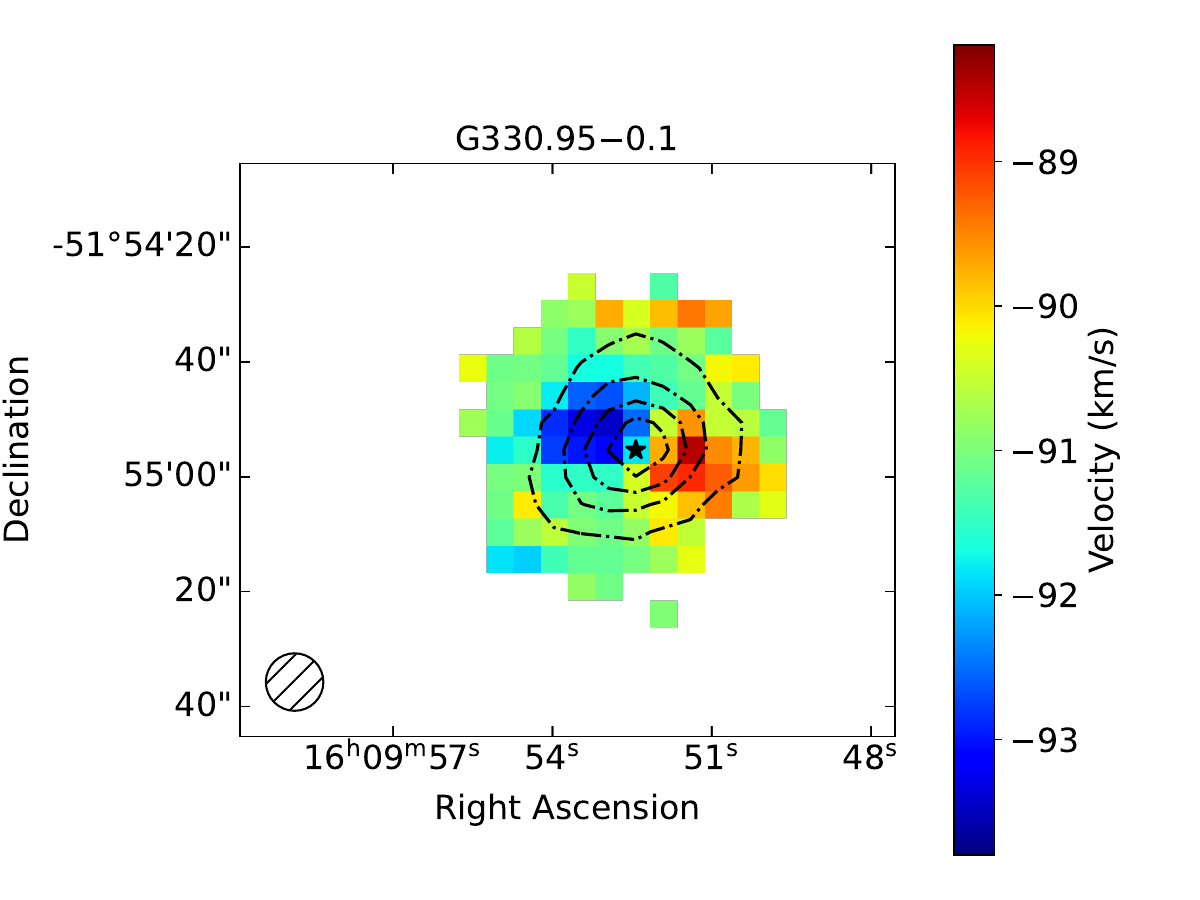}
  \end{subfigure}%
  \begin{subfigure}{0.5\textwidth}
    \centering
    \includegraphics[width=\textwidth]{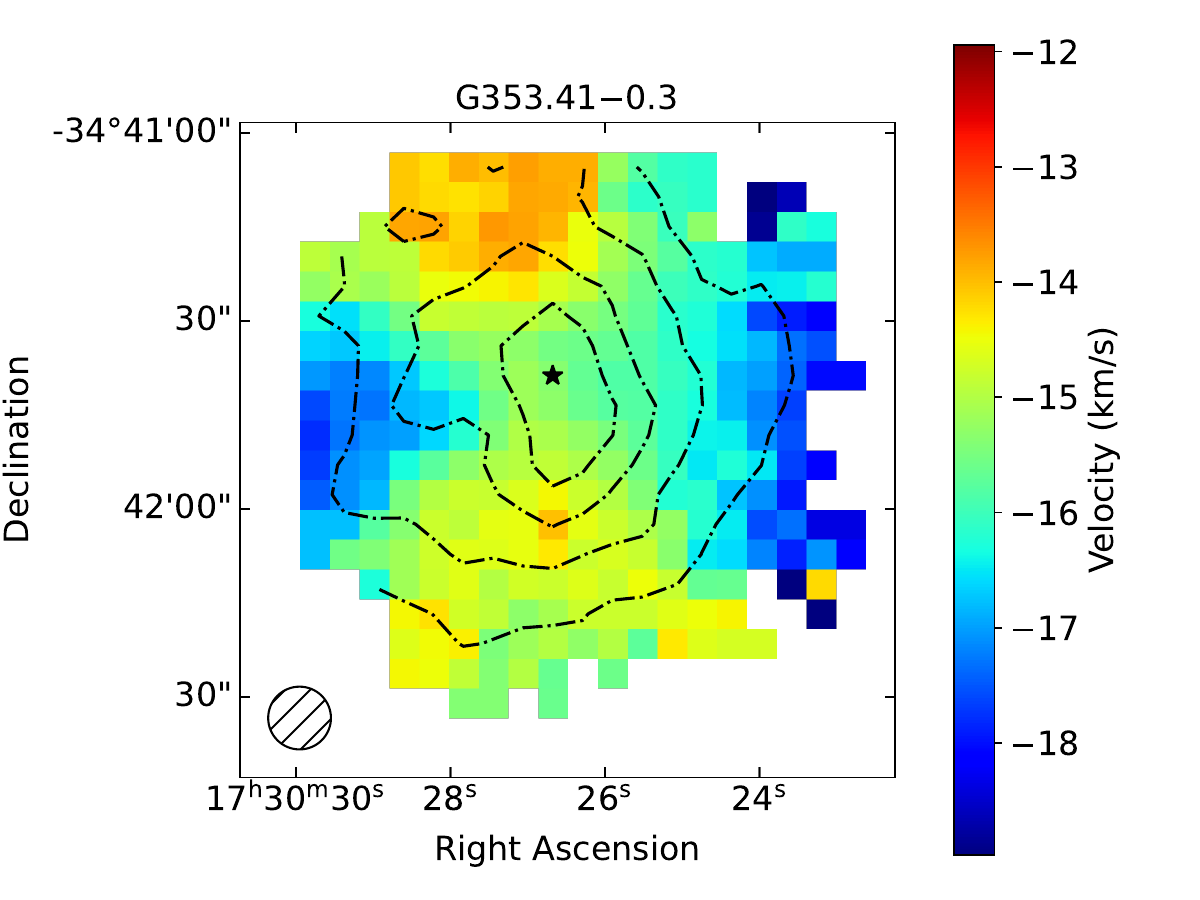}
  \end{subfigure}
  \caption{Examples of various velocity fields in our sample. The \ttcosixfive{} $M_1$ maps from the top left corner in the clockwise direction demonstrate linear, radial, hourglass-like, and outflow-like gradients, respectively. For each source, the integrated intensities are overlaid as contours with levels ranging from 20\% to 80\% of the peak value in steps of 20\%, and the peak position is marked by a star. Finally, the FWHM of the telescope beam is shown by a hatched circle.}
  \label{fig:exampleM1maps}
\end{figure*}

In addition to the spatial distribution, we examined the kinematics of warm molecular gas for the sources of our sample of high-mass star-forming regions by analysing the \ttcosixfive{} first moment maps of our 52 single-core sources. Visual inspections of the \ttcosixfive{} $M_1$ maps reveals a variety of velocity fields, including linear gradients or more complex trends such as radial, hourglass-like, and outflow-like gradients (Fig.\,\ref{fig:exampleM1maps}). Among these various velocity fields, the linear gradients are the most common (44 sources), while the other types occur much less often (eight sources).

To quantify the observed \ttcosixfive{} velocity fields, we estimated the mean velocity gradients (MVGs) by fitting the following equation from \citet{goodman1993dense} to our $M_1$ maps using the Python function \texttt{scipy.optimize.curve\_fit}: 
\begin{equation}
    v_{\text{LSR}} = v_0 + a\Delta \alpha + b\Delta \beta,
\end{equation}
where $v_{\text{LSR}}$ is the velocity at a given pixel, $v_0$ is the velocity at a reference position, $\Delta \alpha$ and $\Delta \beta$ are the right ascension (RA) and declination (Dec) offsets from the reference position measured in radians, and $a$ and $b$ are the projections of the velocity gradient per radian onto the RA and Dec axes. For this fitting, we fixed the reference position at the centre of each source, since the MVGs measure the general direction of gas motions and hence the fitted $a$ and $b$ do not depend on the reference position.

Once $a$ and $b$ were constrained, the magnitude of the MVG, $\mathcal{G}$ (km\,s$^{-1}$\,pc$^{-1}$), was computed as follows: 
\begin{equation}
\label{eq:mvg_g}
\mathcal{G} = \frac{\sqrt{a^2 + b^2}}{D},    
\end{equation}
where $D$ is the distance to the source. In addition, the position angle (PA) of the MVG, $\theta_{\mathcal{G}}$ (degree), was measured counterclockwise from the north to the red-shifted gradient (i.e., red arrows in Fig.\,\ref{fig:3tracers}). Specifically, $\theta_{\mathcal{G}}$ was calculated from \mbox{$\alpha = \left| \text{tan}^{-1} \left(a/b\right) \right|$} as follows:

\begin{equation}
\label{eq:mvg_theta}
    \theta_{\mathcal{G}} =
    \begin{cases}
        \alpha \quad \quad \quad \quad \mathrm{if} \quad a > 0 \quad \mathrm{and} \quad b > 0 \\
        180^{\circ} - \alpha \quad \mathrm{if} \quad a > 0 \quad \mathrm{and} \quad b < 0 \\ 
        180^{\circ} + \alpha \quad \mathrm{if} \quad a < 0 \quad \mathrm{and} \quad b < 0 \\
        360^{\circ} - \alpha \quad \mathrm{if} \quad a < 0 \quad \mathrm{and} \quad b > 0 .
    \end{cases}
\end{equation}

\begin{figure}
    \centering
    \resizebox{0.9\hsize}{!}{\includegraphics{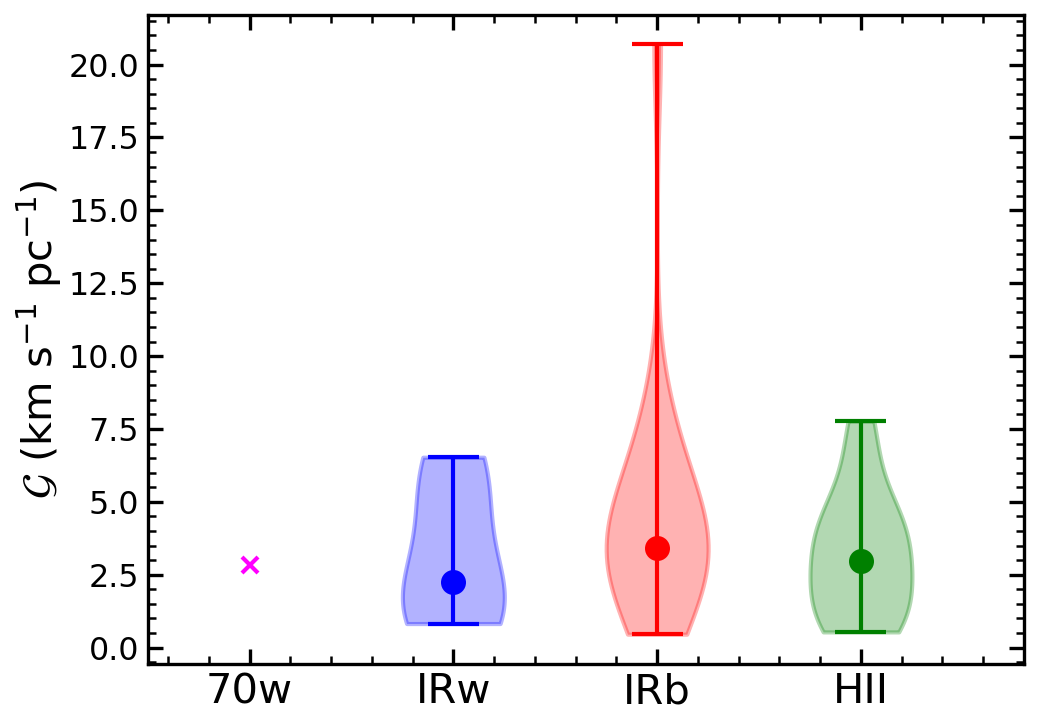}}
    \caption{Distributions of the MVG magnitude, $\mathcal{G}$, for the different evolutionary groups. The colours and symbol schemes are the same as in Fig.\,\ref{fig:linepeaks}.}
    \label{fig:hist_gradient_magnitude}
\end{figure}

We examined the constrained $\mathcal{G}$ for 41 sources (one 70w, 5 IRw, 21 IRb, and 14 HII sources) that have statistically significant gradients $\mathcal{G} \ge 3 \sigma_{\mathcal{G}}$ \citep{goodman1993dense} and found that $\mathcal{G}$ generally ranges from 1\,km\,s$^{-1}$\,pc$^{-1}$ to 8\,km\,s$^{-1}$\,pc$^{-1}$ independent of the evolutionary group (Fig.\,\ref{fig:hist_gradient_magnitude}). An outlier is G351.77$-$0.5, whose large gradient of 20\,km\,s$^{-1}$\,pc$^{-1}$ likely originates from the source's strong outflow. Interestingly, our median $\mathcal{G}$ ($\sim$3\,km\,s$^{-1}$\,pc$^{-1}$) is comparable to that estimated by \citet{tobin2011complex} for a sample of low-mass protostar envelopes (2\,km\,s$^{-1}$\,pc$^{-1}$). A one-to-one comparison of these results, however, is not straightforward, considering that the angular resolution of the study by \citet{tobin2011complex} is three times larger than that of our data.
On the other hand the sources in that study are, at distances of 125--460\,pc, much closer than the sources in our sample. As we demonstrate in Appendix\,\ref{a:mvg_multi_beam_size}, the measurement of the MVGs depends on the physical resolution.

\subsection{Comparison of gas kinematics on different spatial scales} \label{s:velmap}
Section\,\ref{s:mvg} shows that \ttcosixfive{} emission in our single-core sources frequently shows linear velocity gradients. To provide insights into the origin of these velocity gradients, we compared gas kinematics on three different spatial scales, including outflows from small-scale YSO cores that are traced by high velocity \cosixfive{} emission, intermediate-scale envelopes traced by \ttcosixfive{} emission, and large-scale clumps traced by \ttcotwoone{} emission. For sources in the early stage of star formation, we can expect that the intermediate-scale envelopes retain the kinematics of their parental clumps, inducing an alignment between the envelope and clump kinematics \citep[e.g.,][]{smith11}. As star formation progresses, strong outflows may be launched and entrain dense gas from the envelopes, resulting in an alignment between the outflow and envelope kinematics, similar to envelopes of \mbox{Class 0} low-mass protostars \citep{arce2006evolution}. Once the outflows significantly erode into the envelopes, the remaining envelope material could be concentrated near the protostellar disks and be rotating together, leading to a perpendicular alignment between the outflow and envelope kinematics, similar to what is found in envelopes of low-mass \mbox{Class I} protostars by \citet{arce2006evolution}. These possible scenarios demonstrate that the comparison of gas kinematics on different spatial scales can be valuable for understanding the processes of high-mass star formation.

\begin{figure*}
    \centering
    \resizebox{1.0\hsize}{!}{\includegraphics{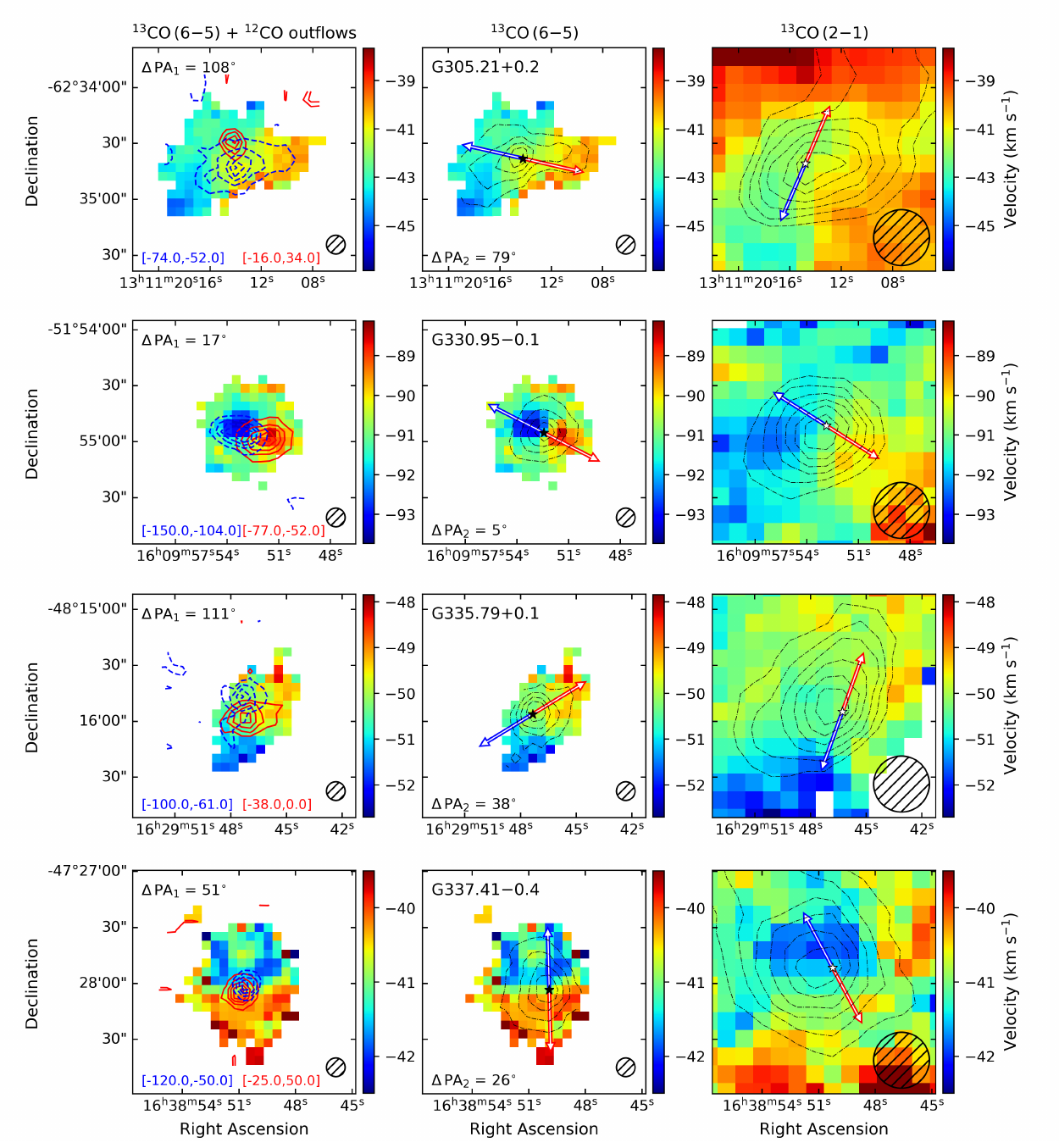}}
    \caption{Comparison of \ttcosixfive{}, \cosixfive{}, and \ttcotwoone{} gas kinematics for four example sources. (Left column) $M_1$ maps of \ttcosixfive{} emission. For each source, the blue and red lobes of \cosixfive{} outflows are overlaid as contours with decreasing levels ranging from 95\% of the peak integrated intensities down to the first level above 3$\sigma$, in steps of 20\% peak values. For G309.38$-$0.1 and G343.76$-$0.1 (Appendix\,\ref{a:15clumps}), a threshold of 2$\sigma$ was used instead. The velocity ranges over which the blue and red lobes were integrated are presented at the bottom left, while the absolute angular difference between the PAs of the \ttcosixfive{} and \cosixfive{} MVGs ($\Delta$\,PA$_1$) is shown at the top left. (Middle column) Same $M_1$ maps as in the left panel, but with the \ttcosixfive{} integrated intensities as contours (20\,\% to 80\,\% of the peak values in steps of 20\,\%). The centre of each source with the maximum integrated intensity is indicated as a black star, while the direction of the \ttcosixfive{} MVG is shown as blue and red arrows. The absolute angular difference between the PAs of the \ttcosixfive{} and \ttcotwoone{} MVGs ($\Delta$\,PA$_2$) is summarised at the bottom left. (Right column) $M_1$ maps of \ttcotwoone{} emission. The \ttcotwoone{} integrated intensities are shown as contours with levels ranging from 40\,\% to 90\,\% of the peak values in steps of 10\,\%, while the directions of the \ttcotwoone{} MVGs are indicated as blue and red arrows. The centres of the sources with the peak integrated intensities are marked as white stars. Finally, for each plot, the FWHM of corresponding telescope beam is displayed by a hatched circle.}
    \label{fig:3tracers}
\end{figure*}

\begin{figure}
    \centering
    \resizebox{0.9\hsize}{!}{\includegraphics{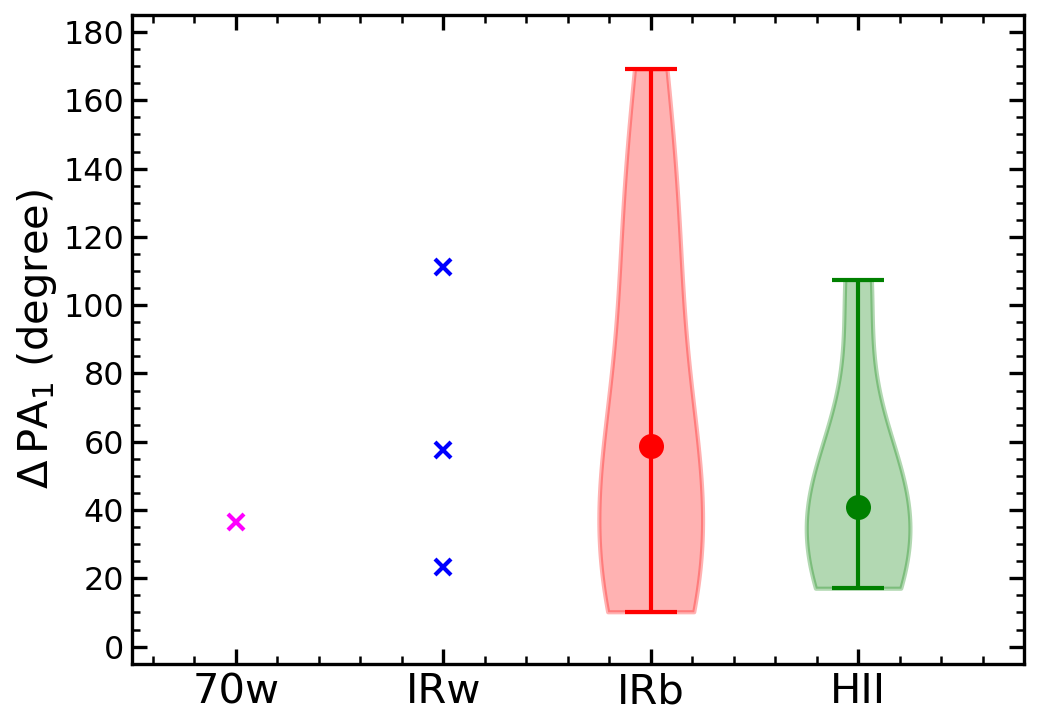}}
    \caption{Distributions of the absolute angular difference $\Delta$PA$_{1}$ for the different evolutionary groups. The colours and symbol schemes are the same as in Fig.\,\ref{fig:linepeaks}.}
    \label{fig:deltaPA_mvg_n_outflw}
\end{figure}

We started our comparison of gas kinematics by evaluating how velocity fields of the outflows and the envelopes are linked to each other. To do so, we utilised \cosixfive{} outflow data from Navarete et al. (in preparation) and examined the absolute angular difference $\Delta$PA$_{1}$ $=$ |\,PA of the \ttcosixfive{} MVGs $-$ PA of the \cosixfive{} outflows\,| (Fig.\,\ref{fig:3tracers} and Appendix\,\ref{a:15clumps}). Navarete et al. identified outflow sources in the Top100 sample by integrating \cosixfive{} emission in high-velocity wings and divided them into several categories, such as single, multiple, unresolved, and resolved (defined as clearly separated blue and red lobes) outflows. For our examination, we computed $\Delta$PA$_{1}$ for 22 sources (one 70w, three IRw, twelve IRb, and six HII sources) that have single resolved outflows along with significant \ttcosixfive{} MVGs and compared the resultant distributions for the different evolutionary groups. Figure \ref{fig:deltaPA_mvg_n_outflw} shows that the $\Delta$PA$_{1}$ distributions support neither the envelope entrainment ($\Delta$PA$_{1} \sim 0^{\circ}$) nor the envelope-disk co-rotation ($\Delta$PA$_{1} \sim 90^{\circ}$) and are not significantly different between the evolutionary groups.

\begin{table*}
\begin{center} 
\caption{Possible interpretations of the observed gas kinematics} 
\label{t:interpreation_gas_kinematics}     
\renewcommand{\footnoterule}{}  % to avoid a line before footnotes
\begin{tabular}{c l c}     % 8 columns 
\hline\hline       
Criteria & Interpretations\tablefootmark{a} & \\ \hline          
$\Delta$PA$_{1} \leq 45^{\circ}$ \& $\Delta$PA$_{2} \leq 45^{\circ}$ & The outflows affect the envelopes, as well as the clumps. & (5)\\
$\Delta$PA$_{1} \leq 45^{\circ}$ \& $\Delta$PA$_{2} > 45^{\circ}$ & The envelopes are impacted by the outflows and are decoupled from the clumps. & (2)\\
45$^{\circ} < \Delta$PA$_{1} \leq 135^{\circ}$ \& $\Delta$PA$_{2} \leq 45^{\circ}$ & The envelopes and clumps rotate along the outflow axes.  & (4)\\ 
45$^{\circ} < \Delta$PA$_{1} \leq 135^{\circ}$ \& $\Delta$PA$_{2} > 45^{\circ}$ & The envelopes rotate along the outflow axes and are decoupled from the clumps. & (3)\\
$\Delta$PA$_{1} > 135^{\circ}$ \& $\Delta$PA$_{2} \leq 45^{\circ}$ & The envelopes are dominated by the clump kinematics. & (0) \\
$\Delta$PA$_{1} > 135^{\circ}$ \& $\Delta$PA$_{2} > 45^{\circ}$ & The outflows, envelopes, and clumps have different kinematics. & (1)\\
\hline
\end{tabular}
\begin{flushleft}
\tablefoot{\tablefoottext{a}{Numbers in brackets present quantities of clumps in each category.}
} 
\end{flushleft}
\end{center} 
\end{table*}

\begin{figure*}[h!]
  \centering
  \resizebox{0.9\hsize}{!}{\includegraphics[width=\textwidth]{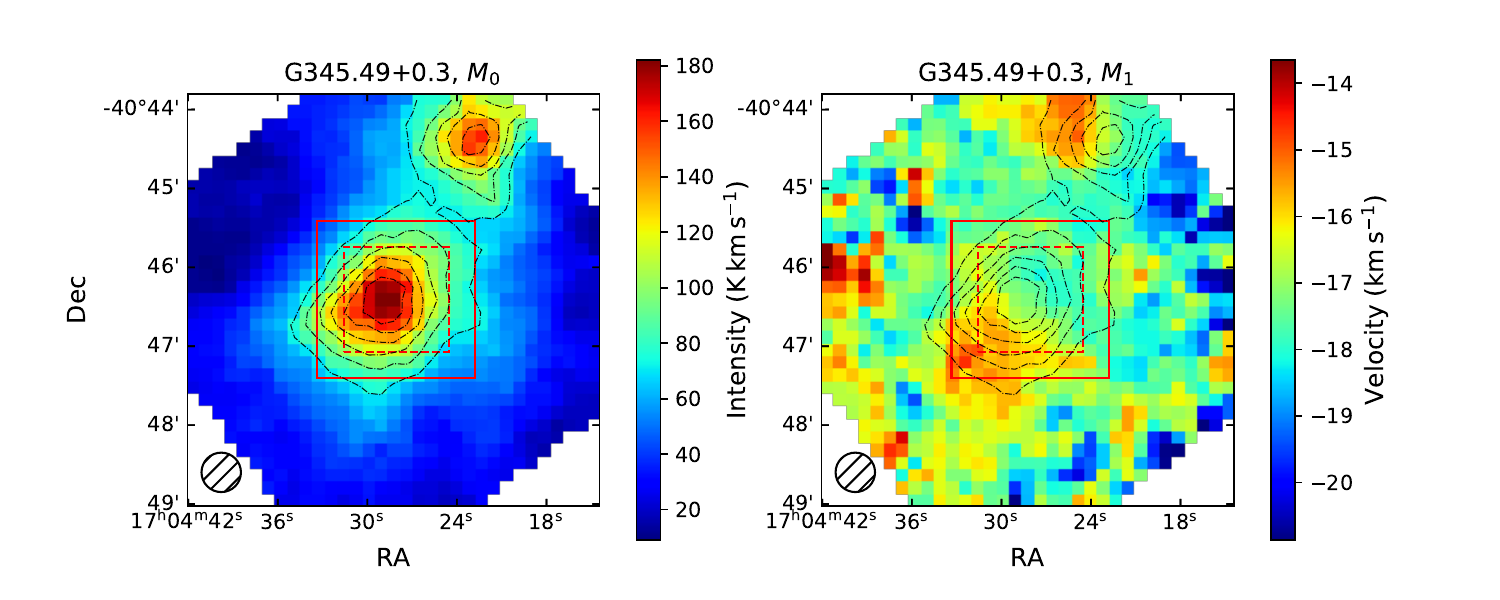}}
  \caption{\ttcotwoone{} $M_0$ (left) and $M_1$ (right) maps of an example source, G345.49+0.3. In both maps, the integrated intensities are overlaid as contours with levels ranging from 40\% to 90\% in steps of 10\% of the largest integrated intensity. The $80^{\prime\prime} \times 80^{\prime\prime}$ coverages of the \ttcosixfive{} data are outlined as dashed red boxes, while the $2^{\prime} \times 2^{\prime}$ regions where we extracted clump-scale kinematics are indicated by solid red boxes. At the left bottom corner of each map, the FWHM of the telescope beam is shown by a hatched circle.}
  \label{fig:34549-13co21}
\end{figure*}

In addition, we probed a relation between the envelope and clump kinematics by comparing the \ttcosixfive{} and \ttcotwoone{} MVGs. To derive \ttcotwoone{} MVGs for our single-core sources, we used data from the SEDIGISM survey \citep{schuller2021sedigism} and produced $2^{\prime} \times 2^{\prime}$ sized $M_0$ and $M_1$ maps (Fig.\,\ref{fig:34549-13co21}). The $2^{\prime} \times 2^{\prime}$ maps, which are $\sim$1.5 times larger than our \ttcosixfive{} maps, were chosen as they include a majority of the regions in which the \ttcotwoone{} line's intensity is higher than 40\% of the peak value. With the resultant $M_0$ and $M_1$ maps, we then estimated MVGs by applying Eqns.\,\ref{eq:mvg_g} and \ref{eq:mvg_theta} and found that 23 sources (two IRw, twelve IRb, and nine HII sources) have statistically significant MVGs in both \ttcotwoone{} and \ttcosixfive{} emission. The estimated \ttcotwoone{} MVGs are presented in Appendix\,\ref{a:table}.

\begin{figure}
    \centering
    \resizebox{0.9\hsize}{!}{\includegraphics{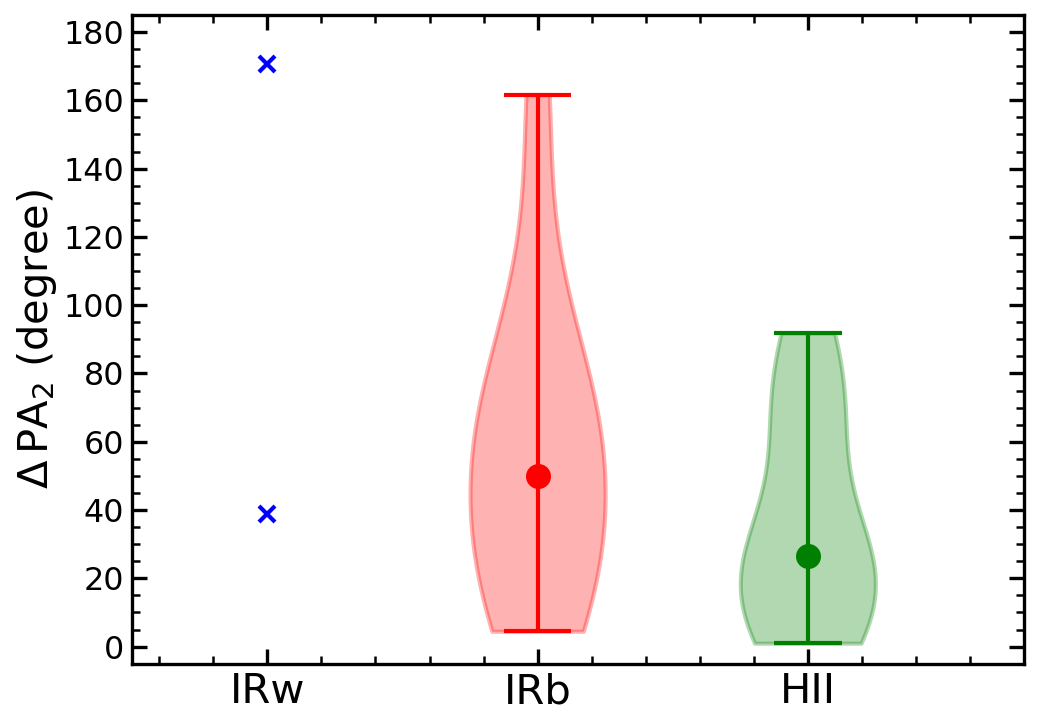}}
    \caption{Distributions of the absolute angular difference $\Delta$PA$_{2}$ for the different evolutionary groups. The colours and symbol schemes are the same as in Fig.\,\ref{fig:M0-fitparas}.}
    \label{fig:deltaPA_mvg21_n_mvg65}
\end{figure}

In Fig.\,\ref{fig:magnitude_mvg21_n_mvg65}, we compared the magnitudes of the \ttcosixfive{} and \ttcotwoone{} MVGs and found that $\mathcal{G}$ is generally larger in \ttcosixfive{} emission. For example, the median magnitudes of the \ttcosixfive{} and \ttcotwoone{} MVGs are 2.8\,km\,s$^{-1}$\,pc$^{-1}$ and 0.7\,km\,s$^{-1}$\,pc$^{-1}$, respectively. The smaller $\mathcal{G}$ for \ttcotwoone{} could partially result from the coarser angular resolution of the SEDIGISM data compared to our \ttcosixfive{} beam (30\arcsec versus 10\arcsec; see Appendix\,\ref{a:mvg_multi_beam_size} for further discussion). In addition, we examined the absolute angular difference $\Delta$PA$_{2}$ $=$ |\,PA of the \ttcosixfive{} MVGs $-$ PA of the \ttcotwoone{} MVGs\,| (Fig.\,\ref{fig:deltaPA_mvg21_n_mvg65}) and found that 13 sources have relatively comparable PAs in both transitions ($\Delta$PA$_{2} \lesssim 45^{\circ}$), while the remaining 10 sources show a wide range of the angular difference. Interestingly, both the magnitude and PA comparisons show no dependence on the evolutionary groups.

Finally, we examined the relationship between the outflow, envelope, and clump kinematics more comprehensively by comparing the \cosixfive{}, \ttcosixfive{}, and \ttcotwoone{} velocity information for a smaller sample of 15 sources. For our comparison, we considered five possibilities based on the measured $\Delta$PA$_{1}$ and $\Delta$PA$_{2}$ (Table 3) and examined them along with the $M_1$ maps. This detailed view on the individual sources revealed that the origin of the \ttcosixfive{} kinematics is relatively clear for seven sources only. For example, the \ttcosixfive{} kinematics is likely decoupled from the surrounding clumps for four sources. Three of them (G330.95$-$0.1, G332.09$-$0.3, and G337.92$-$0.4; two IRb and one HII sources) show a hint of the envelope entrainment, while G305.21$+$0.2 (IRb) has co-rotating envelope and disk. In addition, two sources (G335.79$+$0.1 and G345.00$-$0.2; that is, one IRw and one HII source) show coherent rotations of the envelopes and clumps along the outflow axes. The \ttcosixfive{} kinematics of the remaining source G337.41$-$0.4 (HII) seems to be dominated by the surrounding clump with no obvious impact from outflows.

In summary, we found that our single-core sources have significant linear velocity gradients in \ttcosixfive{} emission. The origin of these velocity gradients is currently unclear, although it is less likely systematically linked to internal star formation processes (that is, $\mathcal{G}$ does not vary much between the different evolutionary groups). The envelope kinematics traced by \ttcosixfive{} emission appears to be highly complex, and probing its origin would be benefited by higher sensitivity, higher resolution observations of a bigger sample of high-mass star-forming regions.

\begin{figure}[h!]
    \hspace{-0.4cm}
    \resizebox{1.1\hsize}{!}{\includegraphics{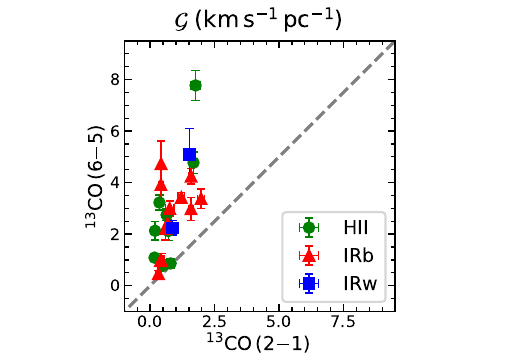}}
    \caption{Comparison between the magnitudes of the \ttcosixfive{} and \ttcotwoone{} MVGs (IRw, IRb, and HII group in blue, red, and green, respectively). A one-to-one relation is indicated by the grey dashed line.}
    \label{fig:magnitude_mvg21_n_mvg65}
\end{figure}

%------------------------
\section{Discussion} \label{s:discussion}
Our observations of \ttcosixfive{} emission towards a statistically significant sample of massive star-forming regions have allowed us to characterise the warm envelopes surrounding high-mass YSOs. In this section, we discuss potential origins of the emission and examine \ttcosixfive{} kinematics in detail for several of our sources.

\subsection{Excitation of \ttcosixfive{} emission in massive star-forming regions} \label{s:warm_gas_tracer}
Key characteristics of the Top100 sources, such as gas temperature and column density, have been examined by various authors based on multi-transition observations of molecular species. For example, \citet{giannetti2017atlasgal} derived temperature by 
analysing emission from CH$_3$CN, CH$_3$CCH, and CH$_3$OH and found a progressive warm-up of the molecular gas with evolution that likely results from stellar feedback. In addition, \citet{giannetti2014atlasgal} showed that the abundance of CO molecules in the gas phase increases with evolution, as warmer dust grains enable frozen CO molecules to sublimate efficiently. A combination of the increased temperature and CO column density in the Top100 sources would excite more CO molecules to mid-$J$ levels, resulting in the observed increase of the \ttcosixfive{} brightness temperature (Sect.\,\ref{s:detection_rate}) and line luminosity (Sect.\,\ref{s:line_luminosity}). This suggests that \ttcosixfive{} emission can be used efficiently to probe the processes of high-mass star formation.

The warm molecular gas traced by mid-$J$ $^{13}$CO emission can be heated by a range of radiative and mechanical feedback processes. For nearby low-mass star-forming regions, these processes can be probed in detail and observational and theoretical studies have suggested that mid-$J$ $^{13}$CO emission originates from passively heated envelopes, UV-heated outflow cavity walls, and outflows themselves \citep{vankempen09II,vankempen09I,yildiz12}. Considering that envelopes and outflows are also prevalent in regions of high-mass star formation, the observed \ttcosixfive{} emission in the Top100 sources could have similar origins.

In the scenario of passively heated envelopes, dust grains are warmed up by absorbing radiation from central protostars and subsequently heat up the surrounding molecular gas via gas-dust collisions \citep{jor02, kris12}. This mechanism has been shown to be responsible for part of the observed \ttcosixfive{} emission in a large sample of low-mass YSOs \citep{yildiz15}. While the same mechanism could certainly work for high-mass star-forming regions, dedicated modelling of gas and dust in massive envelopes is required to assess the exact role of passive heating on the excitation of \ttcosixfive{} emission in such complex systems.

In addition to the passively heated envelopes, \ttcosixfive{} emission could partly arise from UV-heated outflow cavity walls. For low-mass YSOs, UV photons can be created from disk-protostar accretion boundary layers or shocked spots along outflows. These UV photons escape through outflow cavities and are scattered by dust grains on their way, producing PDRs along the cavity walls \citep{spaans95,yildiz15}. While the efficacy of the same mechanism for high-mass star-forming regions has not been confirmed, PDRs have been observed towards many massive YSOs, in particular evolved ones whose intense UV radiation can ionise hydrogen atoms in significant volumes around them. For example, \citet{leurini2013distribution} analysed a number of mid-$J$ CO maps for G327.3$-$0.6 and found that the emission lines, including \ttcosixfive{}, are dominated by the PDR around a local H\textsc{ii} region. In addition, multiple PDR layers have been proposed to reproduce the \ttcosixfive{} observations of several high-mass star-forming regions, such as Cepheus A, DR21, and W49 \citep{graf93,koe94}. Interestingly, our HII sources have \ttcosixfive{} brightness temperatures that are comparable to those measured by \citet{graf93}, implying a non-negligible impact of UV photons on the excitation of \ttcosixfive{} emission in our most evolved sources.

Finally, mechanical processes such as outflows could contribute to the excitation of \ttcosixfive{} emission. For example, \citet{yildiz12} showed that \ttcosixfive{} spectra towards the low-mass protostar NGC 1333 IRAS 4A and 4B have both narrow and broad components with FWHMs of $\sim$2\,km\,s$^{-1}$ and $\sim$10\,km\,s$^{-1}$. The same type of broad component has been found in low-$J$ \citep{yang18} and mid-$J$ $^{13}$CO transitions (Sect.\,\ref{s:line_profiles}) for ATLASGAL sources, implying that outflows are capable of entraining and heating the $^{13}$CO-traced dense gas in massive star-forming regions. Currently, there is a lack of shock modelling effort to study the excitation of \mbox{mid-$J$ $^{13}$CO emission} in outflows, and future studies are needed. In addition to outflows, other mechanical processes during high-mass star formation, such as accretion flows onto clumps and gravitational collapse of clumps \citep{heitsch09,vazquez19}, could also inject a significant amount of energy into the surrounding medium and excite \ttcosixfive{} emission. This mechanism could be vital in particular for early-type clumps without central heating sources where gravitational collapse has just started \citep{pillai23}.

All in all, our study demonstrates that \ttcosixfive{} emission is ubiquitous in a variety of massive clumps, ranging from young sources where protostars have not been yet formed to evolved sources with strong H\textsc{ii} regions. This suggests that \ttcosixfive{} emission could be excited by different processes at different epochs of high-mass star formation.

\subsection{\ttcosixfive{} kinematics in individual clumps} \label{s:kine-fewsrc}
Our velocity analyses for a statistically significant sample suggest the complex nature of \ttcosixfive{} kinematics in high-mass star-forming regions (Sects.\,\ref{s:mvg} and \ref{s:velmap}). In an attempt to shed more light on the processes that drive \ttcosixfive{} kinematics, we present here a detailed examination of several individual sources based on complementary information from literature. Our source selection was based on the availability of kinematics information from other tracers, especially those with high spatial resolutions.

\begin{figure*}[h!]
  \centering
  \begin{subfigure}{0.33\textwidth}
    \centering
    \includegraphics[width=\textwidth]{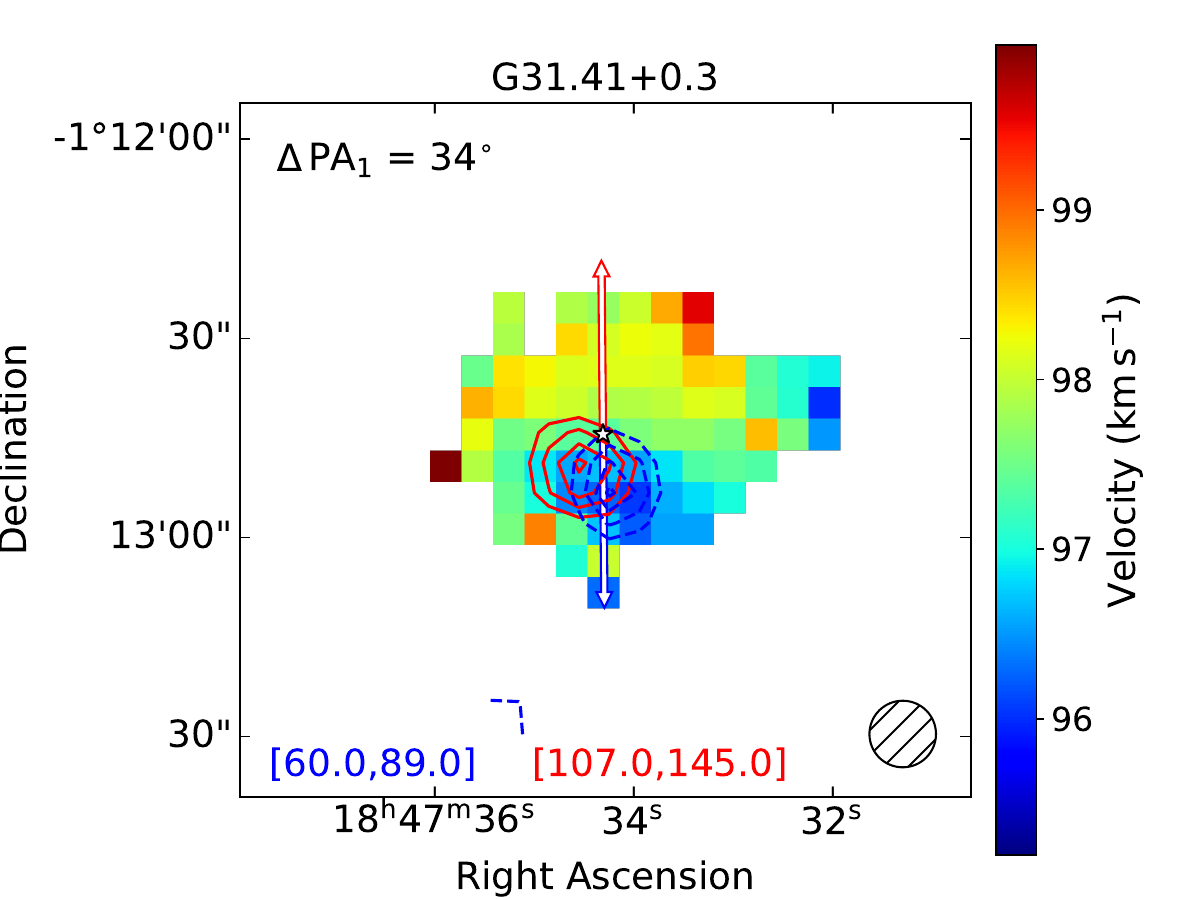}
  \end{subfigure}%
  \begin{subfigure}{0.33\textwidth}
    \centering
    \includegraphics[width=\textwidth]{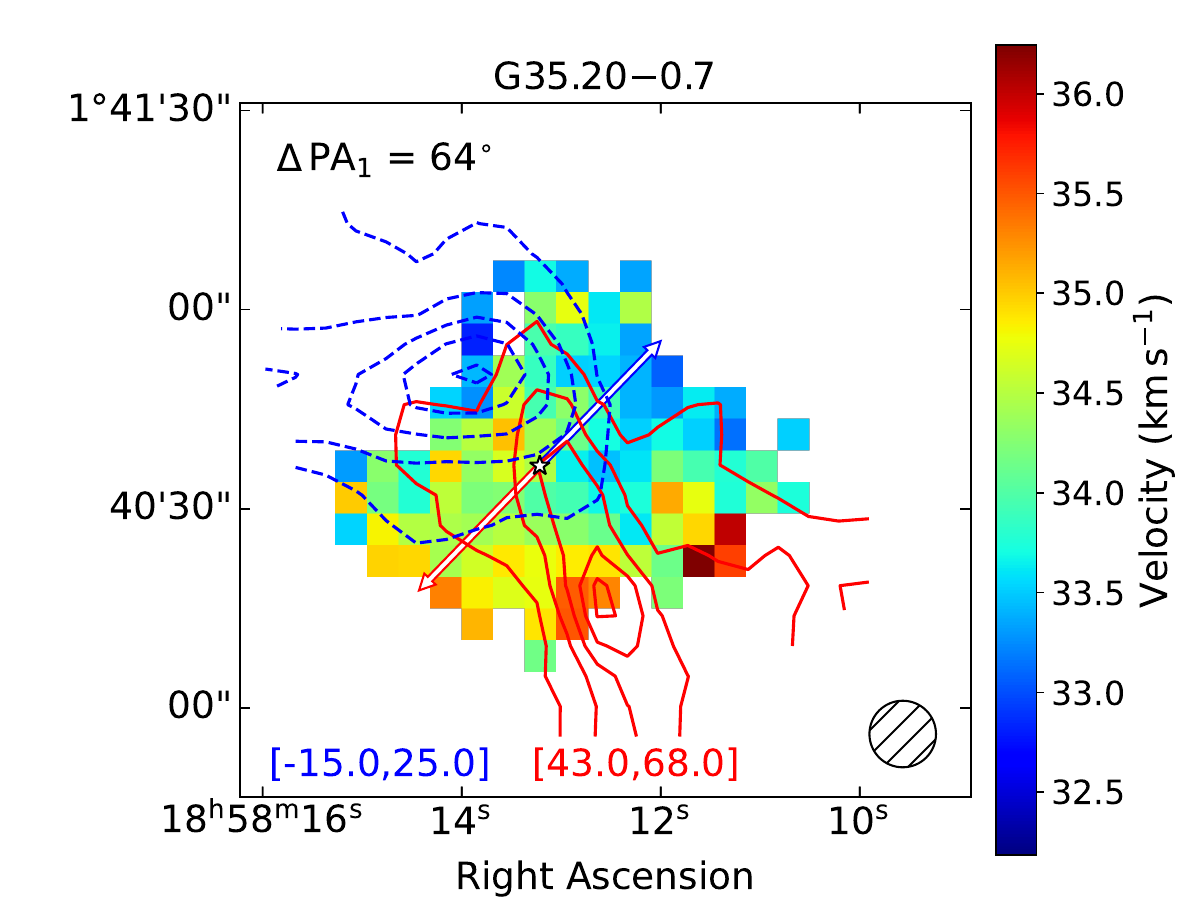}
  \end{subfigure}%
  \begin{subfigure}{0.33\textwidth}
    \centering
    \includegraphics[width=\textwidth]{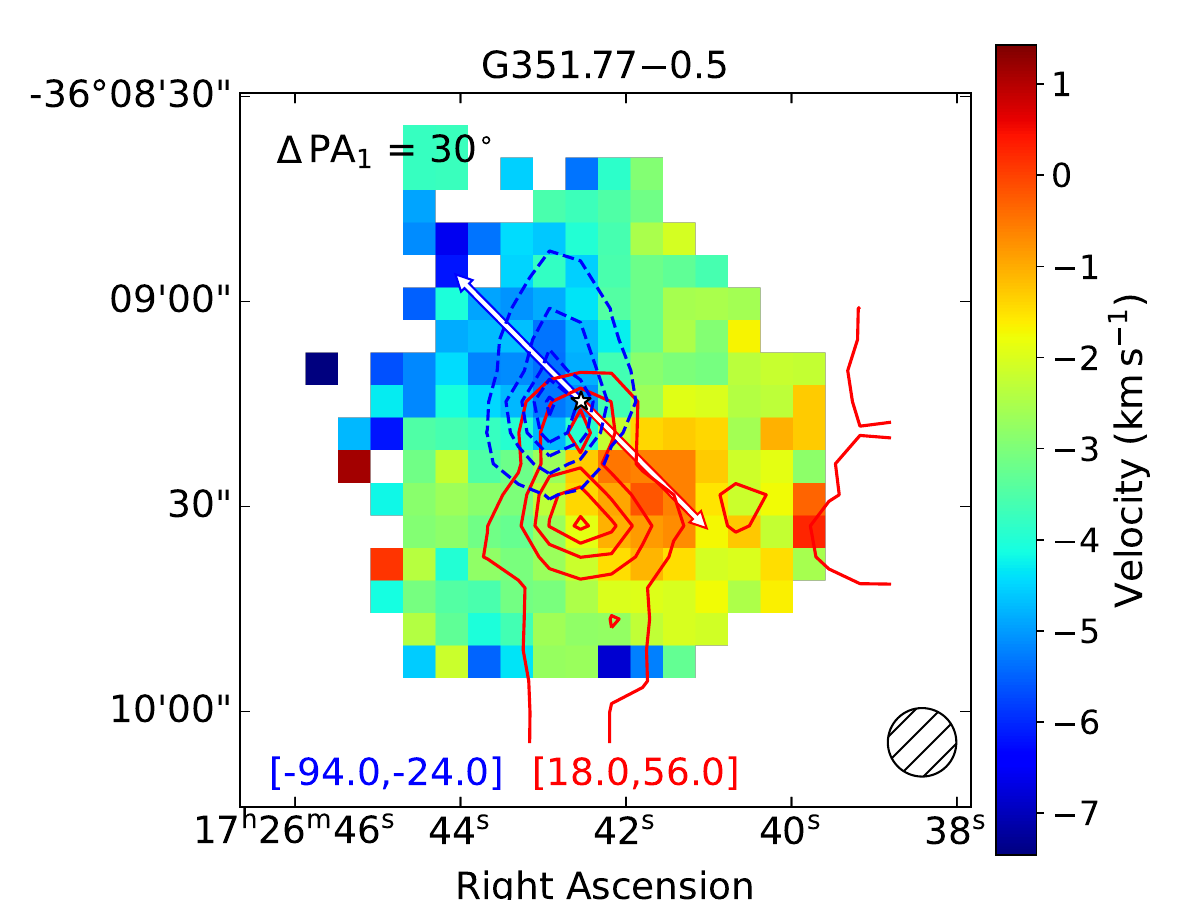}
  \end{subfigure}%
  \caption{\ttcosixfive{} $M_1$ maps of G31.41$+$0.3, G35.20$-$0.7, and G351.77$-$0.5. The descriptions for the overlaid contours, symbols, and labels are the same as in Fig.\ref{fig:3tracers}.}
  \label{fig:addM1maps}
\end{figure*}

\subsubsection*{G43.17+0.0 (HII)}
This HII region source is a part of the well-studied Galactic mini-starburst complex W49A, whose enhanced star formation efficiency has been attributed to various processes, including gravitational collapse \citep{welch87}, a cloud-cloud collision \citep{serabyn93}, expanding shells \citep{peng10}, and re-collapsing shells \citep{rugel19}. In our observations, \ttcosixfive{} emission appears across a region of $S_{\mathrm{eff}}$ $\sim$ 4.4\,pc, which is embedded within a surrounding structure that is at least 10\,pc in diameter. The measured velocity centroids show a radial velocity gradient (Fig.\,\ref{fig:exampleM1maps}) that could result from local collapse considering that the convergence of the gradient points towards the centre of our map. Another possibility is that \ttcosixfive{} emission traces expanding shells. For example, the observed \ttcosixfive{} spectra show double peaks in many places, whose velocities are consistent with those of the shells identified in C$^{18}$O(2--1) emission \citep{peng10}. However, in this case, it is not entirely clear how the expanding shells result in the radial velocity gradient.

\subsubsection*{G31.41+0.3 (HII)}
This HII region source harbours a hot molecular core that has been extensively studied with multiple molecular species. For example, \citet{cesaroni17} performed CH$_3$CN observations and found a rotating and infalling toroid on $\sim$0.1\,pc scales. In addition, \citet{beltran22} identified at least six outflows with lobe sizes of $\sim$0.02--0.2\,pc in SiO\,(5--4) observations. Compared to these small-scale observations, our \ttcosixfive{} data probe a factor of $\sim$10 larger region and show a linear gradient in velocity centroids (Fig.\,\ref{fig:addM1maps}). Interestingly, the direction of the linear velocity gradient is perpendicular to that of the strongest SiO\,(5--4) outflow, implying a rotating \ttcosixfive{} envelope. However, the rotation axis of the envelope is not aligned with that of the small-scale toroid, hampering us from pinpointing the exact origin of \ttcosixfive{} kinematics.

\subsubsection*{G35.20$-$0.7 (IRb)}
As a well-known star-forming region, G35.20$-$0.7 hosts an elongated dust structure that is embedded within a butterfly-shaped reflection nebula \citep{sanchez13}. Compared to the size of the observed \ttcosixfive{} envelope, the elongated structure is roughly one third in length, while the reflection nebula is on a similar scale. \citet{sanchez14} argued that the elongated structure is a filament, rather than an edge-on rotating disk, based on C$^{17}$O\,(3$-$2) and H$^{13}$CO$^+$\,(4$-$3) observations and further suggested that G35.20$-$0.7 is a convergence centre of multiple filaments. Interestingly, the velocity range of 31--37\,km\,s$^{-1}$ along the elongated structure is comparable to the \ttcosixfive{} velocity gradient, implying that \ttcosixfive{} kinematics could trace the gas flow within the supposed converging filaments.

\subsubsection*{G351.77$-$0.5 (IRb)}
The first moment map of G351.77$-$0.5 clearly shows blue-shifted and red-shifted lobes (Fig.\,\ref{fig:addM1maps}), suggesting an outflow origin of \ttcosixfive{} kinematics. Other studies have identified outflows in this region as well. For example, Navarete et al. (in preparation) found an outflow traced by \cosixfive{} emission (blue and red contours in Fig.\,\ref{fig:addM1maps}), whose PA is slightly different from that of our \ttcosixfive{} outflow. In addition, \citet{leurini09,leurini11} detected three outflows in high-resolution ($5.45^{\prime\prime} \times 2.08^{\prime\prime}$) $^{12}$CO\,(2$-$1) observations. Among the three outflows, two are relatively well-aligned with the \cosixfive{} outflow on a similar scale. The angular offset between the \ttcosixfive{} and $^{12}$CO outflows suggests that different transitions and isotopologues of CO molecules probe different parts of the complex outflow system in G351.77$-$0.5. 
\\

In summary, our detailed examination of the four individual clumps supports the previous conclusion that \ttcosixfive{} kinematics has a complex nature. Both small-scale (e.g., outflows) and large-scale (e.g., expanding shells and gas flows along filaments) processes impact on the kinematics of warm envelopes, and understanding their impact would be essential to elucidate how \ttcosixfive{} emission is excited in massive star-forming regions.

%---------------------------
\section{Conclusions} \label{s:conclusions}
In this paper, we characterised \ttcosixfive{} emission towards $\sim$100 massive star-forming clumps at different evolutionary stages. Our APEX observations with high angular and spectral resolutions allowed us to examine the morphology and kinematics of warm envelopes for a statistically significant sample. We summarise our main results and conclusions as follows:

\begin{itemize}
    \item The \ttcosixfive{} emission line is detected towards sources at all stages of high-mass star formation. Key characteristics of the emission line, including the detection rate, line width, and peak brightness temperature, increase with evolutionary stages. In addition, the line luminosity is tightly correlated with the bolometric luminosity and the clump mass. These results suggest that the \ttcosixfive{} emission line is related to star formation processes. 
    
    \item The \ttcosixfive{} integrated intensity images show that a majority of the warm envelopes have a simple single-core morphology and the most evolved clumps (HII) have larger envelopes. The radial intensity profiles of the single-core clumps can be reasonably well described as single power-law functions. Interestingly, the slopes of the most evolved group (HII), tend to be steeper, possibly due to enhancements in density and/or temperature at the central parts of the warm envelopes. 

    \item The kinematics of the warm envelopes presents various forms and is likely not driven by a single physical process. While most of the single-core envelopes show significant linear velocity gradients, the origin of these gradients is unclear in many cases. A detailed examination of individual sources suggests that the gas motions in the warm envelopes could be driven by stellar feedback such as outflows and/or by inheriting the kinematics of large-scale parental clouds. 
\end{itemize}

\begin{acknowledgements}
The authors thank the anonymous referee for the constructive report that helped us improve this paper. We thank MPIfR and APEX staff for operating the observatory and collecting the data. The work of FN is supported by NOIRLab, which is managed by the Association of Universities for Research in Astronomy (AURA) under a cooperative agreement with the National Science Foundation. $Herschel$ was an ESA space observatory with science instruments provided by European-led Principal Investigator consortia and with important participation from NASA.
\end{acknowledgements}

\bibliographystyle{aa} 
\bibliography{ref}

%------------
\appendix

\onecolumn
\section{Tables} 
\label{a:table}

\begin{center}
\begin{longtable}{clcccccccc}
    \caption{\ttcosixfive{} and \cosixfive{} line parameters} \label{t:line_prop} \\
    \hline
    \hline 
    No. & Source & ATLASGAL name & Group & $P$ & $M_0$ & $L_{^{13}\mathrm{CO}}$ & $L_{^{12}\mathrm{CO}}$ & $\tau_{^{13}\mathrm{CO}}$ & Map type \\
        &        &   &             & (K) & (K\,km\,s$^{-1}$) & (W) & (W)   &  &  \\
    \hline 
1 & G8.68$-$0.3 & AGAL008.684$-$00.367 & IRw & 6.5 & 28.2 & $2.5 \times 10^{25}$ & $6.2 \times 10^{25}$ & 0.85 & S \\
2 & G8.71$-$0.4 & AGAL008.706$-$00.414 & IRw & -- & -- & -- & $1.1 \times 10^{25}$ & -- & -- \\
3 & G10.45$-$0.0 & AGAL010.444$-$00.017 & IRw & -- & -- & -- & $6.9 \times 10^{25}$ & -- & -- \\
4 & G10.47+0.0 & AGAL010.472+00.027 & HII & 9.7 & 75.7 & $2.1 \times 10^{26}$ & $5.6 \times 10^{26}$ & 0.45 & S$^{*}$ \\
5 & G10.62$-$0.3 & AGAL010.624$-$00.384 & HII & 37.8 & 350.2 & $3.3 \times 10^{26}$ & $4.4 \times 10^{26}$ & 1.25 & S \\
6 & G12.81$-$0.2 & AGAL012.804$-$00.199 & HII & 28.3 & 275.3 & $6.0 \times 10^{25}$ & $1.2 \times 10^{26}$ & 0.66 & F \\
7 & G13.18+0.0 & AGAL013.178+00.059 & 70w & 5.1 & 17.8 & $3.9 \times 10^{24}$ & $2.1 \times 10^{25}$ & 0.50 & S \\
8 & G13.66$-$0.6 & AGAL013.658$-$00.599 & IRb & 3.3 & 14.2 & $1.1 \times 10^{25}$ & $8.5 \times 10^{25}$ & 0.38 & P \\
9 & G14.11$-$0.5 & AGAL014.114$-$00.574 & IRw & 8.7 & 27.7 & $7.1 \times 10^{24}$ & $1.5 \times 10^{25}$ & 0.88 & S \\
10 & G14.19$-$0.1 & AGAL014.194$-$00.194 & IRw & 2.0 & 7.4 & $4.3 \times 10^{24}$ & $6.7 \times 10^{25}$ & 0.14 & P \\
11 & G14.49$-$0.1 & AGAL014.492$-$00.139 & 70w & -- & -- & -- & $1.7 \times 10^{25}$ & -- & -- \\
12 & G14.63$-$0.5 & AGAL014.632$-$00.577 & IRw & 7.9 & 33.1 & $4.1 \times 10^{24}$ & $1.3 \times 10^{25}$ & 0.42 & S \\
13 & G15.03$-$0.6 & AGAL015.029$-$00.669 & IRb & 49.3 & 399.8 & $6.0 \times 10^{25}$ & $8.9 \times 10^{25}$ & 0.63 & E \\
14 & G18.61$-$0.0 & AGAL018.606$-$00.074 & IRw & 1.6 & 4.3 & $3.0 \times 10^{24}$ & $3.4 \times 10^{25}$ & 0.15 & P \\
15 & G18.73$-$0.2 & AGAL018.734$-$00.226 & IRw & 2.1 & 11.6 & $6.9 \times 10^{25}$ & $5.0 \times 10^{26}$ & 0.23 & P \\
16 & G18.89$-$0.4 & AGAL018.888$-$00.474 & IRw & 3.8 & 14.8 & $1.2 \times 10^{25}$ & $1.1 \times 10^{26}$ & 0.23 & S \\
17 & G19.88$-$0.5 & AGAL019.882$-$00.534 & IRb & 11.5 & 59.8 & $3.1 \times 10^{25}$ & $1.4 \times 10^{26}$ & 0.48 & S \\
18 & G22.37+0.4 & AGAL022.376+00.447 & IRw & -- & -- & -- & $1.7 \times 10^{25}$ & -- & -- \\
19 & G23.21$-$0.3 & AGAL023.206$-$00.377 & IRw & 5.4 & 26.9 & $2.2 \times 10^{25}$ & $1.3 \times 10^{26}$ & 0.54 & P \\
20 & G24.63+0.1 & AGAL024.629+00.172 & IRw & -- & -- & -- & $4.2 \times 10^{25}$ & -- & -- \\
21 & G28.56$-$0.2 & AGAL028.564$-$00.236 & IRw & -- & -- & -- & $2.5 \times 10^{25}$ & -- & -- \\
22 & G28.86+0.0 & AGAL028.861+00.066 & IRb & 10.3 & 55.3 & $1.1 \times 10^{26}$ & $4.9 \times 10^{26}$ & 0.65 & S \\
23 & G30.82$-$0.0 & AGAL030.818$-$00.056 & IRb & 10.0 & 71.5 & $6.5 \times 10^{25}$ & $1.7 \times 10^{26}$ & 0.52 & S \\
24 & G30.85$-$0.0 & AGAL030.848$-$00.081 & 70w & -- & -- & -- & $5.7 \times 10^{25}$ & -- & -- \\
25 & G30.89+0.1 & AGAL030.893+00.139 & 70w & -- & -- & -- & $2.9 \times 10^{25}$ & -- & -- \\
26 & G31.41+0.3 & AGAL031.412+00.307 & HII & 16.0 & 103.4 & $9.4 \times 10^{25}$ & $1.4 \times 10^{26}$ & 1.25 & S \\
27 & G34.26+0.1 & AGAL034.258+00.154 & HII & 38.0 & 272.2 & $2.6 \times 10^{25}$ & $3.5 \times 10^{25}$ & 1.13 & S \\
28 & G34.40+0.2 & AGAL034.401+00.226 & HII & 8.4 & 55.1 & $5.3 \times 10^{24}$ & $2.1 \times 10^{25}$ & 0.35 & S$^{*}$ \\
29 & G34.41+0.2 & AGAL034.411+00.234 & IRb & 4.9 & 23.1 & $2.2 \times 10^{24}$ & $1.3 \times 10^{25}$ & 0.39 & S$^{*}$ \\
30 & G34.82+0.3 & AGAL034.821+00.351 & IRb & 6.3 & 27.9 & $2.7 \times 10^{24}$ & $9.6 \times 10^{24}$ & 0.86 & F \\
31 & G35.20$-$0.7 & AGAL035.197$-$00.742 & IRb & 14.9 & 112.9 & $2.1 \times 10^{25}$ & $6.5 \times 10^{25}$ & 0.49 & S \\
32 & G37.55+0.2 & AGAL037.554+00.201 & IRb & 4.8 & 27.2 & $4.6 \times 10^{25}$ & $1.4 \times 10^{26}$ & 0.76 & P \\
33 & G43.17+0.0 & AGAL043.166+00.011 & HII & 41.7 & 677.6 & $3.2 \times 10^{27}$ & $4.3 \times 10^{27}$ & 1.60 & S \\
34 & G53.14+0.0 & AGAL053.141+00.069 & IRb & 11.7 & 56.6 & $5.5 \times 10^{24}$ & $1.8 \times 10^{25}$ & 0.61 & S \\
35 & G59.78+0.0 & AGAL059.782+00.066 & IRb & 8.6 & 30.0 & $5.5 \times 10^{24}$ & $3.7 \times 10^{25}$ & 0.38 & S$^{*}$ \\
36 & G301.13$-$0.2 & AGAL301.136$-$00.226 & HII & 33.7 & 197.8 & $1.4 \times 10^{26}$ & $5.3 \times 10^{26}$ & -- & S \\
37 & G305.19$-$0.0 & AGAL305.192$-$00.006 & IRw & 3.9 & 16.5 & $9.0 \times 10^{24}$ & $6.5 \times 10^{25}$ & 0.31 & P \\
38 & G305.21+0.2 & AGAL305.209+00.206 & IRb & 30.2 & 230.8 & $1.3 \times 10^{26}$ & $2.3 \times 10^{26}$ & 1.38 & S \\
39 & G305.56+0.0 & AGAL305.562+00.014 & IRb & 21.7 & 105.4 & $5.7 \times 10^{25}$ & $1.5 \times 10^{26}$ & 0.69 & S \\
40 & G305.80$-$0.1 & AGAL305.794$-$00.096 & 70w & -- & -- & -- & $1.2 \times 10^{25}$ & -- & -- \\
41 & G309.38$-$0.1 & AGAL309.384$-$00.134 & IRb & 5.2 & 17.7 & $1.9 \times 10^{25}$ & $1.1 \times 10^{26}$ & 0.47 & S \\
42 & G310.01+0.3 & AGAL310.014+00.387 & IRb & 9.0 & 37.8 & $1.8 \times 10^{25}$ & $6.1 \times 10^{25}$ & 1.17 & S \\
43 & G313.58+0.3 & AGAL313.576+00.324 & IRb & 7.0 & 30.2 & $1.6 \times 10^{25}$ & $7.3 \times 10^{25}$ & 0.59 & S \\
44 & G316.64$-$0.0 & AGAL316.641$-$00.087 & IRb & 2.3 & 6.3 & $3.4 \times 10^{23}$ & $3.5 \times 10^{24}$ & 0.43 & P \\
45 & G317.87$-$0.1 & AGAL317.867$-$00.151 & IRw & 4.2 & 17.6 & $6.0 \times 10^{24}$ & $2.6 \times 10^{25}$ & 0.53 & P \\
46 & G318.78$-$0.1 & AGAL318.779$-$00.137 & IRw & 1.8 & 8.3 & $2.5 \times 10^{24}$ & $1.2 \times 10^{25}$ & 0.32 & P \\
47 & G320.88$-$0.4 & AGAL320.881$-$00.397 & 70w & 2.9 & 6.0 & $2.3 \times 10^{25}$ & $1.4 \times 10^{26}$ & 0.39 & P \\
48 & G326.66+0.5 & AGAL326.661+00.519 & IRb & 26.6 & 77.6 & $9.5 \times 10^{24}$ & $2.2 \times 10^{25}$ & 0.60 & S \\
49 & G326.99$-$0.0 & AGAL326.987$-$00.032 & IRw & -- & -- & -- & $3.8 \times 10^{25}$ & -- & -- \\
50 & G327.12+0.5 & AGAL327.119+00.509 & IRb & 6.9 & 20.9 & $2.4 \times 10^{25}$ & $8.2 \times 10^{25}$ & 0.92 & S \\
51 & G327.39+0.2 & AGAL327.393+00.199 & IRb & 3.5 & 13.9 & $1.8 \times 10^{25}$ & $1.1 \times 10^{26}$ & 0.36 & P \\
52 & G328.81+0.6 & AGAL328.809+00.632 & HII & 33.7 & 274.1 & $9.3 \times 10^{25}$ & -- & -- & S \\
53 & G329.03$-$0.2 & AGAL329.029$-$00.206 & IRw & 4.7 & 27.3 & $1.4 \times 10^{26}$ & $8.9 \times 10^{26}$ & 0.38 & P \\
54 & G329.07$-$0.3 & AGAL329.066$-$00.307 & IRb & 3.4 & 14.2 & $7.2 \times 10^{25}$ & $4.5 \times 10^{26}$ & 0.35 & P \\
55 & G330.88$-$0.3 & AGAL330.879$-$00.367 & HII & 22.3 & 175.8 & $1.2 \times 10^{26}$ & $3.8 \times 10^{26}$ & 0.98 & S \\
56 & G330.95$-$0.1 & AGAL330.954$-$00.182 & HII & 25.8 & 261.3 & $8.5 \times 10^{26}$ & $1.6 \times 10^{27}$ & 1.04 & S \\
57 & G331.71+0.5 & AGAL331.709+00.582 & IRw & 3.0 & 13.2 & $5.5 \times 10^{25}$ & $7.0 \times 10^{26}$ & 0.21 & P \\
58 & G332.09$-$0.4 & AGAL332.094$-$00.421 & IRb & 8.6 & 43.6 & $2.1 \times 10^{25}$ & $9.3 \times 10^{25}$ & 0.43 & S \\
59 & G332.82$-$0.5 & AGAL332.826$-$00.549 & HII & 23.8 & 187.0 & $9.1 \times 10^{25}$ & $2.0 \times 10^{26}$ & 0.76 & S \\
60 & G333.13$-$0.4 & AGAL333.134$-$00.431 & HII & 31.5 & 364.1 & $1.8 \times 10^{26}$ & $3.5 \times 10^{26}$ & 0.72 & E \\
61 & G333.28$-$0.3 & AGAL333.284$-$00.387 & HII & 22.0 & 130.1 & $6.4 \times 10^{25}$ & $1.1 \times 10^{26}$ & 0.72 & S \\
62 & G333.31+0.1 & AGAL333.314+00.106 & IRb & 3.7 & 18.0 & $8.8 \times 10^{24}$ & $7.9 \times 10^{25}$ & 0.29 & S \\
63 & G333.60$-$0.2 & AGAL333.604$-$00.212 & HII & 44.0 & 534.9 & $2.6 \times 10^{26}$ & $5.1 \times 10^{26}$ & 0.63 & F \\
64 & G333.66+0.0 & AGAL333.656+00.059 & 70w & -- & -- & -- & $4.1 \times 10^{25}$ & -- & -- \\
65 & G335.79+0.1 & AGAL335.789+00.174 & IRw & 8.6 & 50.0 & $2.6 \times 10^{25}$ & $1.1 \times 10^{26}$ & 0.57 & S \\
66 & G336.96$-$0.2 & AGAL336.958$-$00.224 & IRw & -- & -- & -- & $2.3 \times 10^{26}$ & -- & -- \\
67 & G337.17$-$0.0 & AGAL337.176$-$00.032 & IRw & 1.1 & 2.9 & $1.3 \times 10^{25}$ & $2.9 \times 10^{26}$ & 0.10 & P \\
68 & G337.26$-$0.1 & AGAL337.258$-$00.101 & IRw & 1.8 & 5.3 & $2.4 \times 10^{25}$ & $2.2 \times 10^{26}$ & 0.30 & P \\
69 & G337.29+0.0 & AGAL337.286+00.007 & 70w & -- & -- & -- & $5.6 \times 10^{25}$ & -- & -- \\
70 & G337.41$-$0.4 & AGAL337.406$-$00.402 & HII & 22.4 & 145.5 & $6.0 \times 10^{25}$ & $1.3 \times 10^{26}$ & 1.15 & S \\
71 & G337.70$-$0.0 & AGAL337.704$-$00.054 & HII & 10.2 & 84.3 & $4.8 \times 10^{26}$ & $9.8 \times 10^{26}$ & 0.84 & S \\
72 & G337.92$-$0.4 & AGAL337.916$-$00.477 & IRb & 30.3 & 260.1 & $1.0 \times 10^{26}$ & $2.5 \times 10^{26}$ & 1.37 & S \\
73 & G338.07+0.0 & AGAL338.066+00.044 & 70w & -- & -- & -- & $3.0 \times 10^{25}$ & -- & -- \\
74 & G338.78+0.4 & AGAL338.786+00.476 & 70w & -- & -- & -- & $1.5 \times 10^{25}$ & -- & -- \\
75 & G338.92+0.5 & AGAL338.926+00.554 & IRb & 9.6 & 76.5 & $5.6 \times 10^{25}$ & $1.8 \times 10^{26}$ & 0.39 & F \\
76 & G339.62$-$0.1 & AGAL339.623$-$00.122 & IRb & 7.3 & 28.9 & $9.8 \times 10^{24}$ & $7.0 \times 10^{25}$ & 0.48 & S \\
77 & G340.37$-$0.3 & AGAL340.374$-$00.391 & IRw & -- & -- & -- & $2.3 \times 10^{25}$ & -- & -- \\
78 & G340.75$-$1.0 & AGAL340.746$-$01.001 & IRb & 6.4 & 22.4 & $6.6 \times 10^{24}$ & $2.9 \times 10^{25}$ & 0.51 & S \\
79 & G340.78$-$0.1 & AGAL340.784$-$00.097 & IRw & 3.9 & 15.9 & $6.0 \times 10^{25}$ & $1.8 \times 10^{26}$ & 0.79 & P \\
80 & G341.22$-$0.2 & AGAL341.217$-$00.212 & IRb & 10.1 & 38.3 & $2.0 \times 10^{25}$ & $8.4 \times 10^{25}$ & 0.55 & S \\
81 & G342.48+0.1 & AGAL342.484+00.182 & IRw & 5.1 & 19.0 & $1.1 \times 10^{26}$ & $5.0 \times 10^{26}$ & 0.58 & S \\
82 & G343.13$-$0.0 & AGAL343.128$-$00.062 & HII & 17.8 & 130.6 & $4.4 \times 10^{25}$ & $2.1 \times 10^{26}$ & 0.79 & S \\
83 & G343.76$-$0.1 & AGAL343.756$-$00.164 & IRw & 6.4 & 34.1 & $1.1 \times 10^{25}$ & $2.5 \times 10^{25}$ & 0.55 & S \\
84 & G344.23$-$0.5 & AGAL344.227$-$00.569 & IRw & 8.1 & 49.7 & $1.2 \times 10^{25}$ & $4.1 \times 10^{25}$ & 0.70 & S \\
85 & G345.00$-$0.2 & AGAL345.003$-$00.224 & HII & 13.3 & 96.4 & $3.3 \times 10^{25}$ & $1.4 \times 10^{26}$ & 0.45 & S \\
86 & G345.49+0.3 & AGAL345.488+00.314 & HII & 24.0 & 163.3 & $3.0 \times 10^{25}$ & $4.5 \times 10^{25}$ & 1.51 & S \\
87 & G345.50+0.3 & AGAL345.504+00.347 & IRb & 20.6 & 116.2 & $2.3 \times 10^{25}$ & $5.9 \times 10^{25}$ & 0.87 & S \\
88 & G345.72+0.8 & AGAL345.718+00.817 & IRb & 8.8 & 30.3 & $2.9 \times 10^{24}$ & $4.8 \times 10^{24}$ & 2.83 & S \\
89 & G351.13+0.7 & AGAL351.131+00.771 & 70w & 3.0 & 6.5 & $8.0 \times 10^{23}$ & $4.8 \times 10^{24}$ & 0.89 & P \\
90 & G351.16+0.7 & AGAL351.161+00.697 & IRb & 32.1 & 249.5 & $3.1 \times 10^{25}$ & $6.0 \times 10^{25}$ & 0.81 & S \\ 
91 & G351.25+0.6 & AGAL351.244+00.669 & IRb & 42.5 & 297.2 & $3.6 \times 10^{25}$ & $7.8 \times 10^{25}$ & 0.71 & E \\
92 & G351.51+0.7 & AGAL351.496+00.691 & IRw & 1.9 & 3.4 & $2.2 \times 10^{23}$ & -- & -- & P \\
93 & G351.57+0.7 & AGAL351.571+00.762 & 70w & 2.4 & 3.2 & $2.0 \times 10^{23}$ & $1.1 \times 10^{24}$ & 0.43 & P \\
94 & G351.58$-$0.3 & AGAL351.581$-$00.352 & IRb & 14.2 & 98.7 & $1.7 \times 10^{26}$ & $4.1 \times 10^{26}$ & 1.20 & S \\
95 & G351.77$-$0.5 & AGAL351.774$-$00.537 & IRb & 34.3 & 329.2 & $1.2 \times 10^{25}$ & $3.2 \times 10^{25}$ & 2.00 & S \\
96 & G353.07+0.4 & AGAL353.066+00.452 & IRw & 3.0 & 7.0 & $2.1 \times 10^{23}$ & $1.8 \times 10^{24}$ & 0.28 & P \\
97 & G353.41$-$0.3 & AGAL353.409$-$00.361 & IRb & 27.7 & 161.5 & $7.0 \times 10^{25}$ & -- & -- & S \\
98 & G353.42$-$0.0 & AGAL353.417$-$00.079 & 70w & -- & -- & -- & $1.7 \times 10^{25}$ & -- & -- \\
99 & G354.94$-$0.5 & AGAL354.944$-$00.537 & 70w & -- & -- & -- & $2.7 \times 10^{24}$ & -- & -- \\
\hline
\end{longtable}
\begin{flushleft}
\tablefoot{The columns are as follows: Source: short version of the ATLASGAL name; ATLASGAL name: source name in the ATLASGAL catalogue; Group: evolutionary group of the source (70w: 70\,\mumeter{} weak, IRw: mid-infrared weak, IRb: mid-infrared bright, HII: H\textsc{ii} region); $P$: \ttcosixfive{} main-beam peak brightness temperature; $M_0$: \ttcosixfive{} integrated intensity; $L_{^{13}\mathrm{CO}}$: \ttcosixfive{} luminosity; $L_{^{12}\mathrm{CO}}$: \cosixfive{} luminosity; $\tau_{^{13}\mathrm{CO}}$: \ttcosixfive{} optical depth (see Appendix\,\ref{a:tau13co} for details); Map type: classification of the source based on its \ttcosixfive{} $M_0$ map (P: poorly resolved, S: single-core, S$^{*}$: single-core in which nearby clumps are masked out, E: extended, F: fragmented-core). The properties of \ttcosixfive{} and \cosixfive{} emission were estimated for the central 20\arcsec and 13\arcsec regions, respectively.} 
\end{flushleft}
\end{center}

\onecolumn
\begin{center}
\begin{longtable}{clcc|cc|cccc}
    \caption{Properties of the $M_0$ maps and the radial profiles for our single-core sources} \label{t:M0map_prop} \\
    \hline 
    \hline 
    No. & Source & $S_{\mathrm{eff}}$ & APR & $r_{\mathrm{b}}$ & $m$ & $r_{\mathrm{b}_1}$ & $r_{\mathrm{b}_2}$ & $m_{\mathrm{i}}$ & $m_{\mathrm{o}}$ \\
    & & (pc) & & (\arcsec) & & (\arcsec) & (\arcsec) & & \\
\hline 
1 & G8.68$-$0.3 & 0.6 & 1.3 & -- & -- & -- & -- & -- & -- \\
2 & G10.47+0.0 & 1.7 & 1.3 & 0.39(0.16) & 1.1(0.1) & 0.81(0.01) & 1.81(0.01) & 1.7(0.0) & 0.7(0.1) \\
3 & G10.62$-$0.3 & 1.8 & 1.2 & 0.84(0.08) & 1.4(0.2) & 0.98(0.02) & 1.78(0.03) & 2.0(0.0) & 1.1(0.2) \\
4 & G13.18+0.0 & 0.4 & 2.4 & 0.88(0.08) & 0.7(0.1) & 0.87(0.09) & 1.40(0.18) & 0.9(0.2) & 0.5(0.0) \\
5 & G14.11$-$0.5 & 0.5 & 1.8 & 0.56(0.11) & 0.6(0.1) & 0.90(0.04) & 1.52(0.04) & 1.1(0.1) & 0.1(0.0) \\
6 & G14.63$-$0.5 & 0.6 & 1.5 & 0.63(0.24) & 0.6(0.1) & 0.74(0.05) & 1.48(0.05) & 1.1(0.1) & 0.6(0.0) \\
7 & G18.89$-$0.4 & 0.9 & 2.4 & -- & -- & -- & -- & -- & -- \\
8 & G19.88$-$0.5 & 0.8 & 1.4 & 0.60(0.07) & 1.4(0.1) & 0.61(0.06) & 1.56(0.02) & 1.6(0.1) & 0.5(0.0) \\
9 & G28.86+0.0 & 1.1 & 1.4 & -- & -- & -- & -- & -- & -- \\
10 & G30.82$-$0.0 & 1.0 & 1.7 & -- & -- & -- & -- & -- & -- \\
11 & G31.41+0.3 & 1.0 & 1.1 & -- & -- & -- & -- & -- & -- \\
12 & G34.26+0.1 & 0.6 & 1.1 & 1.01(0.06) & 1.1(0.2) & 1.00(0.01)$^{*}$ & -- & 1.7(0.0) & -- \\
13 & G34.40+0.2 & 0.3 & 1.5 & -- & -- & -- & -- & -- & -- \\
14 & G34.41+0.2 & 0.2 & 1.2 & -- & -- & -- & -- & -- & -- \\
15 & G35.20$-$0.7 & 0.5 & 1.3 & 0.72(0.05) & 1.4(0.1) & 0.77(0.03) & 1.43(0.08) & 1.6(0.1) & 2.0(0.0) \\
16 & G43.17+0.0 & 4.4 & 1.5 & 0.92(0.03) & 1.2(0.1) & 1.21(0.02)$^{*}$ & -- & 1.9(0.1) & -- \\
17 & G53.14+0.0 & 0.4 & 1.5 & 0.48(0.17) & 0.9(0.1) & 0.68(0.05) & 1.67(0.02) & 1.8(0.1) & 0.4(0.1) \\
18 & G59.78+0.0 & 0.6 & 1.7 & 0.49(0.23) & 0.6(0.1) & 0.79(0.03) & 1.70(0.09) & 1.7(0.1) & 1.3(0.1) \\
19 & G301.13$-$0.2 & 0.8 & 1.0 & 0.92(0.09) & 2.4(0.7) & 0.92(0.01) & 1.75(0.02) & 2.4(0.1) & 1.4(0.1) \\
20 & G305.21+0.2 & 1.1 & 2.4 & 0.45(0.14) & 0.8(0.1) & 0.72(0.06) & 1.51(0.11) & 1.3(0.1) & 0.8(0.1) \\
21 & G305.56+0.0 & 1.2 & 1.5 & 0.52(0.09) & 0.9(0.1) & 0.76(0.08) & 1.35(0.13) & 1.4(0.2) & 0.9(0.0) \\
22 & G309.38$-$0.1 & 0.8 & 2.6 & -- & -- & -- & -- & -- & -- \\
23 & G310.01+0.3 & 0.5 & 1.2 & 0.22(0.20) & 1.0(0.1) & 0.72(0.02) & 1.75(0.02) & 1.6(0.0) & 0.5(0.1) \\
24 & G313.58+0.3 & 0.4 & 1.1 & -- & -- & -- & -- & -- & -- \\
25 & G326.66+0.5 & 0.5 & 1.2 & 0.69(0.11) & 0.8(0.1) & 0.60(0.06) & 1.64(0.04) & 1.3(0.1) & 0.3(0.1) \\
26 & G327.12+0.5 & 0.6 & 1.2 & 0.20(0.14) & 1.0(0.1) & 0.72(0.03) & 1.61(0.02) & 1.9(0.1) & 0.6(0.0) \\
27 & G328.81+0.6 & 0.7 & 1.0 & 0.91(0.02) & 2.3(0.1) & 1.02(0.01) & 1.80(0.02) & 2.4(0.0) & 1.4(0.1) \\
28 & G330.88$-$0.3 & 1.3 & 1.3 & 0.86(0.05) & 1.2(0.1) & 0.87(0.01) & 1.44(0.03) & 1.7(0.0) & 1.2(0.0) \\
29 & G330.95$-$0.1 & 2.2 & 1.1 & 0.93(0.03) & 2.7(0.2) & 1.04(0.01) & 1.75(0.02) & 2.4(0.1) & 0.9(0.1) \\
30 & G332.09$-$0.4 & 0.9 & 1.6 & -- & -- & -- & -- & -- & -- \\
31 & G332.82$-$0.5 & 1.2 & 1.2 & 0.92(0.05) & 1.7(0.2) & 0.99(0.02)$^{*}$ & -- & 1.8(0.0) & -- \\
32 & G333.28$-$0.3 & 1.3 & 1.2 & 0.95(0.08) & 1.1(0.1) & 1.02(0.03)$^{*}$ & -- & 1.5(0.1) & -- \\
33 & G333.31+0.1 & 0.4 & 1.1 & 0.46(0.38) & 1.3(0.5) & 0.64(0.07) & 1.56(0.01) & 1.4(0.1) & 0.2(0.0) \\
34 & G335.79+0.1 & 0.8 & 1.1 & 0.41(0.10) & 1.0(0.1) & 0.84(0.02) & 1.49(0.02) & 1.9(0.1) & 0.9(0.0) \\
35 & G337.41$-$0.4 & 0.9 & 1.3 & 0.88(0.03) & 1.6(0.1) & 0.85(0.01) & 1.70(0.02) & 2.1(0.0) & 1.3(0.1) \\ 
36 & G337.70$-$0.0 & 2.1 & 1.3 & -- & -- & -- & -- & -- & -- \\
37 & G337.92$-$0.4 & 1.1 & 1.3 & 0.69(0.05) & 1.1(0.1) & 0.89(0.01) & 1.57(0.06) & 1.9(0.1) & 1.3(0.1) \\
38 & G339.62$-$0.1 & 0.4 & 1.5 & -- & -- & -- & -- & -- & -- \\
39 & G340.75$-$1.0 & 0.3 & 1.3 & -- & -- & -- & -- & -- & -- \\
40 & G341.22$-$0.2 & 0.6 & 1.2 & 0.35(0.28) & 1.2(0.2) & 0.64(0.05) & 1.55(0.04) & 1.3(0.1) & 0.7(0.1) \\
41 & G342.48+0.1 & 2.1 & 1.9 & -- & -- & -- & -- & -- & -- \\
42 & G343.13$-$0.0 & 0.7 & 1.3 & 0.83(0.02) & 1.8(0.1) & 0.90(0.02) & 1.76(0.02) & 2.2(0.1) & 0.8(0.1) \\
43 & G343.76$-$0.1 & 0.4 & 1.4 & 0.45(0.16) & 1.4(0.2) & 0.56(0.12) & 1.66(0.03) & 1.6(0.1) & 0.3(0.1) \\
44 & G344.23$-$0.5 & 0.5 & 1.5 & -- & -- & -- & -- & -- & -- \\
45 & G345.00$-$0.2 & 0.5 & 1.3 & 0.78(0.05) & 2.0(0.2) & 0.86(0.02) & 1.81(0.02) & 2.0(0.0) & 0.8(0.2) \\
46 & G345.49+0.3 & 0.7 & 1.1 & 0.93(0.03) & 1.0(0.1) & 0.94(0.02)$^{*}$ & -- & 1.7(0.0) & -- \\
47 & G345.50+0.3 & 0.7 & 1.6 & -- & -- & -- & -- & -- & -- \\
48 & G345.72+0.8 & 0.3 & 1.1 & 0.83(0.05) & 1.6(0.2) & 0.95(0.01) & 1.76(0.01) & 1.8(0.0) & 1.1(0.0) \\
49 & G351.16+0.7 & 0.6 & 1.2 & 0.90(0.04) & 1.2(0.1) & 0.95(0.03)$^{*}$ & -- & 1.6(0.1) & -- \\
50 & G351.58$-$0.3 & 1.4 & 1.1 & 0.61(0.14) & 1.0(0.1) & 0.88(0.03)$^{*}$ & -- & 2.0(0.1) & -- \\
51 & G351.77$-$0.5 & 0.3 & 1.1 & 0.94(0.08) & 1.6(0.3) & 1.02(0.01)$^{*}$ & -- & 2.2(0.0) & -- \\
52 & G353.41$-$0.3 & 1.3 & 1.1 & 1.19(0.05) & 1.8(0.4) & 1.00(0.01)$^{*}$ & -- & 1.7(0.0) & -- \\
\hline
\end{longtable}
\begin{flushleft}
\tablefoot{The columns are as follows: $S_{\mathrm{eff}}$: effective size; APR: aspect ratio; $r_{\mathrm{b}}$: break radius of the \ttcosixfive{} radial profile; $m$: power-law index of the \ttcosixfive{} radial profile; $r_{\mathrm{b}_1}$: inner break radius of the 160\,\mumeter{} radial profile. The sources where the single power-law model was used are marked with an asterisk ($^{*}$); $r_{\mathrm{b}_2}$: outer break radius of the 160\,\mumeter{} radial profile; $m_{\mathrm{i}}$: inner power-law index of the 160\,\mumeter{} radial profile; $m_{\mathrm{o}}$: outer power-law index of the 160\,\mumeter{} radial profile. The 1$\sigma$ uncertainties of the fitted parameters are shown in round brackets. Fitting results are only presented for those that have fitted parameters greater than their 1$\sigma$.} 
\end{flushleft}
\end{center}
\twocolumn

\onecolumn
\begin{center}
\begin{longtable}{cl|cc|cc|ccc}
    \caption{Mean velocity gradient (MVG) fitting results} \label{t:mvg_results} \\
    \hline 
    \hline 
    No. & Source & \multicolumn{2}{c|}{\ttcotwoone{}} & \multicolumn{2}{c|}{\ttcosixfive{}} & $\theta_{\mathrm{outflow}}$ & $\Delta \textrm{PA}_1$ & $\Delta \textrm{PA}_2$\\ \cline{3-6}
    & & $\mathcal{G}$ & $\theta_{\mathcal{G}}$ & $\mathcal{G}$ & $\theta_{\mathcal{G}}$ & \\
    & &  ($\times 10^{-2}$ km\,s$^{-1}$\,pc$^{-1}$) & (deg) & (km\,s$^{-1}$\,pc$^{-1}$) & (deg) & (deg) & (deg) & (deg) \\
    \hline 
1 & G14.11$-$0.5 & -- & -- & 1.3(0.4) & 66(20) & -- & -- & -- \\
2 & G14.63$-$0.5 & -- & -- & 0.8(0.2) & 3(19) & -- & -- & -- \\
3 & G31.41+0.3 & -- & -- & 4.1(0.2) & 0(3) & -- & -- & -- \\
4 & G34.26+0.1 & -- & -- & 4.5(0.3) & 12(4) & -- & -- & -- \\
5 & G34.40+0.2 & -- & -- & 6.5(1.4) & 89(11) & -- & -- & -- \\
6 & G34.41+0.2 & -- & -- & 7.2(2.1) & 191(17) & -- & -- & -- \\
7 & G43.17+0.0 & -- & -- & 0.5(0.1) & 48(13) & -- & -- & -- \\
8 & G53.14+0.0 & -- & -- & 7.3(0.7) & 310(6) & -- & -- & -- \\
9 & G326.66+0.5 & -- & -- & 5.1(0.3) & 264(3) & -- & -- & -- \\
10 & G351.16+0.7 & -- & -- & 3.2(0.4) & 129(7) & -- & -- & -- \\
11 & G351.77$-$0.5 & -- & -- & 20.7(1.4) & 225(4) & -- & -- & -- \\
\hline
12 & G13.18+0.0 & -- & -- & 2.8(0.9) & 358(25) & 35 & 37 & -- \\
13 & G19.88$-$0.5 & -- & -- & 3.5(0.4) & 230(7) & 115 & 115 & -- \\
14 & G35.20$-$0.7 & -- & -- & 2.9(0.4) & 136(9) & 200 & 64 & -- \\
15 & G332.82$-$0.5 & -- & -- & 3.9(0.2) & 141(3) & 120 & 21 & -- \\
16 & G340.75$-$1.0 & -- & -- & 4.7(1.2) & 311(15) & 90 & 139 & -- \\
17 & G344.23$-$0.5 & -- & -- & 6.5(0.6) & 12(6) & 35 & 23 & -- \\
18 & G345.72+0.8 & -- & -- & 7.7(0.6) & 255(4) & 190 & 65 & -- \\
\hline
19 & G10.47+0.0 & 79.1(5.1) & 195(4) & 0.9(0.2) & 177(14) & -- & -- & 18 \\
20 & G10.62$-$0.3 & 16.8(3.6) & 11(12) & 1.1(0.1) & 279(6) & -- & -- & 92 \\
21 & G30.82$-$0.0 & 158.7(9.1) & 30(3) & 4.2(0.3) & 89(6) & -- & -- & 59 \\
22 & G301.13$-$0.2 & 18.7(3.2) & 129(10) & 2.1(0.4) & 146(10) & -- & -- & 17 \\
23 & G305.56+0.0 & 31.8(9.0) & 261(16) & 0.5(0.1) & 266(20) & -- & -- & 5 \\
24 & G341.22$-$0.2 & 59.9(7.3) & 340(7) & 2.2(0.4) & 178(10) & -- & -- & 162 \\
25 & G345.49+0.3 & 168.0(9.8) & 140(3) & 4.8(0.4) & 141(5) & -- & -- & 1 \\
26 & G351.58$-$0.3 & 67.2(8.7) & 45(7) & 2.4(0.2) & 332(4) & -- & -- & 73 \\
\hline
27 & G305.21+0.2 & 120.7(11.9) & 337(6) & 3.4(0.2) & 257(5) & 5 & 108 & 80 \\
28 & G309.38$-$0.1 & 43.4(11.1) & 114(15) & 1.0(0.3) & 73(31) & 50 & 23 & 41 \\ 
29 & G310.01+0.3 & 158.1(8.3) & 8(3) & 3.0(0.4) & 352(10) & 335 & 17 & 16 \\
30 & G313.58+0.3 & 41.6(9.2) & 120(13) & 3.9(0.7) & 11(10) & 180 & 169 & 109 \\
31 & G330.95$-$0.1 & 35.7(3.3) & 238(5) & 3.2(0.3) & 243(5) & 260 & 17 & 5 \\
32 & G332.09$-$0.4 & 39.4(7.8) & 302(11) & 1.0(0.2) & 308(12) & 255 & 53 & 6 \\
33 & G333.31+0.1 & 42.1(4.9) & 104(7) & 4.7(0.9) & 139(11) & 150 & 11 & 35 \\
34 & G335.79+0.1 & 85.9(6.3) & 340(4) & 2.2(0.3) & 301(7) & 190 & 111 & 39 \\
35 & G337.41$-$0.4 & 63.8(7.0) & 208(6) & 2.7(0.2) & 181(6) & 130 & 51 & 27 \\
36 & G337.70$-$0.0 & 50.8(4.5) & 234(5) & 0.8(0.2) & 164(20) & 200 & 36 & 70 \\
37 & G337.92$-$0.4 & 75.5(16.0) & 127(12) & 3.0(0.3) & 90(7) & 80 & 10 & 37 \\
38 & G343.13$-$0.0 & 70.6(17.1) & 23(14) & 2.1(0.4) & 314(10) & 360 & 46 & 69 \\
39 & G343.76$-$0.1 & 152.6(18.3) & 67(7) & 5.1(1.0) & 238(10) & 180 & 58 & 171 \\
40 & G345.00$-$0.2 & 176.0(18.5) & 285(6) & 7.8(0.6) & 252(4) & 360 & 108 & 33 \\
41 & G345.50+0.3 & 197.3(13.2) & 43(4) & 3.4(0.4) & 110(7) & 70 & 40 & 67 \\
\hline
\end{longtable}
\begin{flushleft}
\tablefoot{The columns are as follows: $\mathcal{G}$: MVG magnitude; $\theta_{\mathcal{G}}$: position angle (PA) of the MVG direction; $\theta_{\mathrm{outflow}}$: PA of the \cosixfive{} outflow; $\Delta \textrm{PA}_1$: absolute angular offset between $\theta_{\mathrm{outflow}}$ and \ttcosixfive{}'s $\theta_{\mathcal{G}}$; $\Delta \textrm{PA}_2$: absolute angular offset between \ttcosixfive{}'s and \ttcotwoone{}'s $\theta_{\mathcal{G}}$. The 1$\sigma$ uncertainties of the fitted parameters are shown in round brackets. Fitting results are only presented for the sources whose MVG magnitudes are equal or greater than three times their uncertainties ($\mathcal{G} \geq 3\sigma_{\mathcal{G}}$).} 
\end{flushleft}
\end{center}
\twocolumn

%-----------------
\FloatBarrier
\section{Opacity of \ttcosixfive{} emission} \label{a:tau13co}
To assess how optically thick \ttcosixfive{} emission can be in our sources, we calculated the line opacity $\tau_{^{13}\mathrm{CO}}$ based on a ratio of \ttcosixfive{} to \cosixfive{} emission. For our calculation, we obtained averaged spectra from central 20\arcsec{} regions where the CO gas is expected to be densest. Assuming that the CO gas is in local thermodynamic equilibrium (LTE), \cosixfive{} emission is optically thick, and our sources fill the entire beam, $\tau_{^{13}\mathrm{CO}}$ can be estimated by following Eqn. (2) in \citet{leurini2013distribution}: 
\begin{equation}
\tau_{^{13}\mathrm{CO}} = 
- \mathrm{ln} \left( 1 - \frac{T_{^{13}\mathrm{CO}}}{T_{^{12}\mathrm{CO}}} \right),
\end{equation}
where $T_{^{13}\mathrm{CO}}$ and $T_{^{12}\mathrm{CO}}$ are the main-beam peak brightness temperatures of \ttcosixfive{} and \cosixfive{} emission.

Among 78 sources where both \ttcosixfive{} and \cosixfive{} data are available, we found 64 sources that have $\tau_{^{13}\mathrm{CO}} < 1$ and another eleven that have $\tau_{^{13}\mathrm{CO}} < 2$. For G301.13$-$0.2 (HII), $\tau_{^{13}\mathrm{CO}}$ could not be estimated because $T_{^{13}\mathrm{CO}}$ is greater than $T_{^{12}\mathrm{CO}}$. Overall, our results indicate that \ttcosixfive{} emission is generally optically thin and becomes mildly optically thick in a number of our sources. Finally, we note that our estimates are likely upper limits of the actual opacities since $T_{^{12}\mathrm{CO}}$ is 
often underestimated due to self-absorption.

%----------------
\FloatBarrier
\section{Extended $M_0$ maps} \label{a:extM0}
We show the ATLASGAL 870 \mumeter{} and \ttcosixfive{} maps of three example sources to illustrate the morphology of extended sources (Sect.\,\ref{s:M0_classification}).

\onecolumn
\begin{figure*}
    \centering
    \resizebox{0.7\textwidth}{!}{\includegraphics{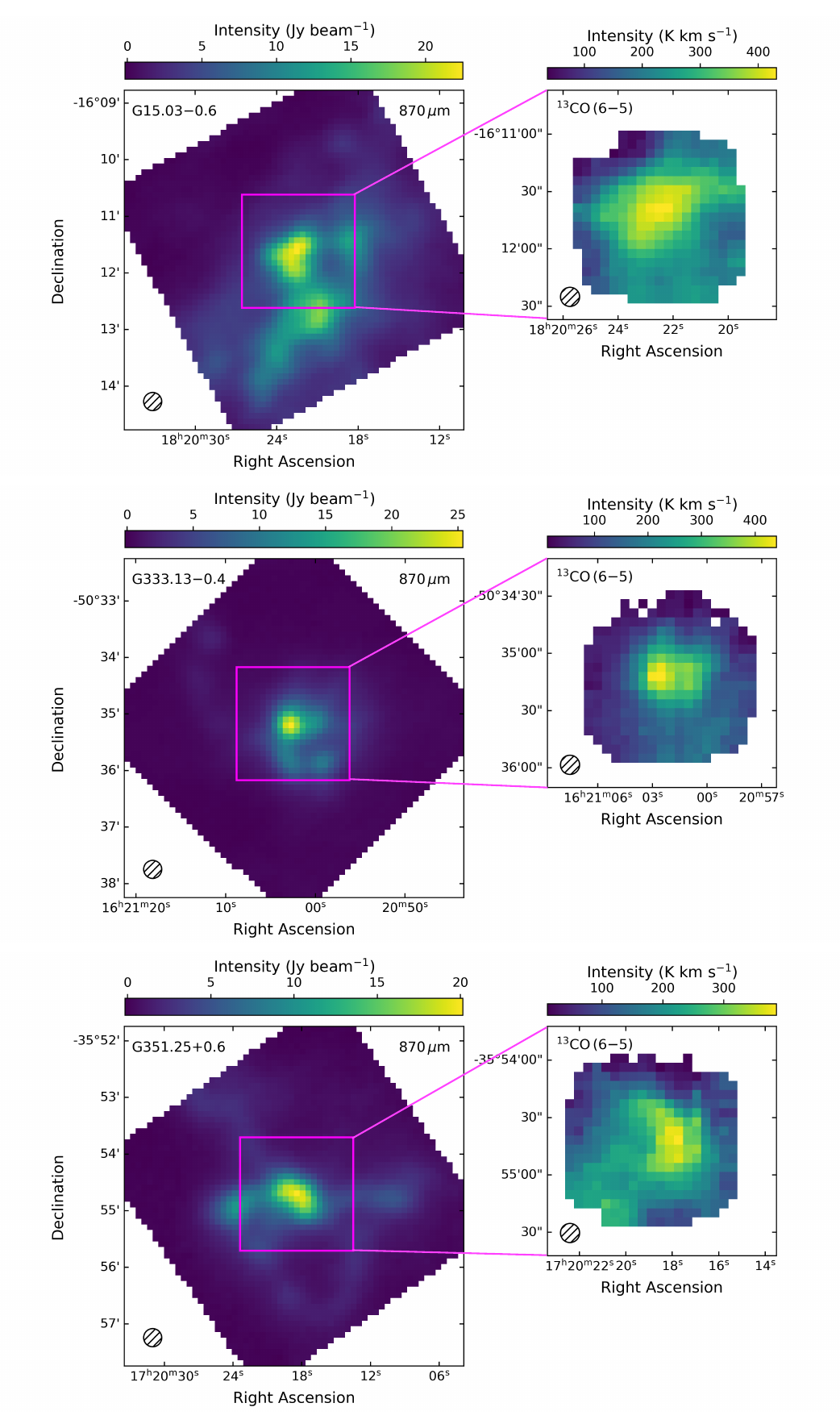}}
    \caption{The ATLASGAL 870\,\mumeter{} (left) and \ttcosixfive{} (right) maps of three example sources: G15.03$-$0.6, G333.13$-$0.4, and G351.25+0.6. The central $2^{\prime} \times 2^{\prime}$ box on the 870\,\mumeter{} image corresponds to the region probed by the \ttcosixfive{} data. The hatched circle at the bottom left of each map illustrates the full width at half maximum (FWHM) of the telescope beam.}
    \label{fig:3extended_clumps}
\end{figure*}
\twocolumn

%----------------
\FloatBarrier
\section{Masking emission from nearby sources} \label{a:mask}
\begin{figure}[h!]
    \resizebox{\hsize}{!}{\includegraphics{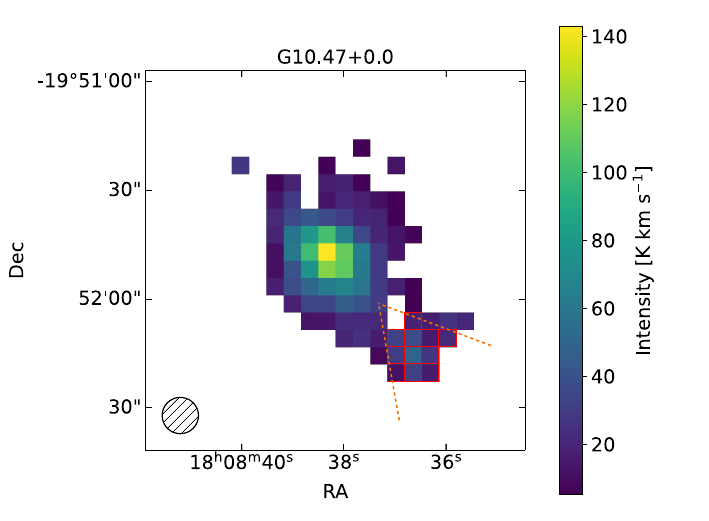}}
    \caption{Illustration of the procedure to mask out potentially contaminating pixels. The orange dashed lines are boundaries of the angular space that cover the nearby object. The masked pixels are shown with red edges.}
    \label{fig:exmpl_masked_core}
\end{figure}

Several of the \ttcosixfive{} and Hi-GAL 160\,\mumeter{} intensity images (four and 17, respectively) contain nearby objects that appear as additional peaks in the vicinity of the main cores. To minimise the contamination from these objects, we masked out all potentially affected pixels by applying the following procedure. Firstly, we visually selected a reference pixel at the edge of the central core that is likely not contaminated. Secondly, we masked out all the pixels within an angular space that starts from the reference pixel and engulfs the nearby object (Fig.\,\ref{fig:exmpl_masked_core}). The boundaries of the angular space were manually determined to fully cover the object.

%-----------------
\FloatBarrier
\section{Impact of angular resolution on the derivation of MVG} \label{a:mvg_multi_beam_size}
We examined the impact of angular resolution on the derivation of MVG by comparing the \ttcosixfive{} MVGs estimated on the 10$^{\prime\prime}$ and 30$^{\prime\prime}$ scales. For our examination, we smoothed the \ttcosixfive{} cube of an example source, G10.62$-$0.3, to an angular resolution of 30\arcsec{} by using the task \texttt{GAUSS\_SMOOTH} in CLASS, then produced a corresponding $M_1$ map in the same manner as in Sect.\,\ref{s:mmtmaps}. A new MVG was then estimated as in Sect.\,\ref{s:mvg} and compared to the original MVG (Sect.\,\ref{s:mvg}).

\begin{figure}    
   \begin{subfigure}{\hsize}
   \hspace{0.5cm}
      \resizebox{0.9\hsize}{!}{\includegraphics{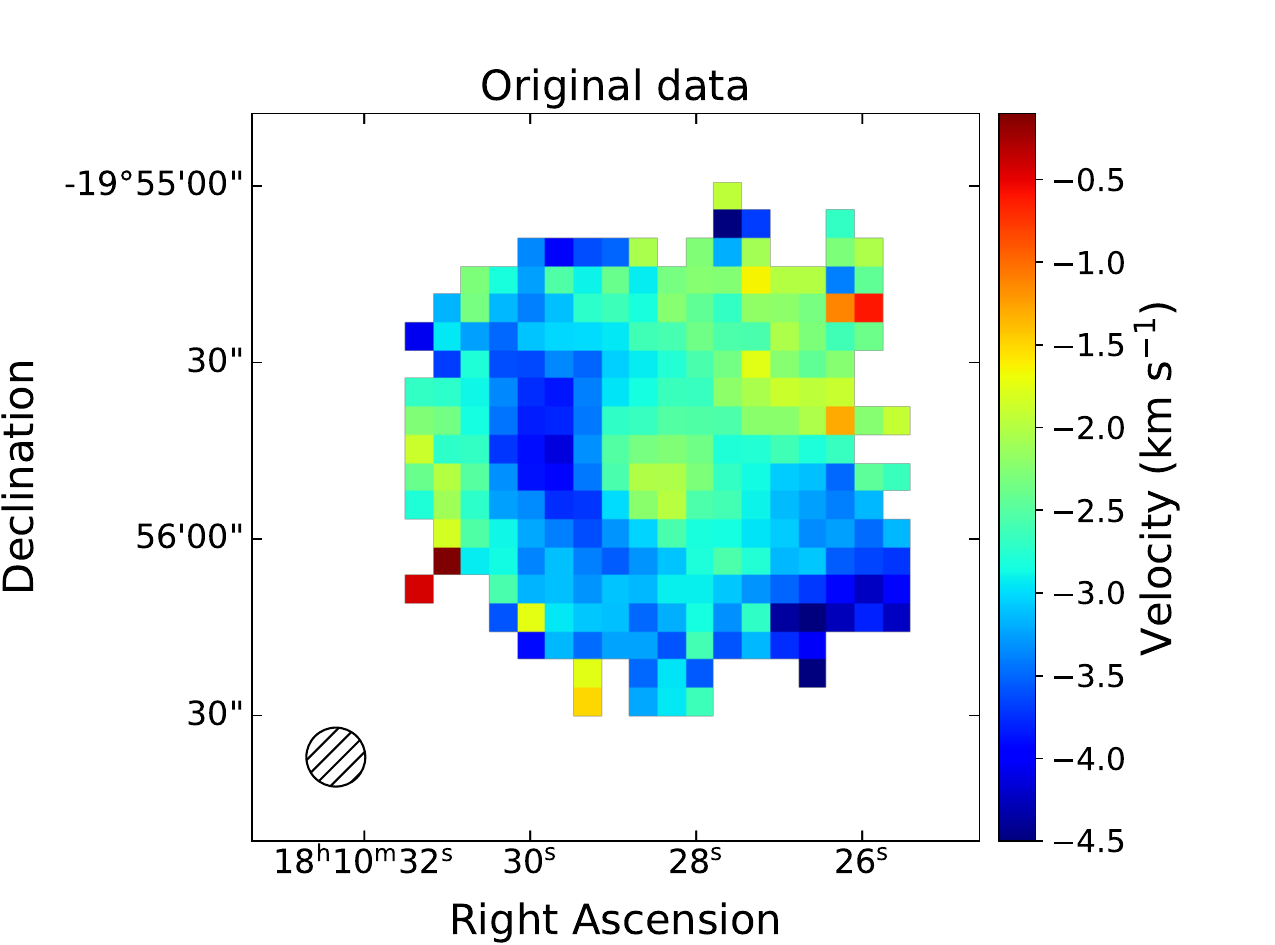}}
   \end{subfigure}
   \begin{subfigure}{\hsize}
   \hspace{0.5cm}
      \resizebox{0.9\hsize}{!}{\includegraphics{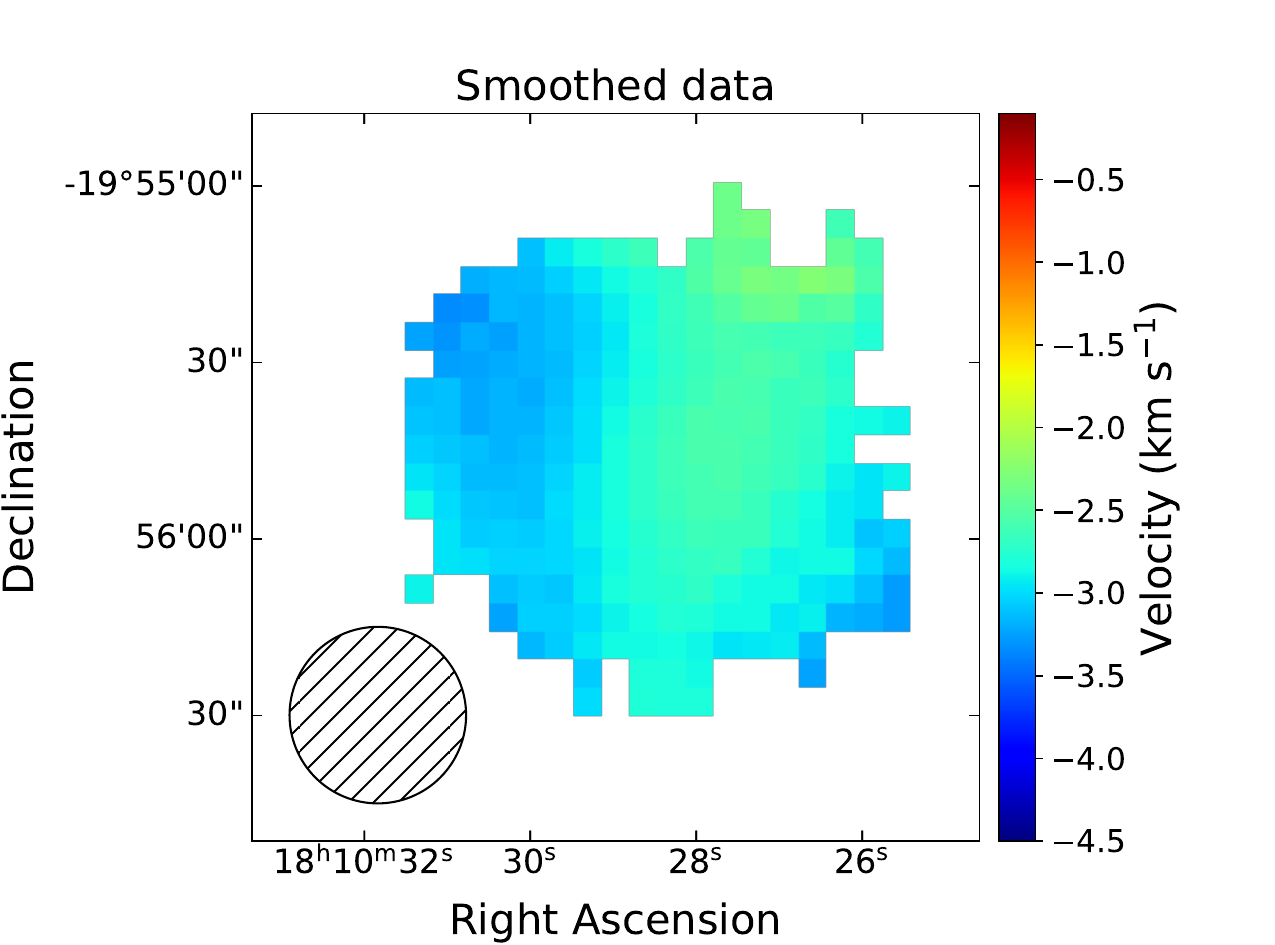}}
   \end{subfigure}
   \caption{The $M_1$ maps of G10.62$-$0.3 at two different angular resolutions: 10$^{\prime\prime}$ (left) and 30$^{\prime\prime}$ (right). The FWHM of the beam size is presented by a hatched circle at the bottom left of each map.}
   \label{fig:M1_maps_2beams}
\end{figure}

Table\,\ref{t:compare_MVG_2beams} shows that the MVG magnitude is decreased by a factor of two at a three times coarser angular resolution, while the MVG direction remains consistent. This result can also be seen in Fig.\ref{fig:M1_maps_2beams} where small-scale velocity variations seem to vanish on the smoothed data, hence the lower magnitude of the MVG.

\begin{table}
\centering
\caption{Comparison of the MVG on 10$^{\prime\prime}$ (original) and 30$^{\prime\prime}$ (smoothed) scales}
\label{t:compare_MVG_2beams}
\begin{tabular}{ c | c c c }
    \hline
    \hline 
    & $v_0$ & $\mathcal{G}$ & $\theta_{\mathcal{G}}$ \\
    & (km s$^{-1}$) & (km s$^{-1}$\,pc$^{-2}$) & (degree) \\
    \hline 
    Original data & $-$2.76(0.03) & 1.09(0.12) & 279(6) \\
    Smoothed data & $-$2.81(0.01) & 0.48(0.02) & 272(3)\\
    \hline
\end{tabular}
\tablefoot{The columns are as follows: $v_0$: velocity at the reference position; $\mathcal{G}$: MVG magnitude; $\theta_{\mathcal{G}}$: PA of the MVG direction. The 1$\sigma$ uncertainty of each parameter is shown in round brackets.}
\end{table}

%------------------
\FloatBarrier
\newpage
\section{Comparison of gas kinematics from three tracers} \label{a:15clumps}
We compare gas kinematics traced by three transitions: \ttcosixfive{}, \cosixfive{}, and \ttcotwoone{}. Among 15 sources where the data for all three lines are available, maps of 11 are presented here, while that of the other four are shown in Fig.\,\ref{fig:3tracers}.

\begin{figure*}
    \centering
    \resizebox{1.0\hsize}{!}{\includegraphics{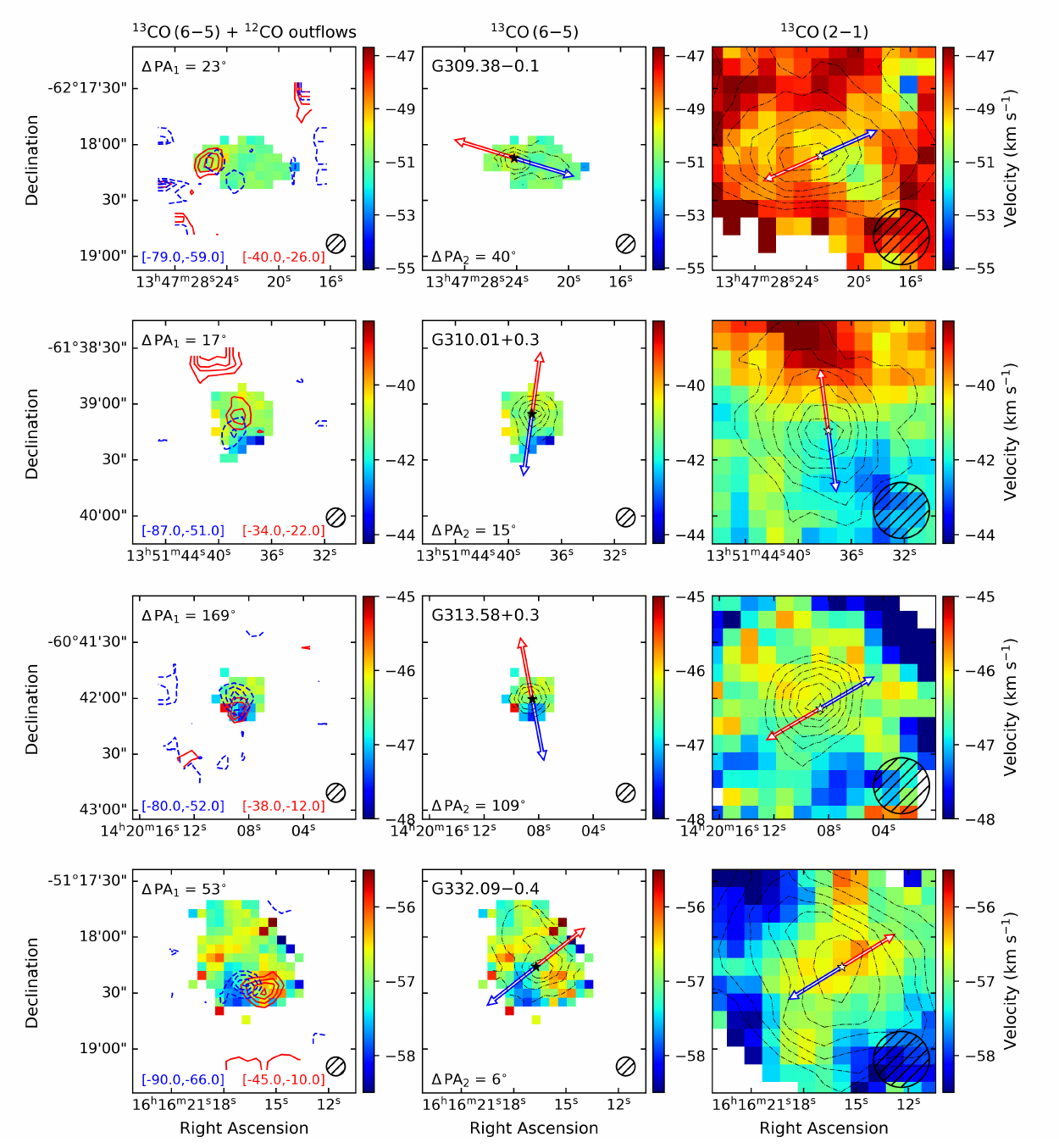}}
    \caption{Comparison of \ttcosixfive{}, \cosixfive{}, and \ttcotwoone{} gas kinematics. The symbols and colours are the same as in Fig.\,\ref{fig:3tracers}.}
    \label{fig:3tracers_app1}
\end{figure*}

\begin{figure*}
    \centering
    \resizebox{1.0\hsize}{!}{\includegraphics{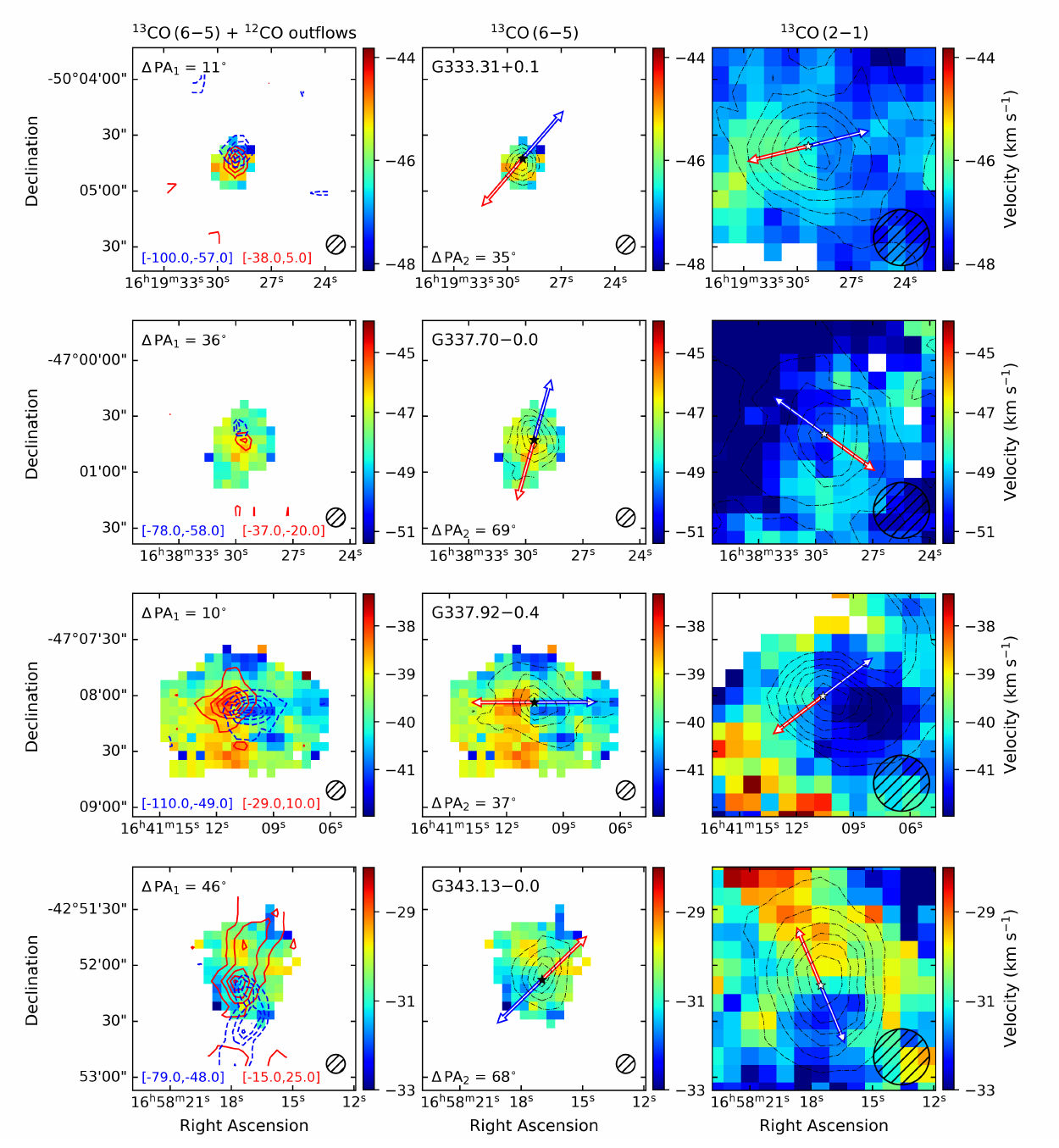}}
    \caption{Comparison of \ttcosixfive{}, \cosixfive{}, and \ttcotwoone{} gas kinematics. The symbols and colours are the same as in Fig.\,\ref{fig:3tracers}.}
    \label{fig:3tracers_app2}
\end{figure*}

\begin{figure*}
    \centering
    \resizebox{1.0\hsize}{!}{\includegraphics{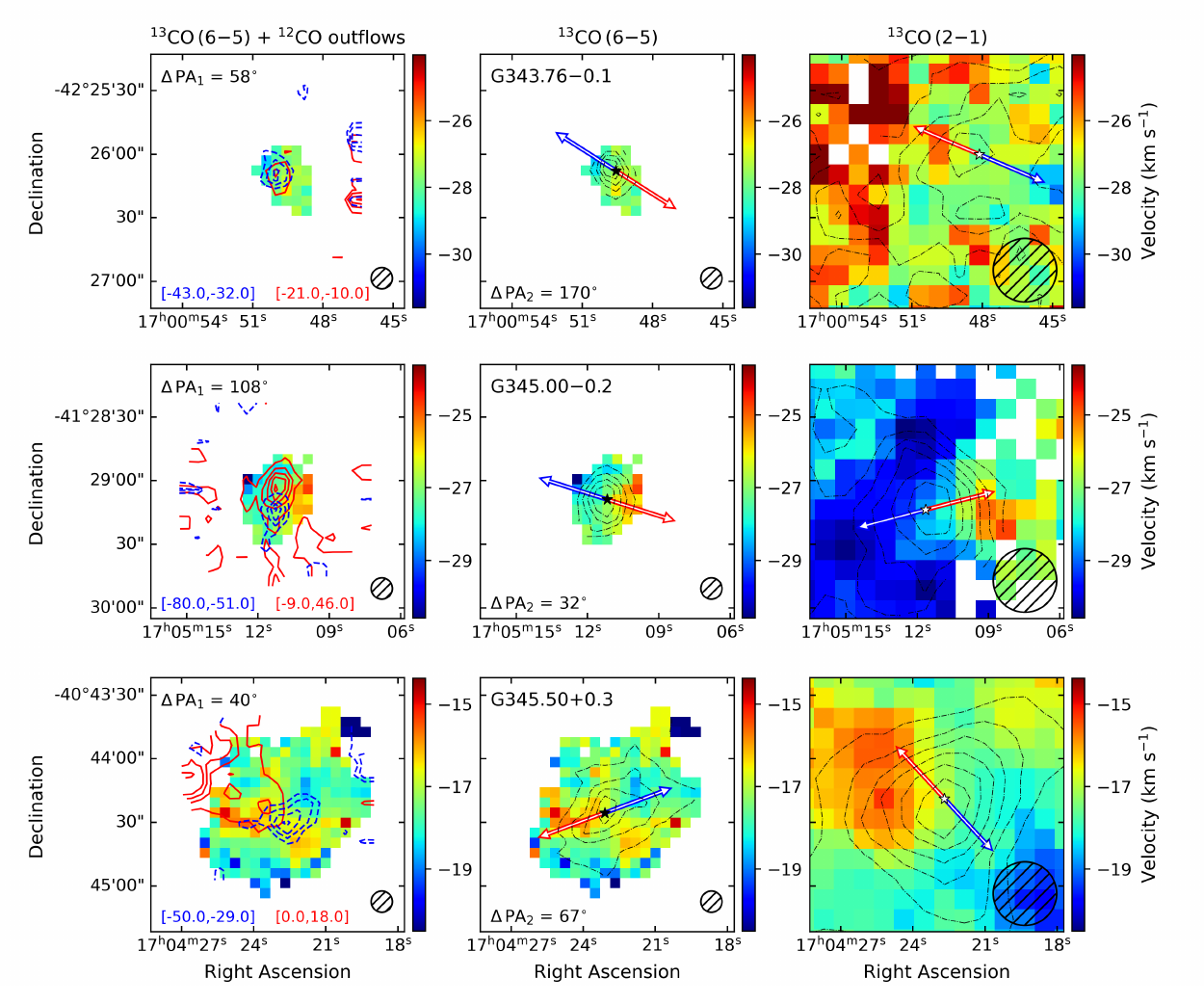}}
    \caption{Comparison of \ttcosixfive{}, \cosixfive{}, and \ttcotwoone{} gas kinematics. The symbols and colours are the same as in Fig.\,\ref{fig:3tracers}.}
    \label{fig:3tracers_app3}
\end{figure*}

\end{document}